\documentclass[aps,prd,twocolumn,10pt,superscriptaddress,amssymb,amsmath,nofootinbib]{revtex4-2}

\usepackage[utf8]{inputenc}
\usepackage{lmodern}
\usepackage[T1]{fontenc}
\usepackage{graphicx} 
\usepackage[pdftex,breaklinks,colorlinks,
linkcolor=Blue,
citecolor=teal,
anchorcolor=red,
urlcolor=cyan,
pdfencoding=auto,
pdftitle={Effective source for second-order self-force calculations: quasicircular orbits in Schwarzschild spacetime},
pdfauthor={Samuel D. Upton, Barry Wardell, Adam Pound, Niels Warburton, Leor Barack},
pdfdisplaydoctitle]{hyperref}
\usepackage[dvipsnames]{xcolor}
\usepackage{mathtools}
\usepackage{mathrsfs}
\usepackage{booktabs}

\usepackage{physics2}
\usepackage[italic=true]{derivative}

\usepackage{orcidlink}

\newcommand{\e}{\varepsilon}
\newcommand{\order}{\mathcal{O}}
\newcommand{\barh}{\bar{h}}
\newcommand{\barT}{\bar{T}}
\newcommand{\adot}{\dot{a}}
\newcommand{\rodot}{\dot{r}_0}

\newcommand{\R}{\mathrm{R}}
\renewcommand{\S}{\mathrm{S}}
\newcommand{\ms}{\mathrm{ms}}
\newcommand{\s}{{\sf s}}
\renewcommand{\r}{{\sf r}}
\newcommand{\geff}{\tilde{g}}
\newcommand{\Teff}{T}
\newcommand{\TeffTR}{\barT}

\newcommand{\calR}{\mathcal{R}}
\newcommand{\calP}{\mathcal{P}}
\newcommand{\force}{f}
\newcommand{\deltar}{r_{1}}
\newcommand{\Ric}{\text{Ric}}
\newcommand{\calJ}{\mathcal{J}}
\newcommand{\hring}{\mathring{h}}

\newcommand{\pdvel}{\vec{\partial}_{\mathcal{V}}}

\DeclareMathOperator{\diag}{diag}
\DeclarePairedDelimiter{\ceil}{\lceil}{\rceil}

\newcommand{\beq}{\begin{equation}}
\newcommand{\eeq}{\end{equation}}

\newcommand{\ilm}{i \ell m}

\begin{document}

\title{Effective source for second-order self-force calculations: quasicircular orbits in Schwarzschild spacetime}

\author{Samuel D. Upton\,\orcidlink{0000-0003-2965-7674}}
\affiliation{School of Mathematical Sciences and STAG Research Centre, University of Southampton, Southampton, United Kingdom, SO17 1BJ}

\author{Barry Wardell\,\orcidlink{0000-0001-6176-9006}}
\affiliation{School of Mathematics and Statistics, University College Dublin, Belfield, Dublin 4, Ireland}

\author{Adam Pound\,\orcidlink{0000-0001-9446-0638}}
\affiliation{School of Mathematical Sciences and STAG Research Centre, University of Southampton, Southampton, United Kingdom, SO17 1BJ}

\author{Niels Warburton\,\orcidlink{0000-0003-0914-8645}}
\affiliation{School of Mathematics and Statistics, University College Dublin, Belfield, Dublin 4, Ireland}

\author{Leor Barack\,\orcidlink{0000-0003-4742-9413}}
\affiliation{School of Mathematical Sciences and STAG Research Centre, University of Southampton, Southampton, United Kingdom, SO17 1BJ}

\begin{abstract}
    Recent years have seen the first production of  ``post-adiabatic'' gravitational-waveform models based on second-order gravitational self-force theory. These models rely on calculations of an effective source in the perturbative second-order Einstein equation. Here, for the first time, we detail the calculation of the effective source in a Schwarzschild background, which underlies the second-order self-force results in [Phys. Rev. Lett. 127, 151102 (2021); ibid. 128, 231101 (2022); ibid. 130, 241402 (2023)]. The source is designed for use in the multiscale form of the Lorenz-gauge Einstein equation, decomposed in tensor spherical harmonics, or in the analogous second-order Teukolsky equation. It involves, among other things, contributions from (i) quadratic coupling of first-order field modes, (ii) the slow evolution of first-order fields, (iii) quadratic products of a first-order puncture field, and (iv) the second-order puncture field. We validate each of these pieces through numerical and analytical tests.
\end{abstract}

\maketitle

\tableofcontents

\section{Introduction}

After several decades of development~\cite{Barack:2018yvs}, gravitational self-force theory has emerged as a practical method of building gravitational-waveform models for compact binaries~\cite{LISAConsortiumWaveformWorkingGroup:2023arg,Abac:2025saz}. Waveform models based on the self-force approach are now fast enough for data analysis~\cite{Katz:2021yft,Islam:2022laz,Nasipak:2023kuf}, and they have proved to be accurate over a large portion of the binary parameter space~\cite{Albertini:2022rfe,LeTiec:2014oez,vandeMeent:2020xgc,Ramos-Buades:2022lgf}. 

The method is based on an expansion of the spacetime metric in powers of the mass ratio $\e\coloneqq m/M$, making it particularly suited to asymmetric binaries, in which one body (the secondary, of mass $m$) is much less massive than the other (the primary, of mass $M$). Most prominently, self-force theory is the standard method of modelling extreme-mass-ratio inspirals (EMRIs) in galactic cores, with mass ratios $\e\sim 10^{-4}$--$10^{-7}$~\cite{LISAConsortiumWaveformWorkingGroup:2023arg}. Traditional scaling arguments~\cite{Rosenthal:2006iy,Hinderer:2008dm} have long suggested that in order to accurately model binary inspirals using this method, the expansion of the metric must be carried to second order in~$\e$. This expectation has been borne out by data analysis studies~\cite{Burke:2023lno} for EMRIs and intermediate-mass-ratio inspirals (IMRIs).

Waveform modelling at both first and second order in~$\e$ has been facilitated by the separation of time scales in these asymmetric systems: the inspiral is slow, occurring on a time scale of order $M/\e$, compared to the orbital period of order $M$. As a consequence, the Einstein equations admit a multiscale expansion~\cite{Hinderer:2008dm,Miller:2020bft,Pound:2021qin,Miller:2023ers,Mathews:2025nyb,Wei:2025lva}, which separates them into field equations on the orbital time scale, together with a small number of ordinary differential equations~(ODEs) that govern the binary's slow evolution. Such a split enables rapid waveform generation by allowing one to pre-compute waveform ingredients in advance by solving the fast-time field equations on a grid of binary parameter values~\cite{Hughes:2021exa,*[Erratum: ]Hughes:2021exaErratum,Katz:2021yft}. The multiscale expansion also leads to more tractable field equations and avoids errors that grow with number of waveform cycles.

In the multiscale expansion, orders in $\e$ are counted relative to the leading, ``adiabatic'' order (0PA). Effects arising from second-order ($\sim\e^2$) metric perturbations first enter into the waveform at first post-adiabatic order (1PA). Calculations of these second-order effects, and construction of the resulting 1PA waveform model, were carried out for the first time in the series of letters~\cite{Pound:2019lzj,Warburton:2021kwk,Wardell:2021fyy}, restricted to the special case of quasicircular orbits of a nonspinning secondary around a nonspinning (or slowly spinning) primary black hole. For that special case, the leading, zeroth-order spacetime is the Schwarzschild metric of the primary as if it were isolated. 

These calculations built on development of the underlying second- (and higher-) order self-force formalism~\cite{Rosenthal:2006iy,Pound:2009sm,Harte:2011ku,Detweiler:2011tt,Pound:2012nt,Gralla:2012db,Pound:2012dk,Pound:2017psq,Upton:2021oxf,Harte:2025tmd} and on the development of numerous practical tools~\cite{Warburton:2013lea,Pound:2014xva,Wardell:2015ada,Pound:2015wva,Miller:2016hjv,Miller:2020bft,Durkan:2022fvm,Spiers:2023mor,Miller:2023ers,Cunningham:2024dog}. In Refs.~\cite{Miller:2020bft,Bonetto:2021exn,Spiers:2023mor,Miller:2023ers,Cunningham:2024dog}, we have detailed many of the specific tools that were used in the second-order calculations in Refs.~\cite{Pound:2019lzj,Warburton:2021kwk,Wardell:2021fyy}. These include the following: our multiscale formulation of the field equations through second order in $\e$~\cite{Miller:2020bft,Miller:2023ers}; our method of solving those field equations~\cite{Miller:2023ers}; our method of computing quadratic source modes from modes of the first-order metric perturbation~\cite{Spiers:2023mor}; our method of computing the slow evolution of the first-order metric perturbation~\cite{Durkan:2022fvm,Miller:2023ers}, which acts as another source for the second-order metric perturbation; our method of deriving physical boundary conditions at large distances~\cite{Cunningham:2024dog}; and our methods of extracting information at the horizon~\cite{Bonetto:2021exn} and future null infinity~\cite{Cunningham:2024dog}. 

In the present paper, we describe another aspect of our second-order calculations: the construction of the effective source. All second-order calculations to date have been based on a puncture scheme, in which the secondary object is replaced by a ``puncture'', a local approximation to the singular field that encodes the object's multipole structure and diverges on a worldline that represents the object's trajectory. This puncture piece of the metric is moved to the right-hand side of the field equations, where it is combined with the physical source to create an effective source, building on methods originally developed for the linear problem~\cite{Barack:2007jh,Vega:2007mc,Wardell:2015ada}. One then solves the field equations for a residual field, which is regular along the secondary's trajectory.

The puncture field was given in covariant form, for generic motion in a generic background spacetime, in Ref.~\cite{Pound:2014xva}. Here we describe its conversion into the concrete form needed for our practical calculations. Our formulation of the field equations, as described in Refs.~\cite{Miller:2020bft,Miller:2023ers}, is in a fully separated form based on a tensor-spherical-harmonic decomposition, which serves to reduce the fast-time field equations to radial ODEs. Putting the puncture in practical form hence involves (i) expanding it in multiscale form, (ii) specializing it to quasicircular orbits in Schwarzschild spacetime, and (iii) decomposing it into tensor spherical harmonics. 

Following this construction of the puncture, we then demonstrate how it is combined with the physical source to obtain the effective source. Generically, the physical second-order source away from the puncture singularity involves two terms: a term (proportional to the second-order Ricci tensor) made up of quadratic products of the first-order metric perturbation; and a term involving the slow evolution of the first-order metric perturbation. The latter is calculated straightforwardly, using as input the data for the field's slow evolution from Refs.~\cite{Durkan:2022fvm,Miller:2023ers}. Over most of the spacetime, the former, quadratic source term is also calculated (relatively) easily; its spherical-harmonic modes are calculated from modes of the first-order metric perturbation using the mode-coupling formulas from Ref.~\cite{Spiers:2023mor}. However, this approach breaks down in a vicinity of the puncture singularity, where it encounters the problem of infinite mode coupling~\cite{Miller:2016hjv}: near the singularity, the first-order field modes converge slowly with increasing spherical-harmonic mode number, such that arbitrarily many first-order modes are required to accurately compute any single mode of the quadratic source. We overcome this obstruction using the method developed for a scalar toy model in Ref.~\cite{Miller:2016hjv}.

Although we focus on describing the source for the second-order Lorenz-gauge field equations, we also carry our calculations of the puncture to higher order (in powers of distance from the singularity) than is strictly necessary for those field equations. This is motivated by a desire to use our results in the second-order Teukolsky equation as well~\cite{Spiers:2023cip,Spiers:2023mor}. Working with the Teukolsky formalism has significant advantages in that fluxes can be calculated from the solution to a single (complex) scalar wave equation, but it involves a more singular source and higher derivatives of the metric perturbations, demanding a higher-order puncture field. Our implementation of the second-order Teukolsky formalism, detailed elsewhere~\cite{Leather:2025InPrep}, will rely on our results here.

Our calculations make extensive use of the formulas in the \textsc{PerturbationEquations} Mathematica package~\cite{PerturbationEquations} associated with Ref.~\cite{Spiers:2023mor}. We also make our numerical infrastructure publicly available in the \textsc{h1Lorenz}~\cite{h1Lorenz}, \textsc{SecondOrderRicci}~\cite{SecondOrderRicci} and \textsc{SecondOrderRicciSS}~\cite{SecondOrderRicciSS} codes.

The paper is organized as follows. In Sec.~\ref{sec:review} we review the field equations through second order, expanded in multiscale form and in a tensor spherical harmonic decomposition. In Sec.~\ref{sec:punctures} we describe the puncture fields, their multiscale expansion, and their decomposition in tensor spherical harmonics. Sections~\ref{sec:d2R} and~\ref{sec:slow_time} describe our calculation of the second-order Ricci tensor and the source term arising from the system's slow evolution. Section~\ref{sec:validation} details many of the checks we perform on the various source contributions, and Sec.~\ref{sec:effsource} presents the total source containing all contributions. We conclude in Sec.~\ref{sec:conclusion} with a summary and a discussion of future work. We use geometric units with $c=G=1$.

\section{Second order self-force theory in a multiscale expansion}
\label{sec:review}

Our calculations are based on the multiscale formulation of the Lorenz-gauge Einstein equations in Refs.~\cite{Miller:2020bft,Miller:2023ers}. However, our derivation of the second-order puncture field, and many of the tests we perform, draw upon the self-consistent formulation of self-force theory~\cite{Pound:2009sm,Pound:2015fma} so we take that as our starting point.

\subsection{Self-consistent self-force theory}

In the self-consistent formulation, the metric is expanded as
\begin{equation}
    {\sf g}_{\mu\nu} = g_{\mu\nu} + \e h^{1}_{\mu\nu} + \e^{2}h^{2}_{\mu\nu} + \order(\e^3). \label{eq:g expansion}
\end{equation}
In most of this paper, the background metric $g_{\mu\nu}$ is the Schwarzschild metric of mass $M$, though in Secs.~\ref{sec:punctures_generic} and~\ref{sec:punctures_twotimescale} we will allow it to be a Kerr metric of generic mass and spin. 

The perturbations $h^n_{\mu\nu}$ satisfy the trace-reversed Einstein field equations $R_{\mu\nu}[{\sf g}]=8\pi \bar T_{\mu\nu}$ in Lorenz gauge in the form
\begin{align}
    E_{\mu\nu}[h^1] ={}& -16\pi \TeffTR^1_{\mu\nu}, \label{eq:EFE1} \\
    E_{\mu\nu}[h^2]  ={}& -16\pi\TeffTR^2_{\mu\nu}+2\delta^{2}R_{\mu\nu}[h^1,h^1]. \label{eq:EFE2}
\end{align}
Here
\begin{equation}
    E_{\mu\nu}[h] \coloneqq \Box h_{\mu\nu} + 2R_{\mu}{}^{\alpha}{}_{\nu}{}^{\beta}h_{\alpha\beta}\label{eq:Edef}
\end{equation}
is the Lorenz-gauge linearized Ricci tensor $\delta R_{\mu\nu}[h] = -\frac{1}{2}E_{\mu\nu}[h]$ for any $h_{\mu\nu}$, with $\Box\coloneqq g^{\mu\nu}\nabla_\mu\nabla_\nu$, with $\nabla_\mu$ the covariant derivative compatible with $g_{\mu\nu}$.  $\delta^{2}R_{\mu\nu}[h,h]$ is the quadratic term in the expansion of the Ricci tensor, given explicitly in Eq.~(10) of Ref.~\cite{Spiers:2023mor} and available in multiple forms in the \textsc{PerturbationEquations} package~\cite{PerturbationEquations}.

In these field equations, the trace-reversed source terms $\TeffTR^n_{\mu\nu}$ are obtained from the Detweiler stress-energy tensor~\cite{Detweiler:2011tt,Upton:2021oxf}, which represents the secondary as a point mass moving on a worldline $z^\alpha$:
\begin{equation}
    \Teff_{\mu\nu} = m \int\tilde{u}_{\mu}\tilde{u}_{\nu}\frac{\delta^4(x^\alpha-z^\alpha(\tilde\tau))}{\sqrt{-\tilde{g}}}\odif{\tilde\tau}; \label{eq:T_def}
\end{equation}
see Appendix~\ref{sec:T_eff_TR_deriv}. The worldline obeys a geodesic equation through order~$\e^2$~\cite{Pound:2012nt,Pound:2017psq},
\begin{equation}
    \frac{\tilde{D}^2z^\alpha}{\odif{\tilde{\tau}}^2} = \order(\e^3), \label{eq:EoM_eff}
\end{equation} 
not in the physical metric but in a certain \emph{effective} metric,
\begin{equation}
    \tilde{g}_{\mu\nu} = g_{\mu\nu} + \e h^{\R1}_{\mu\nu} + \e^{2}h^{\R2}_{\mu\nu} + \order(\e^3), \label{eq:g_eff}
\end{equation}
which is smooth at the particle's worldline and obeys the vacuum Einstein equation there. In the above expressions, geometrical quantities and normalizations are all defined from the effective metric: $\tilde\tau$ is proper time in $\tilde g_{\mu\nu}$, $\tilde u^\alpha \coloneqq dz^\alpha/d\tilde\tau$ is the particle's four-velocity normalized in $\tilde g_{\mu\nu}$, $\tilde u_\mu \coloneqq \tilde g_{\mu\nu}\tilde u^\nu$, and $\tilde D/d\tilde\tau \coloneqq \tilde u^\mu\tilde\nabla_\mu$ is the effective-metric-compatible covariant derivative along the worldline. If we substitute the expansion~\eqref{eq:g_eff}, then the geodesic equation in the effective metric becomes an accelerated equation of motion in the background metric~\cite{Pound:2015fma},
\begin{equation}
    \frac{D^2z^\alpha}{\odif{\tau}^2} = \e \force^\alpha_1 + \e^{2} \force^\alpha_{2} + \order(\e^3), \label{eq:EoM}
\end{equation}
where the self-forces on the right-hand side are constructed from the \emph{regular} fields $h^{\R n}_{\mu\nu}$.

There are several subtleties we highlight in Eqs.~\eqref{eq:EFE1}--\eqref{eq:EFE2}. First, the field equations are written in a specific form of the Lorenz gauge, in which the total metric perturbation $h_{\mu\nu}\coloneqq\sum_{n\geq1} \e^n h^n_{\mu\nu}$ satisfies the condition
\begin{equation}
    Z_{\mu}[\barh] = \order(\e^3), \label{eq:Z_cond}
\end{equation}
where $\barh_{\mu\nu}\coloneqq (\delta^\alpha_\mu\delta^\beta_\nu - \frac{1}{2}g_{\mu\nu}g^{\alpha\beta})h_{\alpha\beta}$ is the trace reversal with respect to the background metric, and
\begin{equation}
    Z_{\mu}[\barh] \coloneqq g^{\alpha\beta}\nabla_{\alpha}\barh_{\mu\beta}. \label{eq:Zdef}
\end{equation}
Crucially, this gauge condition holds for the sum of the perturbations but not for individual perturbations: $Z_\mu[h^n]\neq0$. See Ref.~\cite{Pound:2009sm} for extensive discussion.

Second, the source $\TeffTR^n_{\mu\nu}$ in the $n$th-order field equation is the order-$\e^n$ term in the trace-reversed Detweiler stress-energy tensor, where the expansion~\eqref{eq:g_eff} is substituted into Eq.~\eqref{eq:T_def} but the source orbit $z^\alpha$ is \emph{not} expanded. This approach is referred to as self-consistent because the field equations~\eqref{eq:EFE1}--\eqref{eq:EFE2} and \eqref{eq:EoM} are solved as a coupled system, without perturbatively expanding the particle's trajectory. 

Third, the trace reversal of the stress-energy tensor is taken with respect to the effective metric,
\begin{equation}
    \TeffTR_{\mu\nu} = \left(\delta^\alpha_\mu\delta^\beta_\nu - \frac{1}{2}\geff_{\mu\nu}\geff^{\alpha\beta}\right)\Teff_{\alpha\beta} + \order(\e^3),\label{eq:T_eff_TR_intro}
\end{equation}
rather than with the background metric, $g_{\mu\nu}$. This is derived and demonstrated in Appendix~\ref{sec:T_eff_TR_deriv}.

Fourth, the evolution of the primary black hole must be carefully accounted for in the field equations. Due to absorption of radiation, the primary's mass and spin grow by an amount $\sim\e$ over the course of the inspiral. We write the black hole's physical mass and spin as 
\begin{align}
M_{\rm BH} &= M+\e \delta M,\\
S_{\rm BH} &= \e \delta S. 
\end{align}
While the background metric has a constant mass $M$ and vanishing spin, the mass and spin perturbations lead to terms $h^{1,\delta M}_{\mu\nu}$ and $h^{1,\delta S}_{\mu\nu}$ in the first-order metric perturbations. These can be correctly included in the self-consistent scheme through specification of boundary conditions at the black hole horizon~\cite{Miller:2023ers} or through the introduction of an effective source there~\cite{Lewis:2025ydo}.

Finally, we emphasize that Eq.~\eqref{eq:EFE2} has never been solved directly, nor has there ever been direct use of the stress-energy tensor $\TeffTR^{2}_{\mu\nu}$. This is because the second-order Ricci tensor $\delta^2R_{\mu\nu}[h^1,h^1]$ is too singular to be simply integrated over to obtain the field $h^{2}_{\mu\nu}$; writing the field equation in the form~\eqref{eq:EFE2} requires a subtle distributional definition of $\delta^2R_{\mu\nu}[h^1,h^1]$~\cite{Upton:2021oxf}, which has never been used in practice. Instead, the equations solved in Refs.~\cite{Pound:2021qin,Warburton:2021kwk} were equations for \emph{residual}\footnote{Note that we use two sets of terminology for two closely-related splits of the retarded field. We have singular (S) and regular (R) fields, which are formally defined such that the regular field obeys the vacuum Einstein equation and is perfectly smooth at the particle's worldline. In practice, we only ever have access to a local approximation to the singular field which we denote the puncture ($\calP$) field. Subtracting this from the retarded field yields the residual ($\calR$) field. Many of the equations we give (particularly those in Sec.~\ref{sec:punctures}) hold to all orders in this local approximation. To emphasise this fact, we refer to them as singular and regular fields in that context.} fields $h^{\calR n}_{\mu\nu}\coloneqq h^n_{\mu\nu}-h^{\calP n}_{\mu\nu}$:
\begin{align}
    E_{\mu\nu}[h^{\calR 1}] ={}&-E_{\mu\nu}[h^{\calP 1}], \label{eq:EFE1 res alt} \\
    E_{\mu\nu}[h^{\calR2}]  ={}& 2\delta^{2}R_{\mu\nu}[h^1,h^1] - E_{\mu\nu}[h^{\calP 2}].\label{eq:EFE2 res alt}
\end{align}
Here $h^{\calP n}_{\mu\nu}$ are the \emph{puncture} fields alluded to in the Introduction, and the right-hand sides are the effective sources. The punctures are defined from the field \emph{outside} the secondary and then analytically extended down to the representative worldline $z^\mu$ where they diverge~\cite{Pound:2012dk,Pound:2014xva}. These puncture fields are moved from the left-hand side of the field equations to the right, where they cancel (in particular) the singularity in $\delta^{2}R_{\mu\nu}[h^1,h^1]$. The sources on the right-hand side are first defined at points away from $z^\mu$ and then promoted to the whole domain as integrable functions, as explained most thoroughly in Ref.~\cite{Upton:2021oxf} (see also Ref.~\cite{Gralla:2012db}, for example). 

If we adopt the distributional definition of $\delta^2R_{\mu\nu}[h^1,h^1]$ from Ref.~\cite{Upton:2021oxf}, then we can equivalently write these equations as
\begin{align}
    E_{\mu\nu}[h^{\calR 1}] ={}& -16\pi \TeffTR^1_{\mu\nu}-E_{\mu\nu}[h^{\calP 1}], \label{eq:EFE1 res} \\
    E_{\mu\nu}[h^{\calR2}]  ={}& -16\pi\TeffTR^2_{\mu\nu}+2\delta^{2}R_{\mu\nu}[h^1,h^1] - E_{\mu\nu}[h^{\calP 2}],\label{eq:EFE2 res}
\end{align}
which can be obtained from Eqs.~\eqref{eq:EFE1} and \eqref{eq:EFE2} simply by moving the puncture fields to the right-hand sides. Here, derivatives acting on the punctures are interpreted distributionally, and each term on the right-hand side has a distributional interpretation on the whole domain, as opposed to the treatment in Eqs.~\eqref{eq:EFE1 res alt} and \eqref{eq:EFE2 res alt}, where derivatives are treated as ordinary derivatives on functions at points away from the worldline, and only the total source is promoted to the whole domain.

In this paper, rather than working solely with Eqs.~\eqref{eq:EFE1 res alt} and \eqref{eq:EFE2 res alt}, we make extensive use of Eqs.~\eqref{eq:EFE1} and \eqref{eq:EFE2} as a consistency check on the puncture fields and on the overarching formalism.

\subsection{Multiscale expansion}

We use the multiscale expansion described in Appendix A of Ref.~\cite{Miller:2020bft} and Sec.~II of Ref.~\cite{Miller:2023ers}. 
The expansion begins by foliating the spacetime using a time coordinate
\begin{equation}\label{eq:s def}
    s = t - k(r^*),
\end{equation}
where the height function $k\to r^*$ at future null infinity and $k\to-r^*$ at the future horizon, where $r^*$ is the tortoise coordinate. Slices of constant $s$ then become null at the boundaries; this has numerous advantages in the multiscale expansion, specifically leading to the best-behaved sources at the boundaries and minimizing  oscillations across the domain. Concretely, we adopt ``sharp'' $v$-$t$-$u$ slicing, in which $s=v\coloneqq t+r^*$ in a region $2M\leq r< r_1$; $s=t$ in a region $r_1< r < r_2$ containing the particle; and $s=u\coloneqq t-r^*$ in the region $r>r_2$. This leads to simple field equations in each region and simple junction conditions at each interface~\cite{Miller:2023ers}.

Restricting to the case of a slowly inspiraling, quasicircular binary, we write the particle's worldline in Schwarzschild coordinates $(t,r,\theta,\phi)$ as
\begin{equation}
    z^\alpha(t,\e) = (t,r_p(t,\e),\pi/2,\phi_p(t,\e)).
\end{equation}
The orbit has a slowly evolving frequency
\begin{equation}
    \frac{d\phi_p}{dt} \coloneqq \Omega
\end{equation}
satisfying an evolution equation of the form
\begin{equation}
    \frac{d\Omega}{dt} = \e F_0(\Omega) + \e^2 F_1(J_a) + \order(\e^3),
\end{equation}
where we have introduced $J_a\coloneqq (\Omega,\delta M,\delta S)$ and we suppress dependence on the background mass $M$. $F_0(\Omega)$ is the standard adiabatic (0PA) evolution due to the dissipative fluxes carried off by the first-order metric perturbation, and the 1PA term $F_1(J_a)$ encodes all second-order dissipative effects. The orbital radius is then parametrized in terms of the frequency as
\begin{equation}
    r_p = r_0(\Omega) + \e r_1 (J_a) +\order(\e^2),
\end{equation}
where $r_0 = M(M\Omega)^{-2/3}$ is the standard geodesic relationship between frequency and radius, and $r_1$ represents the correction to that relationship due to the conservative first-order self-force.

Given the above parametrization of the orbital motion, we write the metric throughout the spacetime as
\begin{equation}
    {\sf g}_{\mu\nu} = g_{\mu\nu} + \sum_{n\geq1}\e^n \mathring h^{n}_{\mu\nu}(\phi_p,J_a,x^i).\label{eq:g multiscale expansion}
\end{equation}
Here  $x^i$ are any set of spatial coordinates on slices of constant time $s$. The mechanical variables $(\phi_p,J_a)$ are defined to be constant on these slices, and they encode the complete time dependence of the metric. The mass and spin perturbations evolve according to the standard adiabatic fluxes of energy and angular momentum through the horizon~\cite{Miller:2020bft},
\begin{align}
    \frac{d\delta M}{dt} &= \e \dot{\cal E}^{\cal H}_0(\Omega) + \order(\e^2),\\ 
    \frac{d\delta S}{dt} &= \e \dot{\cal L}^{\cal H}_0(\Omega) + \order(\e^2).
\end{align}

The coefficients $\mathring h^n_{\mu\nu}$ in the  multiscale expansion differ from the perturbations $h^n_{\mu\nu}$ in the self-consistent expansion; $\mathring h^1_{\mu\nu}$ is the leading term in the multiscale expansion of $h^1_{\mu\nu}$, $\mathring h^2_{\mu\nu}$ contains the leading multiscale term in $h^2_{\mu\nu}$ as well as the first subleading multiscale term in $h^1_{\mu\nu}$, and so on~\cite{Miller:2020bft,Lewis:2025ydo}. This leads to a reorganization of the Einstein equations~\eqref{eq:EFE1 res} and \eqref{eq:EFE2 res}. In those field equations, we now apply the chain rule 
\beq\label{eq:chain rule}
\partial_s = \Omega \partial_{\phi_p} + \e \vec{\partial}_{\cal V} + \order(\e^2),
\eeq
where (following Ref.~\cite{Miller:2023ers}) we have introduced the shorthand
\begin{equation}\label{d/dV}
    \vec{\partial}_{\cal V}\coloneqq F_0\partial_\Omega + \dot{\cal E}^{\cal H}_0\partial_{\delta M} + \dot{\cal L}^{\cal H}_0 \partial_{\delta S}
\end{equation}
for the derivative along the system's ``velocity'' $\vec{\cal V}=(F_0, \dot{\cal E}^{\cal H}_0,\dot{\cal L}^{\cal H}_0)$ through parameter space. The resulting field equations read
\begin{align}
    E^0_{\mu\nu}[\mathring h^{\calR 1}] ={}& -16\pi \mathring\TeffTR^1_{\mu\nu}-E^0_{\mu\nu}[\mathring h^{\calP 1}], \label{eq:EFE1 tt} \\
    E^0_{\mu\nu}[\mathring h^{\calR2}] ={}& -16\pi\mathring\TeffTR^2_{\mu\nu}+ 2\delta^{2}R^0_{\mu\nu}[\mathring h^1,\mathring h^1]-E^1_{\mu\nu}[\mathring h^1]\notag\\
    &-E^0_{\mu\nu}[\mathring h^{\calP 2}], \label{eq:EFE2 tt}
\end{align}
where terms labelled with a 0 only involve the leading term in Eq.~\eqref{eq:chain rule}, while $E^1_{\mu\nu}[\mathring h^1]$ contains terms linear in $F_0$, $\dot{\cal E}^{\cal H}_0$, and $\dot{\cal L}^{\cal H}_0$ arising from Eq.~\eqref{eq:chain rule}; see Appendix~\ref{sec:E1Form}. The multiscale expansion of the Detweiler stress-energy tensor is detailed in Appendix~\ref{sec:multiscale T}.

Analogously, the gauge condition~\eqref{eq:Z_cond} becomes 
\begin{align}
    Z^0_{\mu}[\mathring{\bar h}^1] &= 0,\label{eq:Z_cond multiscale 1}\\ 
    Z^0_{\mu}[\mathring{\bar h}^2] &= -Z^1_{\mu}[\mathring{\bar h}^1].\label{eq:Z_cond multiscale 2}
\end{align}

\subsection{Mode decomposition}\label{sec:mode decomp}

The variable $\phi_p$ is periodic, as $\phi_p$ and $\phi_p+2\pi$ represent the same azimuthal angle. Hence, we can expand the metric perturbation in a discrete Fourier series in $\phi_p$. We also take advantage of the spherical symmetry of the background metric by expanding in tensor spherical harmonics. Combining these two expansions, we write
\begin{equation}\label{eq:h modes}
    \mathring h^{n}_{\mu\nu} = \sum_{i\ell m}\frac{a_{i\ell}}{r}h^{n}_{i\ell m}(r,J_a)Y^{i\ell m}_{\mu\nu}(r,\theta,\phi)e^{-im\phi_{p}},
\end{equation}
where $Y^{i\ell m}_{\mu\nu}$, with $i=1,\ldots,10$ are Barack-Lousto-Sago (BLS) tensor spherical harmonics~\cite{Barack:2005nr,Barack:2007tm}; the relationship between these and other common spherical bases is detailed in Ref.~\cite{Spiers:2023mor}. The constants $a_{i\ell}$ are given by
\begin{equation}
    a_{i\ell} = 
    \begin{cases}
        \frac{1}{\sqrt{2}} & \text{for } i=1,2,3,6, \\
        \frac{1}{\sqrt{2\ell(\ell+1)}} & \text{for } i=4,5,8,9, \\
        \frac{1}{\sqrt{2(\ell-1)\ell(\ell+1)(\ell+2)}} & \text{for } i=7,10.
    \end{cases}\label{eq:ail_def}
\end{equation}
Note that our convention here differs from that of Ref.~\cite{Miller:2020bft}, where $h^n_{i\ell m}$ denoted the BLS mode coefficient of $\mathring{\barh}^n_{\alpha\beta}$ rather than of $\mathring{h}^n_{\alpha\beta}$; the two are related by the interchange $i=3\leftrightarrow i=6$. Our $h^{n}_{i\ell m}$ here correspond to the quantities \textsf{hBS}${}^{i\ell m}$ in \textsc{PerturbationEquations}~\cite{PerturbationEquations}.

Given the expansion~\eqref{eq:h modes}, field equations of the form $E^0_{\mu\nu}[\mathring h^n] = S^n_{\mu\nu}$ reduce to radial ordinary differential equations for the mode coefficients: 
\begin{equation}\label{ER=S}
    E^0_{ij\ell m}h^n_{j\ell m} = -\frac{rf}{4a_{i\ell}}S^n_{i\ell m},
\end{equation}
where the index $j=1,\ldots,10$ is summed over, and
\begin{equation}
    f \coloneqq 1-\frac{2M}{r}.
\end{equation}
Here $S^n_{i\ell m}$ are the coefficients in 
\begin{equation}
S^n_{\mu\nu}=\sum_{i\ell m}S^n_{i\ell m} Y^{i\ell m}_{\mu\nu}e^{-im\phi_p}. 
\end{equation}
The operator on the left is 
\begin{equation}\label{eq:E0 def}
E^0_{ij\ell m} \coloneqq \delta_{ij}\Box^0_{\ell m} + {\cal M}^0_{ij},    
\end{equation}
with
\begin{multline}
\Box^0_{\ell m} \coloneqq -\frac{1}{4}\Bigl[\partial_{r^*}^2 + i \omega_m (2 H\partial_{r^*} + H') \\[-2pt]
+ (1-H^2)\omega_m^2 - 4V_\ell(r)\Bigr],  
\end{multline}
where $\omega_m\coloneqq m \Omega$, $H\coloneqq dk/dr^*$ is the derivative of the height function from Eq.~\eqref{eq:s def}, $H'\coloneqq dH/dr^*$, and 
\begin{equation}
V_\ell(r) = \frac{f}{4}\left(\frac{2M}{r^3} + \frac{\ell(\ell+1)}{r^2}\right).    
\end{equation}
The coupling operators ${\cal M}^0_{ij}$, which couple between the different $i$'s, involve at most one radial derivative; they are given in Appendix~\ref{sec:E1Form}. Equations~\eqref{ER=S} can also be loaded in \textsc{PerturbationEquations}\footnote{\label{fn:PertEqScal}\textsc{PerturbationEquations} does not directly provide $E^0_{ij\ell m}$ but does provide the linearized Ricci tensor in Lorenz gauge in BLS harmonics: $\delta R^{0,\text{PertEqs}}_{ij\ell m}=\frac{2a_{i\ell}}{rf}E^{0}_{ij\ell m}$.} \cite{PerturbationEquations}.

In analogy with the decomposition of $E^0_{\mu\nu}[\mathring h^n]=S^n_{\mu\nu}$, we write the decomposition of 
\beq
E^0_{\mu\nu}[\mathring h^{{\cal R}n}] = S^{n}_{\mu\nu} - E^0_{\mu\nu}[\mathring h^{{\cal P}n}]\coloneqq S^{n,\rm eff}_{\mu\nu}
\eeq
as 
\begin{equation}\label{ERres=Seff}
    E^0_{ij\ell m}h^{{\cal R}n}_{j\ell m} = -\frac{rf}{4a_{i\ell}}S^{n,\rm eff}_{i\ell m}.
\end{equation}
In particular, the second-order effective source $S^{2,\rm eff}_{i\ell m}$ is given by the mode decomposition of the right-hand side of Eq.~\eqref{eq:EFE2 tt}: 
\begin{align}\label{S2effilm}
    S^{2, \rm eff}_{i\ell m} &= -16\pi \bar T^2_{i\ell m} + 2\delta^2 R^0_{i\ell m}[\mathring h^1,\mathring h^1] \nonumber  \\
    & \qquad
    + \frac{4a_{i\ell}}{rf}\left(E^1_{ij\ell m}h^1_{j\ell m}
    + E^0_{ij\ell m}h^{{\cal P}2}_{j\ell m}\right).
\end{align}
The puncture field and its mode decomposition are described in Sec.~\ref{sec:punctures}; our calculation of $\delta^2 R^0_{i\ell m}$ in Sec.~\ref{sec:d2R}; and $E^1_{i\ell m}$ in Appendix~\ref{sec:E1Form}.

\subsection{Modified, gauge-damped field equations}

In practice, in Refs.~\cite{Pound:2019lzj,Miller:2020bft,Warburton:2021kwk,Miller:2023ers} we solved a slightly different set of field equations than Eq.~\eqref{ERres=Seff}.  Following Ref.~\cite{Barack:2005nr} and subsequent literature, the equations were modified by the addition of a term proportional to the gauge condition. This modification serves to partially decouple some of the field equations, though the motivation is not particularly relevant in this paper. Because the additional terms in the field equations also serve to damp gauge violations in time-domain calculations, we will refer to these modified equations as the ``gauge-damped'' form of the field equations.

After the addition of the gauge-damping terms, the equations~\eqref{ERres=Seff} become
\begin{equation}\label{EbreveR=breveS}
    \breve E^0_{ij\ell m}h^{{\cal R}n}_{j\ell m} = -\frac{rf}{4a_{i\ell}}\breve S^{n, \rm eff}_{i\ell m}.
\end{equation}
On the left-hand side, the wave operator is now $\breve E^0_{ij\ell m} = \delta_{ij}\Box^0_{\ell m} + \breve{\cal M}^0_{ij}$, which differs from $E^0_{ij\ell m}$ only in the form of the coupling operators $\breve{M}^0_{ij}$. These coupling operators are given in Appendix~A of Ref.~\cite{Miller:2023ers}, for example, where they are denoted by ${\cal M}^{(0)}_{ij}$ (noting the relations $h_{i\ell m}=\bar h_{i\ell m}$ for $i\neq3,6$, $h_{3\ell m}=\bar h_{6\ell m}$, and $h_{6\ell m}=\bar h_{3\ell m}$, along with $\breve{\cal M}^0_{6j}={\cal M}^{(0)}_{3j}$ and $\breve{\cal M}^0_{3j}={\cal M}^{(0)}_{6j}$). On the right-hand side of the field equations, only the operators $E^n_{i\ell m}$ are altered relative to Eq.~\eqref{S2effilm}:
\begin{align}
    \breve S^{2, \rm eff}_{i\ell m} &= -16\pi \bar T^2_{i\ell m} + 2\delta^2 R^0_{i\ell m}[\mathring h^1,\mathring h^1]
    \nonumber \\
    &\qquad 
    + \frac{4a_{i\ell}}{rf}\left(\breve E^1_{ij\ell m}h^1_{j\ell m} 
    + \breve E^0_{ij\ell m}h^{{\cal P}2}_{j\ell m}\right).
\end{align}
$\breve E^1_{ij\ell m}$ is given by Eq.~(65) of Ref.~\cite{Miller:2023ers}, where it is denoted $E^{(1)}_{ij\ell m}$. For convenience we reproduce it in Eq.~\eqref{breveE1ijlm} below. 

Although we used the gauge-damped form of the field equations in previous papers, we focus on the non-damped form in our checks of the effective source in this paper. This is because, as we explain below, various pieces of the singular field satisfy separate non-damped field equations, which can be checked individually; in the gauge-damped equations, these pieces must be combined to perform an analogous check.

\section{Singular and regular fields}
\label{sec:punctures}

The expressions for the singular field through second order were originally derived in Refs.~\cite{Pound:2009sm,Pound:2012dk} in a local (Fermi--Walker) coordinate system in a neighbourhood of the trajectory. These were converted to a covariant form in Ref.~\cite{Pound:2014xva}. In all cases, the expressions were derived in the context of a self-consistent expansion~\cite{Pound:2009sm,Pound:2012nt,Pound:2012dk,Pound:2015tma}, rather than a multiscale expansion.
We first recap the form of the singular field in the self-consistent expansion before showing how this changes in the multiscale framework. In Sec.~\ref{sec:punctures_twotimescale} we provide the multiscale expansion of the singular field for a generic orbit in Kerr spacetime; in Sec.~\ref{sec:punctures in coordinates} we then specialize it to quasicircular orbits in Schwarzschild spacetime. Section~\ref{sec:int_lm} then discusses the decomposition into tensor-spherical-harmonic modes. Finally, Sec.~\ref{sec:res_field} describes the calculation of the first-order residual field, which is required as input for the second-order singular field as well as for the second-order Ricci tensor in later sections.

\subsection{Self-consistent form in a generic spacetime}\label{sec:punctures_generic}

In the self-consistent expansion, the first- and second-order singular fields have the schematic form
\begin{align}
    h^{\S1}_{\mu\nu} &\sim \frac{m}{\lambda} +m\lambda^0 + {\cal O}(\lambda),\label{hS1 schematic}\\
    h^{\S2}_{\mu\nu} &\sim \frac{m^2}{\lambda^2} + \frac{m^2 + m h^{\R1} + \delta m}{\lambda} + {\cal O}(\log\lambda),\label{hS2 schematic}
\end{align}
where $\lambda$ is a formal order-counting parameter that we use to count powers of spatial distance from the particle's worldline (and set to unity at the end of the calculation), and the tensorial structure of coefficients is left deliberately unspecified. As described in Ref.~\cite{Pound:2014xva}, the second-order singular field can naturally be split into three pieces:
\begin{equation}
    h^{\S2}_{\mu\nu} = h^{\S\S}_{\mu\nu} + h^{\S\R}_{\mu\nu} + h^{\delta m}_{\mu\nu}. \label{eq:hS2_sc}
\end{equation}
Schematically, the $h^{\S\S}_{\mu\nu}$ field is the `singular times singular' piece with leading order form $\sim m^2/\lambda^2$, $h^{\S\R}_{\mu\nu}$ is the `singular times regular' piece that behaves as $\sim m h^{\R1}_{\mu\nu}|_{\gamma}/\lambda$, and $h^{\delta m}_{\mu\nu} \sim \delta m_{\mu\nu}/\lambda$ tracks the correction to the object's monopole, where  we use  $|_\gamma$ to denote evaluation on the worldline. Explicit covariant expressions for the individual fields were derived by one of us in Ref.~\cite{Pound:2014xva} and are presented in Eqs.~(130)--(132) of the same reference. 

In the $h^{\delta m}_{\mu\nu}$ term, the quantity $\delta m_{\mu\nu}$ is given by~\cite{Pound:2014xva}
\begin{align}
    \delta m_{\mu\nu} ={}& \frac{m}{3}(2h^{\R1}_{\mu\nu} + g_{\mu\nu}g^{\alpha\beta}h^{\R1}_{\alpha\beta}) + 4mu_{(\mu}h^{\R1}_{\nu)\alpha}u^{\alpha} \nonumber \\
        & + m(g_{\mu\nu} + 2u_{\mu}u_{\nu})h^{\R1}_{\alpha\beta}u^{\alpha}u^{\beta}, \label{eq:dm_sc}
\end{align}
with the components of $h^{\R1}_{\mu\nu}$ evaluated on the worldline.
We note that this expression for $\delta m_{\mu\nu}$ implies $h^{\delta m}_{\mu\nu}$ has the same form as $h^{\S\R}_{\mu\nu}$, $\sim m h^{\R1}_{\mu\nu}|_\gamma/\lambda$. However, it is distinguished from $h^{\S\R}_{\mu\nu}$ by its spherical symmetry around the particle. 

The three fields satisfy separate field equations,
\begin{align}
    E_{\mu\nu}[h^{\S\S}] ={}& 2\delta^{2}R_{\mu\nu}[h^{\S1},h^{\S1}] & \text{for } x\notin\gamma, \label{eq:EhSS_sc_off} \\
    E_{\mu\nu}[h^{\S\R}] ={}& 4Q^{\text{Ric}}_{\mu\nu}[h^{\S1}] & \text{for } x\notin\gamma, \label{eq:EhSR_sc_off} \\
    E_{\mu\nu}[h^{\delta m}] ={}& 0 & \text{for } x\notin\gamma, \label{eq:Ehdm_sc_off}
\end{align}
where 
\begin{equation}\label{eq:Qdef}
Q^{\text{Ric}}_{\mu\nu}[h] \coloneqq \frac{1}{2}\delta^{2}R_{\mu\nu}[h^{\R1},h] + \frac{1}{2}\delta^{2}R_{\mu\nu}[h,h^{\R1}] 
\end{equation}
is a smooth linear operator acting on $h_{\mu\nu}$ constructed from the second-order Ricci tensor and the first-order regular field.
Motivated by the distributional analysis in Ref.~\cite{Upton:2021oxf}, we can promote Eqs.~\eqref{eq:EhSS_sc_off}--\eqref{eq:Ehdm_sc_off} to include the worldline; see Appendix~\ref{sec:T_eff_TR_deriv} for more details.
This gives us the following equations:
\begin{align}
    E_{\mu\nu}[h^{\S\S}] ={}& 2\delta^{2}R_{\mu\nu}[h^{\S1},h^{\S1}], \label{eq:EhSS_sc_inc} \\
    E_{\mu\nu}[h^{\S\R}] ={}& -16\pi\TeffTR^{Q^{\text{Ric}}}_{\mu\nu} + 4Q^{\text{Ric}}_{\mu\nu}[h^{\S1}], \label{eq:EhSR_sc_inc} \\
    E_{\mu\nu}[h^{\delta m}] ={}& -16\pi\TeffTR^{\delta m}_{\mu\nu}, \label{eq:Ehdm_sc_inc}
\end{align}
where $\TeffTR^{\delta m}_{\mu\nu}$ and $\TeffTR^{Q^{\text{Ric}}}_{\mu\nu}$ are certain parts of the (trace-reversed) Detweiler stress-energy tensor and are given by Eqs.~\eqref{eq:TeffTR_dm} and~\eqref{eq:TeffTR_Q}, respectively.
The sum of the first- and second-order singular fields additionally satisfies the gauge condition~\eqref{eq:Z_cond}.

\subsection{Multiscale expansion for generic orbits}\label{sec:punctures_twotimescale}

\subsubsection{Singular field}\label{sec:ms_punc}

We adopt the multiscale expansion from Ref.~\cite{Pound:2021qin}, which is valid for generic orbits in Kerr spacetime. However, we more closely follow the notation of Ref.~\cite{Mathews:2025nyb}. 

The particle's spatial trajectory is parametrized by action angles $\mathring\varphi^i=(\mathring\varphi^r,\mathring\varphi^\theta,\mathring\varphi^\phi)$ and by slowly evolving variables $\mathring\varpi_J = (\mathring\pi_j,\delta M, \delta S)$, where the orbital elements $\mathring\pi_i=(\mathring p, \mathring e, \mathring \iota)$ are averaged versions of the semi-latus rectum $p$, eccentricity $e$, and maximum inclination angle $\iota$. These variables satisfy differential equations of the form
\begin{align}
    \frac{d\mathring\varphi^i}{dt} &= \Omega^i(\mathring\pi_i),\\
    \frac{d\mathring\pi_i}{dt} &= \e F^{(0)}_i(\mathring\pi_j) + \e^2 F^{(1)}_i(\mathring\varpi_J) + \order(\e^3),
\end{align}
where $\Omega^i(\mathring\pi_j)$ are the Kerr-geodesic orbital frequencies. The forcing functions $F^{(n)}_i$ are constructible from the local self-forces $f^\mu_1$ and $f^\mu_2$ on the particle. In the special case of quasicircular orbits, the action angles reduce to the single angle $\phi_p$, and we adopt the frequency $\Omega$ as our sole slowly evolving orbital element.

Written in terms of these variables, the multiscale expansion of the spatial trajectory $z^i$ reads
\beq\label{z expansion}
z^i = z_0^i(\mathring\varphi^j,\mathring\pi_j) + \e z_1^i(\mathring\varphi^j,\mathring\varpi_j) +\order(\e^2),
\eeq
such that $z^\mu_0=(t,z^i_0)$ (and $z^t_{1}=0$ trivially). This is the expansion of the accelerated trajectory $z^i$ about another accelerated trajectory, $z_0^i$. $z_0^i(\mathring\varphi^j,\mathring\pi_j)$ has the same functional dependence on $(\mathring\varphi^j,\mathring\pi_j)$ as a geodesic has on the geodesic action angles $\varphi^i$ and geodesic orbital elements $\pi_i$, but the on-shell time dependence (and $\e$ dependence) of $(\mathring\varphi^j,\mathring\pi_j)$ is not that of a geodesic.  

In Refs.~\cite{Pound:2021qin,Mathews:2025nyb}, multiscale expansions of the particle's worldline were carried out entirely in coordinate form. This is not ideal for the singular field, which is (initially) in covariant form. To obtain a covariant multiscale expansion of the singular field, we follow the covariant worldline-expansion methods from Ref.~\cite{Pound:2015fma}. 

Since we only require an approximation to the singular field through second order in $\e$, the expansion of $h^{\S2}_{\mu\nu}$ becomes trivial; we simply replace $z^\mu$ with $z^\mu_{0}$, $u^\mu$ with $\mathring u^\mu_0$ (the leading term in the multiscale expansion of the four-velocity, explained below), and $a^\mu$ with 0. Hence, our goal in this section is to derive the multiscale expansion of $h^{\S1}_{\mu\nu}$,
\begin{equation}\label{hS1 multiscale expansion}
    h^{\S1}_{\mu\nu} = \mathring h^{\S1}_{\mu\nu} + \e\mathring h^{\S(1,1)}_{\mu\nu} + \order(\e^2),
\end{equation}
by substituting the multiscale expansion of the worldline into $h^{\S1}_{\mu\nu}$. The first subleading term will then be added to $ h^{\S2}_{\mu\nu}$ to give us our second-order singular field in the multiscale framework,
\begin{equation}
    \mathring h^{\S2}_{\mu\nu} = h^{\S2}_{\mu\nu}\bigr|_{\gamma\to\gamma_0} + \mathring h^{\S(1,1)}_{\mu\nu}.
\end{equation}

The first-order singular field in covariant form, as given in Ref.~\cite{Pound:2014xva}, is
\begin{align}\label{h1 covariant}
h^{\S1}_{\mu\nu} ={}& \frac{2 m}{\lambda\s} g^{\alpha'}_{\mu} g^{\beta'}_{\nu} \left(g_{\alpha'\beta'} + 2 u_{\alpha'} u_{\beta'}\right) \nonumber \\
    & + \frac{m\lambda^0}{\s^3} g^{\alpha'}_\mu g^{\beta'}_{\nu} \Bigl[\left(\s^2- \r^2\right) a_{\sigma} (g_{\alpha'\beta'}+2u_{\alpha'} u_{\beta'})\nonumber\\
        &\qquad\qquad\qquad+8\r\s^2 a_{(\alpha'} u_{\beta')}\Bigr] \nonumber\\
    & + \frac{m\lambda}{3\s^3}g^{\alpha'}_\mu g^{\beta'}_{\nu} \Big[\!\left(\r^2 - \s^2\right)
			\left(g_{\alpha'\beta'}+2u_{\alpha'} u_{\beta'}\right)R_{u\sigma u\sigma} \nonumber\\
        &\qquad\qquad\qquad - 12\s^4 R_{\alpha' u\beta' u}- 12\r\s^2 u_{(\alpha'}R_{\beta')u\sigma u}\nonumber\\
        &\qquad\qquad\qquad +12\s^2 (\r^2 + \s^2)\dot{a}_{(\alpha'}u_{\beta')} \nonumber\\
        &\qquad\qquad\qquad + \r (3\s^2-\r^2)\dot{a}_{\sigma}(g_{\alpha'\beta'}+ 2u_{\alpha'} u_{\beta'})\Big] \nonumber\\
    & + \order(\lambda |a|^2,\lambda^2).
\end{align}
Here primed indices refer to tensors at a point $z^\mu$ on $\gamma$; $a^\mu\coloneqq Du^\mu/d\tau$ is the covariant acceleration of $\gamma$, and an overdot denotes $D/d\tau_0$; $\tau_0$ is proper time on $z^\mu_0$; $g^{\mu'}_{\mu}$ is the parallel propagator along the unique geodesic connecting $z^\mu$ to the field point $x^\mu$; $\s(z,x)\coloneqq\sqrt{P_{\mu'\nu'}\sigma^{\mu'}\sigma^{\nu'}}$ and $\r(z,x)\coloneqq\sigma_{\mu'}u^{\mu'}$ are, respectively, measures of the proper spatial and temporal distances between $z^\mu$ and $x^\mu$; Synge's world function $\sigma(z,x)$ is one-half the squared geodesic distance between $z^\mu$ and $x^\mu$, and $\sigma_{\alpha'}\coloneqq\nabla_{\alpha'}\sigma$ is a directed measure of the geodesic distance between the two points; we use the notation $a_\sigma\coloneqq a^{\alpha'}\sigma_{\alpha'}$ and $R_{u\sigma u\sigma}\coloneqq R_{\mu'\nu'\alpha'\beta'}u^{\mu'}\sigma^{\nu'}u^{\alpha'}\sigma^{\beta'}$; and $P_{\mu\nu}=g_{\mu\nu}+u_\mu u_\nu$ projects orthogonally to $\gamma$.

We consider expansions of the trajectory, four-velocity, and acceleration in sequence, in each case assessing how the expansions contribute to the expansion of $h^{\rm S1}_{\mu\nu}$. First, we note that geometrically, $z_1^\mu(\mathring\varphi^j,\mathring\varpi_j)  =\frac{\partial}{\partial\e}z^\mu(\mathring\varphi^j,\mathring\varpi_J,\e)\big|_{\e=0}$ is a deviation vector defined on $z^\mu_0$. From Eq.~(B5) in Ref.~\cite{Pound:2015fma}, it satisfies 
\beq\label{commutation}
z^\mu_1\nabla_\mu u^\nu\bigr|_{\gamma_0} = P_0^\nu{}_\mu \frac{Dz^\mu_1}{d\tau_0},
\eeq
where $P_0^{\mu\nu}=g^{\mu\nu}+u^{\mu}_0u^{\nu}_0$,  $u^\mu_0=\frac{dz^\mu_0}{d\tau_0}$, and $\frac{D}{d\tau_0}=u_0^\mu\nabla_\mu$.

Substituting \eqref{z expansion} into \eqref{h1 covariant} gives us
\begin{align}\label{hS1 z expansion}
h^{\S1}_{\mu\nu} &= 2m \biggl[g_{\mu}^{ \mu'}g_{\nu}^{\nu'}\frac{g_{\mu'\nu'}+2u_{0\mu'}u_{0\nu'}}{\lambda \s_0} \nonumber\\
&\qquad\ +\e \lambda^{-2}z_1^{\alpha'}\nabla_{\alpha'}\biggl(g_{\mu}^{ \mu'}g_{\nu}^{\nu'}\frac{g_{\mu'\nu'}+2u_{0\mu'}u_{0\nu'}}{\s}\biggr)\biggr|_{\gamma_0} \nonumber\\
&\qquad\qquad\qquad\qquad\qquad\qquad\quad+ \order(\e^2,\lambda^0)\biggr],
\end{align}
where $\s_0=\s(z_0,x)$, and primed indices now refer to tensors at $z^\mu_0$. Here and below we truncate expressions at order $\lambda^0$, but we provide them through order $\lambda$ online \cite{PuncturesRepository}. Using $\sigma^{\alpha'}_{\ \beta'}=g^{\alpha'}_{\ \beta'}+\order(\lambda^2)$ and $g_{\alpha}^{\alpha'}{}_{;\beta'}=\order(\lambda)$~\cite{Poisson:2011nh}, along with Eq.~\eqref{commutation} and $\frac{Du_0^\mu}{d\tau_0}=\order(\e)$, we get
\begin{align}
h^{\S1}_{\mu\nu} ={}& 2m g_{\mu}^{ \mu'}g_{\nu}^{\nu'}\biggl\{\frac{g_{\mu'\nu'}+2u_{0\mu'}u_{0\nu'}}{\lambda \s_0} + \e\biggl[4\frac{u^{(\mu'}_0}{\lambda\s_0}\frac{Dz_{1\perp}^{\nu')}}{d\tau_0} \nonumber \\
    & -\frac{g_{\mu'\nu'}+2u_0^{\mu'}u_0^{\nu'}}{\lambda^2\s_0^3}\biggl(z^{\alpha'}_{1\perp}\sigma_{\alpha'}+\frac{Dz_{1\perp}^{\alpha'}}{d\tau_0}\sigma_{\alpha'}\r_0\biggr)\biggr] \nonumber \\
    & + \order(\e^2,\lambda^0)\biggr\},\label{hS1 part 1}
\end{align}
where $\r_0=u_0^{\mu'}\sigma_{\mu'}$ and $z^{\mu'}_{1\perp}=P_0^{\mu'}{}_{\nu'}z^{\nu'}_1$.

The order-$\e$ correction in \eqref{hS1 part 1} is a mass-dipole-moment term in the singular field -- cf. Eqs.~(145) and~(146) in Ref.~\cite{Pound:2014xva}. To complete the expansion, we must also expand $u^\mu_0$. We have
\beq\label{u0}
u^\mu_0 = \frac{dz_0^\mu}{d\tau_0} = \frac{dt}{d\tau_0}\left[\Omega^i\partial_{\mathring\varphi^i}+\e F^{(0)}_i\partial_{\mathring\pi_i}+\order(\e^2)\right]z^\mu_0.
\eeq
We write $dt/d\tau_0\coloneqq U=U_0(\mathring\varphi^i,\mathring\pi_i)+\e U_1(\mathring\varphi^i,\mathring\varpi_J)+\order(\e^2)$, noting that $U_1$ vanishes in the special case of quasicircular orbits~\cite{Miller:2020bft}.
Equation~\eqref{u0} then becomes
\beq
u^\mu_0 = \mathring{u}^\mu_0 + \e v^\mu +\order(\e^2), \label{eq:u0def}
\eeq
where 
\beq\label{u0check}
\mathring{u}^\mu_0=U_0\Omega^i\partial_{\mathring\varphi^i} z_0^\mu,
\eeq
and
\beq\label{v}
v^\mu = U_0 F^{(0)}_i\partial_{\mathring\pi^i}z^\mu_0 + U_1 \mathring u^\mu_0.
\eeq

We must also expand $\frac{Dz_{1\perp}^\mu}{d\tau_0}$ in a similar way. We define operators $\frac{D^{(n)}}{d\tau_0}$ that act on tensor functions of $(\mathring\varphi^i,\mathring\pi_i)$ as
\begin{align}
    \frac{D^{(0)}V^\mu}{d\tau_0} ={}& U_0\Omega^i\partial_{\mathring\varphi^i} V^\mu +\Gamma^\mu_{\beta\gamma}\mathring{u}^\beta_0 V^\gamma, \label{eq:D0tau0} \\ 
    \frac{D^{(1)}V^\mu}{d\tau_0} ={}& U_0F^{(0)}_i\partial_{\mathring\pi_i}V^\mu + U_{1}\Omega^i\partial_{\mathring\varphi^i}V^\mu + \Gamma^\mu_{\beta\gamma}v^\beta V^\gamma,
\end{align}
with the obvious extension to covectors and tensors, such that
\begin{equation}
    \frac{D}{d\tau_0} = \frac{D^{(0)}}{d\tau_0}+\e\frac{D^{(1)}}{d\tau_0} + \order(\e^2). \label{eq:Dtau_exp}
\end{equation}

Substituting these expansions into \eqref{hS1 part 1} yields
\begin{align}
h^{\S1}_{\mu\nu} ={}& 2m g_{\mu}^{ \mu'}g_{\nu}^{\nu'}\biggl\{\frac{g_{\mu'\nu'}+2\mathring{u}_{0\mu'}\mathring{u}_{0\nu'}}{\lambda\mathring{\s}_0} +\e\biggl[\frac{4\mathring{u}^{(\mu'}_0\mathring{u}^{\nu')}_1}{\lambda\mathring{\s}_0} \nonumber \\
    &\qquad -\frac{(g_{\mu'\nu'}+2\mathring{u}_{0\mu'}\mathring{u}_{0\nu'})\bigl(\mathring{z}^{\alpha'}_{1\perp}\sigma_{\alpha'}+\mathring{u}^{\alpha'}_1\sigma_{\alpha'}\mathring{\r}_0\bigr)}{\lambda^2\mathring{\s}_0^3}\biggr] \nonumber\\
    &\qquad\qquad\qquad\qquad\qquad+ \order(\e^2,\lambda^0)\biggr\},
\end{align}
where $\mathring{u}_1^{\mu'}\coloneqq\frac{D^{(0)}\mathring{z}_{1\perp}^{\mu'}}{d\tau_0}+v^{\mu'}$, $\mathring{z}^{\alpha'}_{1\perp}\coloneqq\mathring{P}_0^{\alpha'}{}_{\beta'}z^{\beta'}_1$, $\mathring{P}^0_{\mu'\nu'}=g_{\mu'\nu'}+\mathring{u}^0_{\mu'}\mathring{u}^0_{\nu'}$, $\mathring{\s}_0=\sqrt{\mathring{P}^0_{\mu'\nu'}\sigma^{\mu'}\sigma^{\nu'}}$, and $\mathring{\r}_0=\mathring{u}_0^{\mu'}\sigma_{\mu'}$.

At order $\lambda^0$, acceleration terms appear in $h^{\S1}_{\mu\nu}$. Note that quantities such as $\e z^{\mu'}_1\nabla_{\!\mu'}(\cdots)_{\alpha'} a^{\alpha'}$ are $\order(\e^2)$, and we can neglect them. Hence, for acceleration-dependent terms, rather than working through the full details of Eq.~\eqref{hS1 z expansion}, we may simply replace $a^{\mu'}$ with $a^{\mu'}_0 + \e z^{\nu'}_1\nabla_{\!\nu'}a^{\mu'}|_{\gamma_0}+\order(\e^2)$, where  $a^\mu_0=\frac{Du^\mu_0}{d\tau_0}$. From Eq. (57) of Ref.~\cite{Pound:2015fma}, we have 
\beq
a^\mu = a_0^\mu + \e\left(\frac{D^2z_{1\perp}^\mu}{d\tau^2}+R^\mu{}_{\alpha\beta\gamma}u^\alpha_0 z^\beta_{1\perp}u^\gamma_0\right) +\order(\e^2).
\eeq
Expanding the four-velocities and derivatives in multiscale form then gives us
\begin{align}
    a^\mu ={}& \e\biggl(\frac{D^{(0)}\mathring{u}_{1}^\mu}{d\tau_0}+\frac{D^{(1)}\mathring{u}_{0}^\mu}{d\tau_0}+R^\mu{}_{\alpha\beta\gamma}\mathring{u}^\alpha_0 \mathring{z}^\beta_{1\perp}\mathring{u}^\gamma_0\biggr) + \order(\e^2) \nonumber \\
	   ={}& \e\biggl(\frac{D^{(0)}v^\mu}{d\tau_0}+\frac{D^{(1)}\mathring{u}_{0}^\mu}{d\tau_0}+\frac{(D^{(0)})^2\mathring{z}_{1\perp}^\mu}{d\tau_0^2} \nonumber \\
        & + R^\mu{}_{\alpha\beta\gamma}\mathring{u}^\alpha_0 \mathring{z}^\beta_{1\perp}\mathring{u}^\gamma_0\biggr) + \order(\e^2) \nonumber \\
        ={}& \e(a_{\ms}^\mu + a_{\delta z}^\mu) + \order(\e^2), \label{eq:a}
\end{align}
where we have used $\frac{D^{(0)}\mathring{u}^\mu_0}{d\tau_0}=0$ and defined
\begin{align}
    a_{\ms}^\mu \coloneqq{}& \frac{D^{(0)}v^\mu}{d\tau_0} + \frac{D^{(1)}\mathring{u}_{0}^\mu}{d\tau_0}, \label{eq:a_ms} \\
    a_{\delta z}^\mu \coloneqq{}& \frac{(D^{(0)})^2\mathring{z}_{1\perp}^\mu}{d\tau_0^2} + R^\mu{}_{\alpha\beta\gamma}\mathring{u}^\alpha_0 \mathring{z}^\beta_{1\perp}\mathring{u}^\gamma_0, \label{eq:a_dz}
\end{align}
to separately keep track of terms $\sim v^{\mu'}$ and those $\sim \mathring{z}^\mu_{1\perp}$.\footnote{\label{fn:GW}This split is mainly done for historical reasons, as the terms $\sim \mathring{z}^\mu_{1\perp}$ were derived first in the context of a Gralla--Wald-type puncture scheme and not a multiscale scheme; see Refs.~\cite{Gralla:2008fg,*[Erratum: ]Gralla:2008fgErratum,Gralla:2012db, Pound:2014xva, Pound:2015fma} for more information on Gralla--Wald expansions. In such a scheme, one expands the worldline, $\gamma$, around a background geodesic, $\gamma_{0,\rm GW}$, as
\begin{equation*}
    z^\mu(s,\e) = z_{0,\rm GW}^\mu(s) + \e z_{1, \rm GW}^\mu(s) + \order(\e^2),
\end{equation*}
where $s$ can be any parameter on $\gamma$ and $\gamma_{0,\rm GW}$. This is not the same as the multiscale expansion that we perform in Eq.~\eqref{z expansion}, as our $z_0^\mu$ is \emph{not} a geodesic. However, the \emph{form} of $h^{\delta z}_{\mu\nu}$ is the same in both the multiscale and Gralla--Wald schemes; we just make the replacement $z_{1,\rm GW}^\mu(s) \to \mathring{z}_{1}^\mu(\mathring{\varphi}^i,\mathring\pi_i)$.}

Similarly, at order $\lambda$, $Da^\mu/d\tau$ appears. We can again simply replace $\frac{Da^\mu}{d\tau}$ with $\frac{Da_0^\mu}{d\tau_0} + \e  z^\nu_1\nabla_{\!\nu}\frac{Da^\mu}{d\tau}|_{\gamma_0}+\order(\e^2)$. Using the equations in Appendix B of Ref.~\cite{Pound:2015fma}, one finds
\begin{align}
    z_1^\beta\nabla_{\!\beta}\frac{Da^\alpha}{d\tau}\Bigr|_{\gamma_0} ={}& \frac{D^3 z^\alpha_{1\perp}}{d\tau_0^3} + R^\alpha{}_{\beta\gamma\delta;\mu}u^\beta_0z^\gamma_{1\perp}u^\delta_0u^\mu_0\nonumber\\
	   & + R^\alpha{}_{\beta\gamma\delta}u^\beta_0\frac{Dz^\gamma_{1\perp}}{d\tau_0}u^\delta_0 +\order(\e).
\end{align}
Substituting the multiscale expansions as above, we obtain
\begin{align}
\frac{Da^\alpha}{d\tau} ={}& \e\biggl(\frac{(D^{(0)})^2v^\alpha}{d\tau_0^2}+\frac{D^{(0)}D^{(1)}\mathring{u}_0^\alpha}{d\tau^2_0}+\frac{D^{(1)}D^{(0)}\mathring{u}_0^\alpha}{d\tau^2_0} \nonumber \\
    & + \frac{(D^{(0)})^3 \mathring{z}^\alpha_{1\perp}}{d\tau_0^3} + R^\alpha{}_{\beta\gamma\delta;\mu}\mathring{u}^\beta_0\mathring{z}^\gamma_{1\perp}\mathring{u}^\delta_0\mathring{u}^\mu_0 \nonumber \\
    & + R^\alpha{}_{\beta\gamma\delta}\mathring{u}^\beta_0\frac{D^{(0)}\mathring{z}^\gamma_{1\perp}}{d\tau_0}\mathring{u}^\delta_0\biggr) + \order(\e^2) \nonumber \\
    ={}& \e(\adot_{\ms}^\alpha + \adot_{\delta z}^\alpha) + \order(\e^2), \label{eq:a_dot}
\end{align}
where similarly to before, we define
\begin{align}
    \adot_{\ms}^\mu \coloneqq{}& \frac{(D^{(0)})^2v^\alpha}{d\tau_0^2}+\frac{D^{(0)}D^{(1)}\mathring{u}_0^\alpha}{d\tau^2_0}+\frac{D^{(1)}D^{(0)}\mathring{u}_0^\alpha}{d\tau^2_0}, \label{eq:adot_ms} \\
    \adot_{\delta z}^\mu \coloneqq{}& \frac{(D^{(0)})^3 \mathring{z}^\alpha_{1\perp}}{d\tau_0^3} + R^\alpha{}_{\beta\gamma\delta;\mu}\mathring{u}^\beta_0\mathring{z}^\gamma_{1\perp}\mathring{u}^\delta_0\mathring{u}^\mu_0 \nonumber \\
        & + R^\alpha{}_{\beta\gamma\delta}\mathring{u}^\beta_0\frac{D^{(0)}\mathring{z}^\gamma_{1\perp}}{d\tau_0}\mathring{u}^\delta_0. \label{eq:adot_dz}
\end{align}

Our final result for the multiscale expansion of $h^{\S1}_{\mu\nu}$ takes the form~\eqref{hS1 multiscale expansion}. The leading term is given by $h^{\S1}_{\mu\nu}$ with the replacements $z^\mu\to z^\mu_0$, $u^\mu\to\mathring u^\mu_0$, and $a^\mu\to0$:
\begin{align}\label{h1 leading covariant}
\mathring h^{\S1}_{\mu\nu} &= \frac{2 m}{\lambda\mathring{\s}_0} g^{\alpha'}_{\mu} g^{\beta'}_{\nu} \left(g_{\alpha'\beta'} + 2 \mathring{u}^0_{\alpha'} \mathring{u}^0_{\beta'}\right) \nonumber \\
    & + \frac{m\lambda}{3\mathring{\s}^3}g^{\alpha'}_\mu g^{\beta'}_{\nu} \Big[\!\left(\mathring{\r}^2 - \mathring{\s}^2\right)
			\left(g_{\alpha'\beta'}+2\mathring{u}^0_{\alpha'} \mathring{u}^0_{\beta'}\right)R_{\mathring{u}_0\sigma \mathring{u}_0\sigma} \nonumber\\
        &\quad - 12\mathring{\s}^4 R_{\alpha' \mathring{u}_0\beta' \mathring{u}_0}- 12 \mathring{\r} \mathring{\s}^2 \mathring{u}^0_{(\alpha'}R_{\beta')\mathring{u}_0\sigma \mathring{u}_0}\Big] + \order(\lambda^2).
\end{align}
We provide the $\order(\lambda^2)$ term in the online repository \cite{PuncturesRepository} and note that it (along with the $\order(\lambda^3)$ and $\order(\lambda^4)$ terms) is also given in trace-reversed form in Eq.~(4.7) of Ref.~\cite{Heffernan:2012su}.
The subleading term is the sum of two pieces:
\begin{equation}
    \mathring h^{\S(1,1)}_{\mu\nu} = \mathring{h}^{\delta z}_{\mu\nu} + \mathring{h}^{\ms}_{\mu\nu}.
\end{equation}
$\mathring{h}^{\delta z}_{\mu\nu}$ is given by $h^{\delta z}_{\mu\nu}$ from Ref.~\cite{Pound:2014xva} (obtained in the context of a Gralla--Wald expansion; see footnote~\ref{fn:GW}), with the replacement $u^\mu\to\mathring u^\mu_0$.\footnote{This is not quite true in the case that $U_1$ is nonzero. Generically, $U_1$ contains terms proportional to $z^\mu_{1\perp}$, which should appear in $\mathring{h}^{\delta z}_{\mu\nu}$ but are instead placed in $\mathring{h}^{\ms}_{\mu\nu}$ in Eq.~\eqref{eq:hms} through the vector $v^\mu$.} The second term, which we refer to as the ``multiscale piece'' of the second-order singular field, is new to this paper. It takes the form
\begin{align}
    \mathring{h}^{\ms}_{\mu\nu} ={}& \frac{m}{\mathring{\s}_0^3}g^{\alpha'}_{\mu}g^{\beta'}_{\nu}\Bigl[\frac{2}{\lambda}\Bigl(\mathring{\r}_{0}v_{\gamma'}\sigma^{\gamma'}(g_{\alpha'\beta'} + 2\mathring{u}^0_{\alpha'}\mathring{u}^0_{\beta'}) \nonumber \\
        & - 4\mathring{\s}_0^2\mathring{u}^0_{(\alpha'}v_{\beta')}\Bigr) + \lambda^{0} a^{\gamma'}_{\ms}\Bigl(8\mathring{\r}_0\mathring{\s}_0^2g_{\gamma'(\alpha'}\mathring{u}^0_{\beta')} \nonumber \\
        & + \sigma_{\gamma'}(\mathring{\s}_0^2-\mathring{\r}_0^2)(g_{\alpha'\beta'} + 2\mathring{u}^0_{\alpha'}\mathring{u}^0_{\beta'})\Bigr)\Bigr] + \order(\lambda^1). \label{eq:hms}
\end{align}
As noted earlier, we provide the $\order(\lambda)$ term in the online repository \cite{PuncturesRepository}. In that repository we provide the full set of multiscale-expanded covariant punctures, valid for generic orbits in Kerr spacetime.

\subsubsection{Field equations}\label{sec:field eqs}

With the multiscale expansion of the singular field completed, the second-order singular field~\eqref{hS2 schematic} now becomes
\begin{multline}\label{hS2 multiscale schematic}
    \mathring h^{\S2}_{\mu\nu} \sim \frac{m^2 + m \mathring{z}_{1\perp} }{\lambda^2} + \frac{m^2 +m \mathring{u}_1 +m v + m h^{\R1}_{\mu\nu} + \mathring{\delta m}_{\mu\nu}}{\lambda} \\
    + {\cal O}(\log\lambda).
\end{multline}
This can be split into five pieces, 
\begin{equation}
    \hring^{\S2}_{\mu\nu} = \hring^{\S\S}_{\mu\nu} + \hring^{\S\R}_{\mu\nu} + \hring^{\delta m}_{\mu\nu} + \hring^{\delta z}_{\mu\nu} + \hring^{\ms}_{\mu\nu}. \label{eq:hS2_split}
\end{equation}
Here $\hring^{\S\S/\S\R}_{\mu\nu} \coloneqq h^{\S\S/\S\R}_{\mu\nu}|_{\gamma\to\gamma_{0}}$ contain all the terms of the form $m^2/\lambda^n$ and $m h^{\R1}_{\mu\nu}/\lambda$ in Eq.~\eqref{hS2 multiscale schematic}; $\hring^{\delta z}_{\mu\nu}$ contains the terms proportional to $m \mathring{z}_{1\perp}$ and $m\mathring{u}_1$; $\hring^{\ms}_{\mu\nu}$ contains the terms proportional to $m v$; and the mass monopole correction is altered by the term discussed in the previous section, so that
\begin{align}
    \mathring{\delta m}_{\mu\nu} ={}& \frac{m}{3}(2\hring^{\R1}_{\mu\nu} + g_{\mu\nu}\hring^{\R1}) + m(g_{\mu\nu}+2\mathring{u}^0_\mu\mathring{u}^0_\nu)\hring^{\R1}_{\alpha\beta}\mathring{u}_0^\alpha\mathring{u}_0^\beta \nonumber \\
        & + 4m\mathring{u}^0_{(\mu}(\hring^{\R1}_{\nu)\alpha}\mathring{u}_0^\alpha + 2\dot{\mathring{z}}^{1\perp}_{\nu)}). \label{eq:dm_ms}
\end{align}

As in the self-consistent scheme, each piece of the multiscale second-order singular field satisfies its own field equation.
These are
\begin{align}
    E^0_{\mu\nu}[\hring^{\S\S}] ={}& 2\delta^{2}R^0_{\mu\nu}[\hring^{\S1},\hring^{\S1}] & \text{for } x\notin\gamma, \label{eq:E_hSS} \\
    E^0_{\mu\nu}[\hring^{\S\R}] ={}& 4\mathring{Q}^{\Ric}_{\mu\nu}[\hring^{\S1}] & \text{for } x\notin\gamma, \label{eq:E_hSR} \\
    E^0_{\mu\nu}[\hring^{\delta m}] ={}& 0 & \text{for } x\notin\gamma, \label{eq:E_hdm} \\
    E^0_{\mu\nu}[\hring^{\delta z}] ={}& 0 & \text{for } x\notin\gamma, \label{eq:E_hdz} \\
    E^0_{\mu\nu}[\hring^{\ms}] ={}& -E^1_{\mu\nu}[\hring^{\S1}] & \text{for } x\notin\gamma, \label{eq:E_hms}
\end{align}
where $E^n_{\mu\nu}$ is the $n$th-order piece in the multiscale expansion of the Lorenz gauge wave operator, and $\mathring{Q}^{\Ric}_{\mu\nu}[h]$ has the same form as Eq.~\eqref{eq:Qdef} but with the replacements $\delta^{2}R_{\mu\nu}\to\delta^{2}R^0_{\mu\nu}$ and $h^{\R1}_{\mu\nu}\to\hring^{\R1}_{\mu\nu}$.
Following Appendix~\ref{sec:T_eff_TR_deriv}, we may promote these to the full domain where the field equations are then given by
\begin{align}
    E^0_{\mu\nu}[\hring^{\S\S}] ={}& 2\delta^{2}R^0_{\mu\nu}[\hring^{\S1},\hring^{\S1}], \label{eq:E_hSS_T} \\
    E^0_{\mu\nu}[\hring^{\S\R}] ={}& -16\pi\mathring{\TeffTR}^{\Ric}_{\mu\nu} + 4\mathring{Q}^{\Ric}_{\mu\nu}[\hring^{\S1}], \label{eq:E_hSR_T} \\
    E^0_{\mu\nu}[\hring^{\delta m}] ={}& -16\pi\mathring{\TeffTR}^{\delta m}_{\mu\nu}, \label{eq:E_hdm_T} \\
    E^0_{\mu\nu}[\hring^{\delta z}] ={}& -16\pi\mathring{\TeffTR}^{\delta z}_{\mu\nu}, \label{eq:E_hdz_T} \\
    E^0_{\mu\nu}[\hring^{\ms}] ={}& -16\pi\mathring{\TeffTR}^{\ms}_{\mu\nu} - E^1_{\mu\nu}[\hring^{\S1}]. \label{eq:E_hms_T}
\end{align}
Here the stress-energy tensors are individual pieces of the multiscale Detweiler stress-energy tensor with $\mathring{\TeffTR}^{\Ric}_{\mu\nu}$, $\mathring{\TeffTR}^{\delta m}_{\mu\nu}$, $\mathring{\TeffTR}^{\delta z}_{\mu\nu}$ and $\mathring{\TeffTR}^{\ms}_{\mu\nu}$ given by Eqs.~\eqref{eq:TRic_ms}, \eqref{eq:Tdm_ms}, \eqref{eq:Tdz} and~\eqref{eq:Tms}, respectively.

The terms proportional to $m\times m$ and the terms proportional to $m\times h^{\R1}$ satisfy separate gauge conditions:
\begin{align}
    Z^0_{\mu}[\bar{\hring}^{\S\S}] ={}& 0, \label{eq:Z_hSS} \\
    Z^0_{\mu}[\bar{\hring}^{\S\R} + \bar{\hring}^{\delta m} + \bar{\hring}^{\delta z} + \bar{\hring}^{\ms}] ={}& - Z^1_{\mu}[\bar{\hring}^{\S1}]. \label{eq:Z_hSR}
\end{align}
We highlight that the ``singular times regular'' pieces of the singular field must be combined in order to satisfy the gauge condition, while the individual pieces satisfy the individual field equations~\eqref{eq:E_hSR_T}--\eqref{eq:E_hms_T}. This is our primary motivation for focusing on the field equations without gauge damping in our checks of the punctures; since the gauge-damped field equations differ from the non-damped ones by terms proportional to the gauge condition, they effectively mix the separate pieces of the puncture.

\subsection{Coordinate expressions in Schwarzschild spacetime}\label{sec:punctures in coordinates}

The singular fields that have been derived so far are valid for any vacuum spacetime compatible with matched asymptotic expansions.
We continue our calculation by specialising to a quasicircular orbit in the Schwarzschild spacetime.
The multiscale framework for such a system is laid out in Appendix A of~\cite{Miller:2020bft}; we summarise the key results in the first part of this section.
We then move on to describe the rotated coordinate system we adopt in order to facilitate the decomposition into BLS harmonic modes.

\subsubsection{Quasicircular orbits in Schwarzschild spacetime: multiscale expansion}

To perform the multiscale expansion, we expand all quantities at fixed mechanical parameters $J^a\coloneqq (\Omega,\delta M,\delta S)$.
Here $\Omega\coloneqq \odv{\phi_p}{t}$ is the orbital azimuthal frequency associated with the small object's azimuthal position, $\phi_p$, $\delta M$ is the correction to the central body's mass, and $\delta S$ is the correction to the central body's spin.

For the orbital configuration being considered, our expanded worldline from Eq.~\eqref{z expansion} is given (in Schwarzschild coordinates) by
\begin{align}
    z^\alpha_0 ={}& (t,r_0(\Omega),\pi/2,\phi_p), \label{eq:z0} \\
    z^\alpha_1 ={}& (0,\deltar(J^a),0,0), \label{eq:z1}
\end{align}
where $\deltar$ is the correction to the worldline's radial position at fixed frequency. 
The expansion of the $4$-velocity from Eq.~\eqref{eq:u0def} is then given by
\begin{align}
    \mathring{u}_0^\alpha ={}& U_0 (1,0,0,\Omega), \label{eq:uc0_comp} \\
    v^\alpha ={}& U_0 (0,\rodot,0,0). \label{eq:v_comp}
\end{align}
In Eqs.~\eqref{eq:z0}--\eqref{eq:v_comp}
\begin{align}
    r_0 ={}& \frac{M}{(M\Omega)^{2/3}}, \label{eq:r0_def}\\
    U_0 ={}& \frac{1}{\sqrt{1-3(M\Omega)^{2/3}}}, \label{eq:U0} \\
    \deltar ={}& -\frac{\force^r_1}{3U^2_0\Omega^{2}f_0}, \label{eq:deltar} \\
    \rodot ={}& F_0\odv{r_0}{\Omega} \nonumber \\
        ={}& \frac{2f_0}{M^{1/3}U_0^4\Omega^{4/3}(1-6(M\Omega)^{2/3})}\force^t_1, \label{eq:r0dot}
\end{align}
where $f_0 \coloneqq 1-2(M\Omega)^{2/3}$, and $\force_1^\mu$ is the first-order self-force with components
\begin{align}
    \force_1^t ={}& - \frac{1}{2}f^{-1}_0 h^{\R1}_{\mathring{u}_{0}\mathring{u}_{0},t} \nonumber \\
        ={}& -\frac{U_0^2}{2f_0}(h^{\R1}_{tt,t}+2\Omega h^{\R1}_{t\phi,t} + \Omega^{2}h^{\R1}_{\phi\phi,t}), \\
    \force_1^r ={}& \frac{1}{2}f_0 h^{\R1}_{\mathring{u}_{0}\mathring{u}_{0},r} \nonumber \\
        ={}& \frac{1}{2}U_0^2f_0(h^{\R1}_{tt,r}+2\Omega h^{\R1}_{t\phi,r} + \Omega^{2}h^{\R1}_{\phi\phi,r}).
\end{align}
Equations~\eqref{eq:r0_def}--\eqref{eq:r0dot} were obtained by substituting Eqs.~\eqref{z expansion},~\eqref{eq:z0} and~\eqref{eq:z1} into Eq.~\eqref{eq:EoM} and solving order by order; see Ref.~\cite{Miller:2020bft}.

To calculate $h^{\delta z}_{\mu\nu}$, we require expressions for the projection of $z_1^\mu$, and of its time derivatives,  orthogonal to the worldline.
These can be easily found by contracting $\mathring{z}_{1}^\mu$ with $\mathring{P}^0_{\mu\nu}$ and taking time derivatives as in~\eqref{eq:D0tau0}, so that
\begin{align}
    \mathring{z}_{1\perp}^\mu ={}& \mathring{P}^{\mu}_0{}_{\nu}z^\nu_1 =z^\mu_1, \label{eq:z1perp} \\
    \dot{\mathring{z}}_{1\perp}^\mu ={}& \frac{U_0\Omega^{4/3}\deltar}{f_0 M^{1/3}}(M^{2/3},0,0,f_0\Omega^{1/3}), \\
    \ddot{\mathring{z}}_{1\perp}^\mu ={}& (0,-\Omega^2\deltar,0,0), \\
    \dddot{\mathring{z}}_{1\perp}^\mu ={}& -\frac{U_0\Omega^{10/3}\deltar}{f_0 M^{1/3}}(M^{2/3},0,0,f_0\Omega^{1/3}).
\end{align}
We additionally require expressions for the acceleration and its time derivative for the $h^{\ms}_{\mu\nu}$ piece of the singular field.
These are calculated using Eqs.~\eqref{eq:a_ms} and~\eqref{eq:adot_ms} and are given by
\begin{align}
    a^\mu_{\ms} ={}& \frac{U_0^{4}\Omega^{4/3}(1-6(M\Omega)^{2/3})\rodot}{2f_0 M^{1/3}}(M^{2/3},0,0,f_{0}\Omega^{1/3}) \nonumber \\
        ={}& \frac{\force^t_1}{M^{2/3}}(M^{2/3},0,0,f_0\Omega^{1/3}), \label{eq:a_BL} \\
    \adot^\mu_{\ms} ={}& \Bigr(0,0,-\frac{U_{0}^{3}\Omega^{2}}{2}(1-6(M\Omega)^{2/3})\rodot,0\Bigl) \nonumber \\
        ={}& \Bigr(0,0,-\frac{f_0\Omega^{2/3}}{M^{1/3}U_0}\force^t_1,0\Bigl). \label{eq:a_dot_BL}
\end{align}

\subsubsection{Rotated and Riemann normal coordinates}
\label{sec:rotatedcoordinates}

\begin{figure}[tb]
    \centering
    \includegraphics[width=0.6\linewidth]{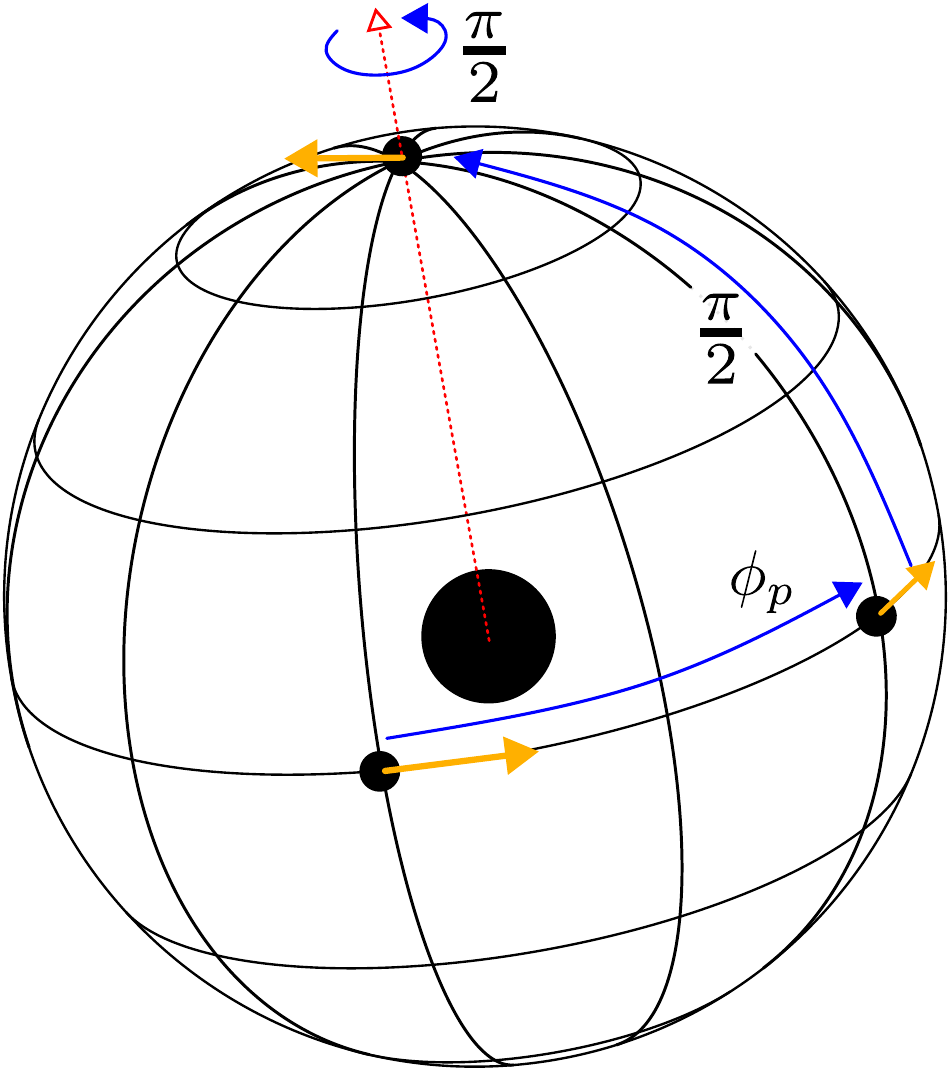}
    \caption{Visualisation of the transformation between rotated and unrotated coordinates, with blue arrows depicting rotations and solid orange arrows indicating the particle's velocity. The particle is rotated from $\phi=\phi_{p}$ to $\phi=0$, moved from the equator to the north pole and then has its velocity vector aligned with the curve $\beta=0$. }
    \label{fig:rotation}
\end{figure}

As in many previous self-force calculations, e.g.~\cite{Barack:2001gx,Barack:2002mha,Barack:2002bt,Detweiler:2002gi,Haas:2006ne,Heffernan:2012su,Heffernan:2012vj,Warburton:2013lea,Wardell:2015ada}, we take advantage of the spherical symmetry of the problem and, instead of using standard Schwarzschild coordinates, $(t,r,\theta,\phi)$, we instead adopt rotated polar coordinates, $(t,r,\alpha,\beta)$, that instantaneously place the small object at the north pole, $\alpha=0$.
The two coordinate systems are related via
\begin{align}
    \sin\theta\cos(\phi-\phi_p) ={}& \cos\alpha, \\
    \sin\theta\sin(\phi-\phi_p) ={}& \sin\alpha\cos\beta, \\
    \cos\theta ={}& \sin\alpha\sin\beta,
\end{align}
which is equivalent to a $z-y-z$ counter-clockwise Euler angle rotation of $(\phi_p,\pi/2,\pi/2)$; see Fig.~\ref{fig:rotation}.

We then introduce Riemann normal coordinates on the $2$-sphere, $(w_1,w_2)$, which are defined by
\begin{align}
    w_1 ={}& 2\sin\Bigl(\frac{\alpha}{2}\Bigr)\cos\beta, \\
    w_2 ={}& 2\sin\Bigl(\frac{\alpha}{2}\Bigr)\sin\beta,
\end{align}
so that the line element becomes~\cite{Heffernan:2012su}
\begin{align}
    \odif{s}^2 ={}& -f(r)\odif{t}^2 + f(r)^{-1}\odif{r}^2 \nonumber \\
        & + r^{2}\biggl[\biggl(\frac{16-w_2^2(8-w_1^2-w_2^2)}{4(4-w_1^2-w_2^2)}\biggr)\odif{w}_1^2 \nonumber \\
        & + 2\biggl(\frac{w_1 w_2(8-w_1^2-w_2^2)}{4(4-w_1^2-w_2^2)}\biggr)\odif{w}_1\odif{w}_2 \nonumber \\
        & + \biggl(\frac{16-w_1^2(8-w_1^2-w_2^2)}{4(4-w_1^2-w_2^2)}\biggr)\odif{w}_2^2\biggr]. \label{eq:metric_w1w2}
\end{align}
The reason that the Riemann normal coordinates are introduced is that the $(\alpha,\beta)$ coordinates are ill-defined on the worldline of the small object.
Near to the particle, the $(w_1,w_2)$ coordinates act as coordinates on a two-dimensional Cartesian plane which is tangent to a sphere of constant Schwarzschild radius as
\begin{align}
    w_1 \sim{}& \alpha\cos\beta, \label{eq:w1_def} \\
    w_2 \sim{}& \alpha\sin\beta. \label{eq:w2_def}
\end{align}

The fact that the particle is always at $\alpha=0$ in the rotated $(\alpha,\beta)$ coordinates has the advantage of reducing the number of azimuthal modes that need to be calculated.
The exact number required is related to the order of the puncture and the specific $i$ mode; see Sec.~\ref{sec:int_lm} for further details.
Azimuthal modes in the rotated coordinates, denoted by $m'$, are related to those in the unrotated coordinates, denoted by $m$, via
\begin{equation}
    f_{\ell m} = \sum_{m'=-\ell}^{\ell}D^{\ell}_{mm'}(\phi_p,\pi/2,\pi/2)f_{\ell m'}, \label{eq:flm_lmp}
\end{equation}
where $D^{\ell}_{mm'}(\alpha,\beta,\gamma)$ is the Wigner-D matrix~\cite{Wigner1931}.

We obtain coordinate expansions of Synge's world function, the parallel propagator, \(\mathring{r}_0\) and \(\mathring{s}_0\) using the techniques described in  Refs.~\cite{Ottewill:2008uu,Heffernan:2012su,Upton:2023tcv}.
Full details are available in the provided references but as an example, through order $\lambda$, we have
\begin{align}
    \sigma_{\alpha'} ={}& - \lambda\Delta x_{\alpha'} + \order(\lambda^2), \\
    g^{\beta'}_{\alpha} ={}& \delta^{\beta'}_{\alpha'} + \lambda \Gamma^{\beta'}_{\alpha'\gamma'}\Delta x^{\gamma'} + \order(\lambda^2), \\
    \mathring{r}_0 ={}& -\lambda \mathring{u}^{\alpha'}_{0}\Delta x_{\alpha'} + \order(\lambda^2), \\
    \mathring{s}_0 ={}& \lambda\rho + \order(\lambda^2),
\end{align}
where
\begin{equation}
    \rho^{2} = \mathring{P}^0_{\mu'\nu'}\Delta x^{\mu'}\Delta x^{\nu'}. \label{eq:rho_cov_def}
\end{equation}
These coordinate expansions are calculated in Riemann normal coordinates in terms of the coordinate difference,
\begin{align}
    \Delta x^{\mu'} ={}& (0,\Delta r, \Delta w_1, \Delta w_2) \nonumber \\
        ={}& \Bigl(0,r-r_0, 2\sin\Bigl(\frac{\alpha}{2}\Bigr)\cos\beta, 2\sin\Bigl(\frac{\alpha}{2}\Bigr)\sin\beta\Bigr).
\end{align}
Here, we have chosen $\Delta t=0$ so that the base and field points are evaluated on the same time slice.

These coordinate expansions, along with expressions for the four-velocity, Christoffel symbols and Riemann tensor (all evaluated on the worldline) are then substituted into the expressions for the singular field and truncated at the appropriate order to produce a puncture field of that order.
As an example,
\begin{equation}
    \rho^{2} = 2r_0^2 f_{0} U_0^2\chi(\delta^{2}+1-\cos\alpha), \label{eq:rho_def}
\end{equation}
with
\begin{equation}
    \delta^2 \coloneqq \frac{\Delta r^2}{2r_0^2 f_0^2 U_0^2 \chi}\label{eq:delta_def}
\end{equation}
and
\begin{equation}
    \chi \coloneqq 1 - \frac{M}{r_0 f_0}\sin^{2}\beta, \label{eq:chi_def}
\end{equation}
as in Refs.~\cite{Warburton:2013lea,Wardell:2015ada}.

With the singular field written in the Riemann normal coordinate basis, we convert to the polar coordinates $(\alpha,\beta)$ using the Jacobian between the two coordinate systems.
To make the integrals over $\alpha$ analytically tractable, we replace $\cos\bigl(\frac{\alpha}{2}\bigr)$ and odd powers of $\sin\bigl(\frac{\alpha}{2}\bigr)$ by expanding them in terms of $\sin\alpha$ and even powers of $\sin\bigl(\frac{\alpha}{2}\bigr)$:
\begin{align}
    \cos\Bigl(\frac{\alpha}{2}\Bigr) ={}& 1 - \frac{\lambda^{2}}{2}\sin^{2}\Bigl(\frac{\alpha}{2}\Bigr) - \frac{\lambda^{4}}{8}\sin^{4}\Bigl(\frac{\alpha}{2}\Bigr) + \order(\lambda^6), \label{eq:cosa2approx} \\
    \sin^{j}\Bigl(\frac{\alpha}{2}\Bigr) ={}& \sin\alpha\sin^{j-1}\Bigl(\frac{\alpha}{2}\Bigr)\Bigl[\frac{1}{2} + \frac{\lambda^2}{4}\sin^{2}\Bigl(\frac{\alpha}{2}\Bigr) \nonumber \\
        & + \frac{3\lambda^4}{16}\sin^{4}\Bigl(\frac{\alpha}{2}\Bigr) + \order(\lambda^6)\Bigr], \label{eq:sina2approx}
\end{align}
with $j$ odd.
As can be seen from Eqs.~\eqref{eq:w1_def} and~\eqref{eq:w2_def}, $\alpha \sim w_{1/2} \sim \lambda$ so we can use $\alpha$ to count powers of distance, which dictates the order of expansion needed in both Eqs.~\eqref{eq:cosa2approx} and~\eqref{eq:sina2approx}.
We then contract the metric perturbations with the approximated Jacobian and repeatedly apply Eqs.~\eqref{eq:cosa2approx}--\eqref{eq:sina2approx} to the resulting expressions, truncating at the appropriate order in distance.
This results in expressions for the singular field in $(t,r,\alpha,\beta)$ coordinates that are ready to be decomposed into BLS modes. Structurally, they have the form
\begin{widetext}
\begin{align}
     \mathring{h}^{\S 1}_{\mu\nu} &\sim h_{\mu\nu}^{\text{coeff}}\Biggl[\frac{1}{\lambda}\Biggl(c_{-1,0}^{\calP1}\frac{1}{\rho}\Biggr)
     + \lambda^0\Biggl(c_{0,1}^{\calP1}\frac{\Delta r}{\rho}
     + c_{0,3}^{\calP1}\frac{\Delta r^3}{\rho^3} + \biggl(d_{0,0}^{\calP1}\frac{1}{\rho}\biggr)\sin \alpha \cos \beta\Biggr)
     \nonumber \\ &
     +  \lambda^1 \Biggl(c_{1,0}^{\calP1}\rho+c_{1,2}^{\calP1}\frac{\Delta r^2}{\rho} + c_{1,4}^{\calP1}\frac{\Delta r^4}{\rho^3}+c_{1,6}^{\calP1}\frac{\Delta r^6}{\rho^5} + \biggl(d_{1,1}^{\calP1}\frac{\Delta r}{\rho}
     + d_{1,3}^{\calP1}\frac{\Delta r^3}{\rho^3}\biggr)\sin \alpha \cos \beta\Biggr)
     \nonumber \\ &
     +  \lambda^2 \Biggl(c_{2,1}^{\calP1}\Delta r\, \rho  +c_{2,3}^{\calP1}\frac{\Delta r^3}{\rho} + c_{2,5}^{\calP1}\frac{\Delta r^5}{\rho^3}+c_{2,7}^{\calP1}\frac{\Delta r^7}{\rho^5}+c_{2,9}^{\calP1}\frac{\Delta r^9}{\rho^7}
     \nonumber \\ & \quad
     + \biggl( d_{2,0}^{\calP1}\,\rho+d_{2,2}^{\calP1}\frac{\Delta r^2}{\rho} + d_{2,4}^{\calP1}\frac{\Delta r^4}{\rho^3}+d_{2,6}^{\calP1}\frac{\Delta r^6}{\rho^5}\biggr)\sin \alpha \cos \beta\Biggr)
     + \order(\lambda^{3}) \biggr],  \label{eq:hP1coord}
     \displaybreak[0] \\
     \mathring{h}^{\S\S}_{\mu\nu} &\sim h_{\mu\nu}^{\text{coeff}}\Biggl[\frac{1}{\lambda^2}\Biggl(c_{-2,0}^{\S\S}\frac{1}{\rho^2} + c^{\S\S}_{-2,2}\frac{\Delta r^2}{\rho^4} + \bigg(d^{\S\S}_{-2,1} \frac{\Delta r}{\rho^4}\biggr)\sin \alpha \cos \beta\Biggr)
     \nonumber \\ &
     + \frac{1}{\lambda}\Biggl(c_{-1,1}^{\S\S}\frac{\Delta r}{\rho^2}
     + c_{-1,3}^{\S\S}\frac{\Delta r^3}{\rho^4}+c_{-1,5}^{\S\S}\frac{\Delta r^5}{\rho^6} +\biggl(d_{-1,0}^{\S\S}\frac{1}{\rho^2} + d^{\S\S}_{-1,2}\frac{\Delta r^2}{\rho^4} + d^{\S\S}_{-1,4}\frac{\Delta r^4}{\rho^6}\biggr)\sin \alpha \cos \beta\Biggr)
     \nonumber \\ &
     + \lambda^0 \log \lambda \Biggl(c_{L,0}^{\S\S} \log \rho \Biggr)
     +  \lambda^0 \Biggl(c_{0,0}^{\S\S}+c_{0,2}^{\S\S}\frac{\Delta r^2}{\rho^2} + c_{0,4}^{\S\S}\frac{\Delta r^4}{\rho^4}+c_{0,6}^{\S\S}\frac{\Delta r^6}{\rho^6}+c_{0,8}^{\S\S}\frac{\Delta r^8}{\rho^8}
     \nonumber \\ & \quad
     + \bigg(d_{0,1}^{\S\S}\frac{\Delta r}{\rho^2} + d_{0,3}^{\S\S}\frac{\Delta r^3}{\rho^4}+d_{0,5}^{\S\S}\frac{\Delta r^5}{\rho^6}+d_{0,7}^{\S\S}\frac{\Delta r^7}{\rho^8} \bigg)\sin \alpha \cos \beta\Biggr)
     + \lambda^1 \log \lambda \Biggl(c_{L,1}^{\S\S} \Delta r \, \log \rho+d_{L,0}^{\S\S}  \, \log \rho \sin \alpha \cos \beta\Biggr)
     \nonumber \\ &
     +  \lambda^1 \Biggl(c_{1,1}^{\S\S}\Delta r+c_{1,3}^{\S\S}\frac{\Delta r^3}{\rho^2} + c_{1,5}^{\S\S}\frac{\Delta r^5}{\rho^4}+c_{1,7}^{\S\S}\frac{\Delta r^7}{\rho^6}+c_{1,9}^{\S\S}\frac{\Delta r^9}{\rho^8}+c_{1,11}^{\S\S}\frac{\Delta r^{11}}{\rho^{10}}
     \nonumber \\ & \quad + \biggl(d_{1,2}^{\S\S}\frac{\Delta r^2}{\rho^2} + d_{1,4}^{\S\S}\frac{\Delta r^4}{\rho^4}+d_{1,6}^{\S\S}\frac{\Delta r^6}{\rho^6}+d_{1,8}^{\S\S}\frac{\Delta r^8}{\rho^8}+d_{1,10}^{\S\S}\frac{\Delta r^{10}}{\rho^{10}}\biggr)\sin \alpha \cos \beta\Biggr) + \order(\lambda^{2}) \biggr], \label{eq:hSScoord}
    \displaybreak[0] \\
    \mathring{h}^{\S\R}_{\mu\nu} &\sim h_{\mu\nu}^{\text{coeff}} \Biggl[\frac{1}{\lambda}\Biggl(b^{\S\R}_{-1,0}\frac{1}{\rho} + b^{\S\R}_{-1,2} \frac{\Delta r^2}{\rho^3}+ b^{\S\R}_{-1,1}\frac{\Delta r}{\rho^3} \sin\alpha \cos \beta \Biggr)
    \nonumber \\ &
    +\lambda^0 \Biggl(b^{\S\R}_{0,1}\frac{\Delta r}{\rho} + b^{\S\R}_{0,3}\frac{\Delta r^3}{\rho^3} + b^{\S\R}_{0,5}\frac{\Delta r^5}{\rho^5} + \biggl(b^{\S\R}_{0,0} \frac{1}{\rho}+b^{\S\R}_{0,2} \frac{\Delta r^2}{\rho^3}+b^{\S\R}_{0,4} \frac{\Delta r^4}{\rho^5}\biggl) \sin\alpha \cos \beta \Biggr)
    \nonumber \\ & 
    +\lambda^1 \Biggl(b^{\S\R}_{1,0}\rho+b^{\S\R}_{1,2}\frac{\Delta r^2}{\rho} + b^{\S\R}_{1,4}\frac{\Delta r^4}{\rho^3} + b^{\S\R}_{1,6}\frac{\Delta r^6}{\rho^5} + b^{\S\R}_{1,8}\frac{\Delta r^8}{\rho^7} + \biggl(b^{\S\R}_{1,1}\frac{\Delta r}{\rho} + b^{\S\R}_{1,3}\frac{\Delta r^3}{\rho^3} + b^{\S\R}_{1,5}\frac{\Delta r^5}{\rho^5} + b^{\S\R}_{1,7}\frac{\Delta r^7}{\rho^7}\biggl) \sin\alpha \cos \beta \Biggr)
    \nonumber \\ &
    + \order(\lambda^2)\Biggr], \label{eq:hSRcoord}
    \displaybreak[0] \\
    \mathring{h}^{\delta m}_{\mu\nu} &\sim h_{\mu\nu}^{\text{coeff}} \Biggl[\frac{1}{\lambda}\Biggl(b^{\delta m}_{-1,0}\frac{1}{\rho}\Biggr)
    +\lambda^0 \Biggl(b^{\delta m}_{0,1}\frac{\Delta r}{\rho} + b^{\delta m}_{0,3}\frac{\Delta r^3}{\rho^3} + \biggl(b^{\delta m}_{0,0} \frac{1}{\rho}\biggl) \sin\alpha \cos \beta \Biggr)
    \nonumber \\ &
    +\lambda^1 \Biggl(b^{\delta m}_{1,0}\rho+b^{\delta m}_{1,2}\frac{\Delta r^2}{\rho} + b^{\delta m}_{1,4}\frac{\Delta r^4}{\rho^3} + b^{\delta m}_{1,6}\frac{\Delta r^6}{\rho^5} + \biggl(b^{\delta m}_{1,1}\frac{\Delta r}{\rho} + b^{\delta m}_{1,3}\frac{\Delta r^3}{\rho^3}\biggl) \sin\alpha \cos \beta \Biggr)
    + \order(\lambda^2)\Biggr], \label{eq:hdmcoord}
    \displaybreak[0] \\
    \mathring{h}^{\delta z}_{\mu\nu} &\sim h_{\mu\nu}^{\text{coeff}} \Biggl[\frac{1}{\lambda^2}\Biggl(c^{\delta z}_{-2,1}\frac{\Delta r}{\rho^3}\Biggr)
    +\frac{1}{\lambda} \Biggl(c^{\delta z}_{-1,0}\frac{1}{\rho} + c^{\delta z}_{-1,2}\frac{\Delta r^2}{\rho^3} + c^{\delta z}_{-1,4}\frac{\Delta r^4}{\rho^5} + \biggl(d_{-1,1}^{\delta z}\frac{\Delta r}{\rho^3}\biggr)\sin \alpha \cos \beta\Biggr)
    \nonumber \\ &
    +\lambda^0 \Biggl(c^{\delta z}_{0,1}\frac{\Delta r}{\rho} + c^{\delta z}_{0,3}\frac{\Delta r^3}{\rho^3} + c^{\delta z}_{0,5}\frac{\Delta r^5}{\rho^5} + c^{\delta z}_{0,7}\frac{\Delta r^7}{\rho^7} + \bigg( d^{\delta z}_{0,0}\frac{1}{\rho} + d^{\delta z}_{0,2}\frac{\Delta r^2}{\rho^3} + c^d{\delta z}_{0,4}\frac{\Delta r^4}{\rho^5} \bigg)\sin \alpha \cos \beta \Biggr)
    \nonumber \\ &
    +\lambda^1 \Biggl(c^{\delta z}_{1,0}\rho + c^{\delta z}_{1,2}\frac{\Delta r^2}{\rho} + c^{\delta z}_{1,4}\frac{\Delta r^4}{\rho^3} + c^{\delta z}_{1,6}\frac{\Delta r^6}{\rho^5} + c^{\delta z}_{1,8}\frac{\Delta r^8}{\rho^7} + c^{\delta z}_{1,10}\frac{\Delta r^{10}}{\rho^9} 
    \nonumber \\ & \quad
    + \biggl(d^{\delta z}_{1,1}\frac{\Delta r}{\rho} + d^{\delta z}_{1,3}\frac{\Delta r^3}{\rho^3} + d^{\delta z}_{1,5}\frac{\Delta r^5}{\rho^5} + d^{\delta z}_{1,7}\frac{\Delta r^7}{\rho^7}\biggr) \sin \alpha \cos \beta\Biggr)
    + \order(\lambda^2)\Biggr], \label{eq:hdzcoord}
    \displaybreak[0] \\
    \mathring{h}^{\ms}_{\mu\nu} &\sim h_{\mu\nu}^{\text{coeff}} \Biggl[\frac{1}{\lambda}\Biggl(d^{\ms}_{-1,0}\frac{1}{\rho} + \biggl(c^{\ms}_{-1,1} \frac{\Delta r}{\rho^3} \biggr) \sin \alpha \cos \beta \Biggr)
    \nonumber \\ &
    + \lambda^0 \Biggl(d^{\ms}_{0,1}\frac{\Delta r}{\rho} + d^{\ms}_{0,3}\frac{\Delta r^3}{\rho^3} + \biggl(c^{\ms}_{0,0} \frac{1}{\rho}+c^{\ms}_{0,2} \frac{\Delta r^2}{\rho^3}+c^{\ms}_{0,4} \frac{\Delta r^4}{\rho^5} \biggr) \sin \alpha \cos \beta \Biggr)
    \nonumber \\ &
    +\lambda^1 \Biggl(d^{\ms}_{1,0}\, \rho + d^{\ms}_{1,2}\frac{\Delta r^2}{\rho} + d^{\ms}_{1,4}\frac{\Delta r^4}{\rho^3} + d^{\ms}_{1,6}\frac{\Delta r^6}{\rho^5}  + \biggl(c^{\ms}_{1,1} \frac{\Delta r}{\rho}+c^{\ms}_{1,3} \frac{\Delta r^3}{\rho^3}+c^{\ms}_{1,5} \frac{\Delta r^5}{\rho^5} +c^{\ms}_{1,7} \frac{\Delta r^7}{\rho^7} \biggr) \sin \alpha \cos \beta \Biggr)
    \nonumber \\ &
    + \order(\lambda^2)\Biggr], \label{eq:hmscoord}
\end{align}
where
\begin{equation}
    h_{\mu\nu}^{\text{coeff}} =
    \begin{bmatrix}
    1 & 1 & \cos \beta & \sin\alpha \sin \beta\\
    1 & 1 & \cos \beta & \sin \alpha \sin \beta\\
    \cos \beta & \cos \beta & 1 & \sin \alpha \cos\beta\sin\beta \\
    \sin \alpha \sin \beta & \sin \alpha  \sin \beta & \sin \alpha \cos \beta \sin \beta &\sin^2 \alpha
\end{bmatrix}
\end{equation}
\end{widetext}
captures the non-smooth behaviour of the spherical coordinate system and where the $b_{i,j}^\cdot$, $c_{i,j}^\cdot$ and $d_{i,j}^\cdot$ depend on the component being considered (with $c_{i,j}^\cdot$ only non-zero for $h_{tt}$, $h_{t\alpha}$, $h_{t\beta}$, $h_{rr}$, $h_{\alpha\alpha}$, $h_{\alpha\beta}$ and $h_{\beta\beta}$, and $d_{i,j}^\cdot$ only non-zero for $h_{tr}$, $h_{r\alpha}$ and $h_{r\beta}$) and are a finite power series in $\chi$ and $1/\chi$. The coefficients in the power series depend on the orbital parameters and, with the exception of $c^{\calP1}_{i,j}$ and $c^{\S\S}_{i,j}$, the regular field  and its derivatives evaluated on the worldline.

\subsection{Decomposition into \texorpdfstring{$(\ell m')$}{(ℓm')} modes}\label{sec:int_lm}

A procedure for calculating BLS modes of the singular field was previously described in Ref.~\cite{Miller:2016hjv} for the case of a scalar field and in Ref.~\cite{Wardell:2015ada} for the first-order gravitational case, both involving only even values of $m'$ and odd powers of $\rho$.
The extension of the method to the remaining cases needed for the second-order gravitational singular field (including vector and tensor sectors, odd $m'$ modes, even powers of $\rho$ and terms featuring $\log\rho$) was outlined in Ref.~\cite{Pound:2021qin} but the full details were never presented.
Here, we recap the general strategy from both references before going on to fully expand and provide details for the cases discussed in Ref.~\cite{Pound:2021qin}.

Before proceeding with the mode decomposition, we note that the transformation to $(\alpha,\beta)$ coordinates introduces spurious, $\beta$-dependent directional non-smoothness in $\rho$ at the south pole, and that this non-smoothness leads to slow convergence with $\ell$.
To handle this, as in a number of previous self-force calculations, e.g.~\cite{Vega:2009qb,Warburton:2013lea,Wardell:2015ada,Miller:2016hjv}, we introduce a window function that smooths out the non-physical behaviour at the south pole while keeping the singular fields unchanged near the worldline.
This is discussed extensively in Ref.~\cite{Miller:2016hjv}, and we refer interested readers to that reference for full details.
The main takeaway is that we require a window function that, at the north pole, acts as $\mathcal{W}_{\mathbf{m}}^n = 1 + \order(\alpha^n)$, and at the south pole, acts as $\mathcal{W}_{\mathbf{m}}^n=\order[(\pi-\alpha)^{\mathbf{m}}]$ (recalling that powers of $\alpha$ count powers of $\lambda$).
To satisfy these conditions, we use the window function from Ref.~\cite{Miller:2016hjv},
\begin{align}
    \mathcal{W}^n_{\mathbf{m}} \coloneq{}& 1-\frac{n}{2}\binom{(\mathbf{m}+n-2)/2}{n/2} \nonumber \\
        & \times B\biggl(\frac{1-\cos\alpha}{2};\frac{n}{2},\frac{\mathbf{m}}{2}\biggr), \label{eq:windowfunc}
\end{align}
where $\displaystyle\binom{p}{q}$ is the binomial coefficient and $B(z;a,b)$ is the incomplete beta function; see Sec.~8.17 of Ref.~\cite{NIST:DLMF}.
When $n$ and $\mathbf{m}$ are positive integers and $\mathbf{m}$ is even, then $\mathcal{W}^n_{\mathbf{m}}$ takes the form of a polynomial in $\frac{1-\cos\alpha}{2}$ which can in turn be converted to an even polynomial in $\rho$ using Eq.~\eqref{eq:rho_def}.
As we have four total orders in our punctures, we choose $n=4$ and we let $\mathbf{m}$ be $|m'|+|s|$ for even $m'$ or $|m'|+|s|+1$ for odd $m'$, where $s$ is the spin weight of the field ($s=0$ for $i=1,2,3,6$, $|s|=1$ for $i=4,5,8,9$ and $|s|=2$ for $i=7,10$).

With the window function incorporated, we can continue on with the mode decomposition.
The BLS mode coefficients of $\mathring{h}^n_{\mu\nu}$ are then obtained by calculating
\begin{equation}\label{Rnilm}
    h^n_{\ilm'} = \frac{r}{a_{i\ell}\kappa_{i}}\oint\mathcal{W}^{4}_{2\ceil[\big]{\tfrac{|m'|+|s|}{2}}}\mathring{h}^n_{\alpha\beta}\eta^{\alpha\mu}\eta^{\beta\nu}Y^{\ilm'*}_{\mu\nu}\odif{\Omega},
\end{equation}
where $\oint\odif{\Omega} \coloneq \int_0^{2\pi}\odif{\beta}\int_0^\pi\sin\alpha\odif{\alpha}$, the BLS tensor harmonics $Y^{\ilm}_{\mu\nu}$ are given explicitly in App.~B of Ref.~\cite{Miller:2020bft} (with the replacement $(\theta,\phi)\to(\alpha,\beta)$), $a_{i\ell}$ is given by Eq.~\eqref{eq:ail_def},
\begin{equation}
    \kappa_{i} \coloneq \begin{cases}
        f(r)^{2} & \text{for } i=3, \\
        1 & \text{otherwise},
    \end{cases}
\end{equation}
and
\begin{equation}
    \eta^{\mu\nu} \coloneq \diag(1,f(r)^{2},r^{-2},r^{-2}\sin^{-2}\alpha). \label{eq:nu_def}
\end{equation}

As discussed in Ref.~\cite{Pound:2021qin}, to reduce the number of integrals we need to perform, we note that the parity of the expressions in $\beta$ will mean that certain integrals over $\beta$ will be identically zero for certain $m'$.
That is, the integral is only ever non-vanishing when the integrand is an even function of $\beta$. We can split this into four cases:
\begin{enumerate}
    \item Even power of both $\sin\beta$ and $\cos\beta$. \label{list:betacase1}
    \item Odd power of both $\sin\beta$ and $\cos\beta$. \label{list:betacase2}
    \item Odd power of $\cos\beta$ and even power of $\sin\beta$. \label{list:betacase3}
    \item Odd power of $\sin\beta$ and even power of $\cos\beta$. \label{list:betacase4}
\end{enumerate}
When integrated against the factor of $e^{-im'\beta}$ appearing in the tensor spherical harmonics, cases \ref{list:betacase1} and \ref{list:betacase2} will be non-vanishing for $m'$ even, and cases \ref{list:betacase3} and \ref{list:betacase4} will be non-vanishing for $m'$ odd.
Additionally, cases \ref{list:betacase1} and \ref{list:betacase3} only require the real part of $e^{-im'\beta}$, and cases \ref{list:betacase2} and \ref{list:betacase4} only require the imaginary part.
Taking this into account, we dramatically reduce the number of integrals over $\alpha$ that we must perform.
\begin{table}[tb]
    \centering
    \begin{ruledtabular}
    \begin{tabular}{cccccc}
         & $h^{\S\S}_{\mu\nu}$ & $h^{\S\R}_{\mu\nu}$ & $h^{\delta z}_{\mu\nu}$ & $h^{\delta m}_{\mu\nu}$ & $h^{\ms}_{\mu\nu}$ \\
         \midrule
         $i=1,3,6$ & 1 & 1,3 & 1 & 1,3 & 3,4 \\
         $i=2$ & 3 & 1,3 & 3 & 1,3 & 1 \\
         $i=4,8$ & 3,4 & all & 3,4 & all & 1,2 \\
         $i=5,7,9,10$ & 1,2 & all & 1,2 & all & 3,4 \\
    \end{tabular}
    \end{ruledtabular}
    \caption{The different cases, as described under Eq.~\eqref{eq:nu_def}, encountered in each of the second-order singular fields when performing the integration over $\beta$. By taking these cases into account, we can dramatically simplify and reduce the number of integrals over $\alpha$ that need to be performed.}
    \label{tab:beta_cases}
\end{table}
Table~\ref{tab:beta_cases} details the cases in which $i$ modes of second-order singular fields appear.

\begin{figure}[!htp]
    \centering
    \includegraphics[width=.85\linewidth]{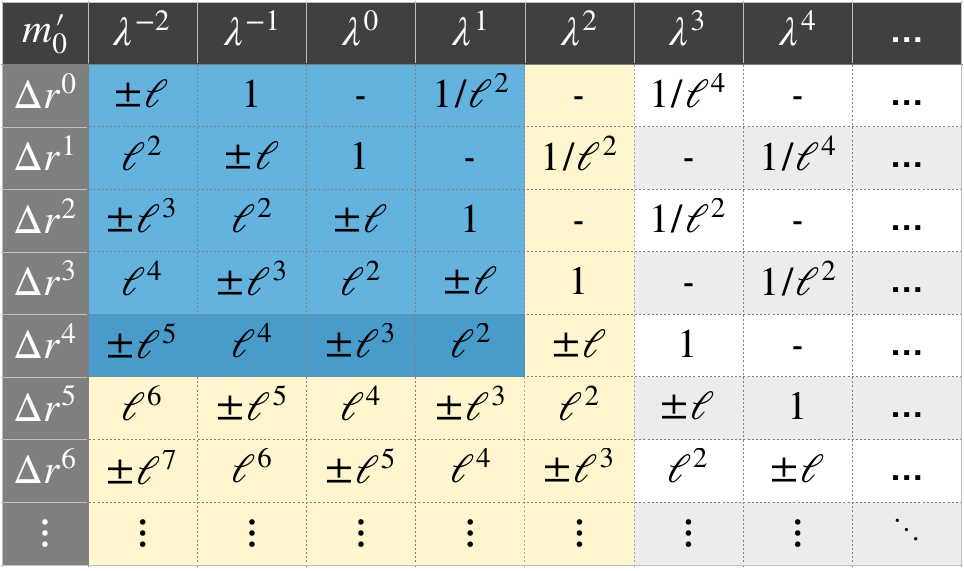}\quad
    \includegraphics[width=.85\linewidth]{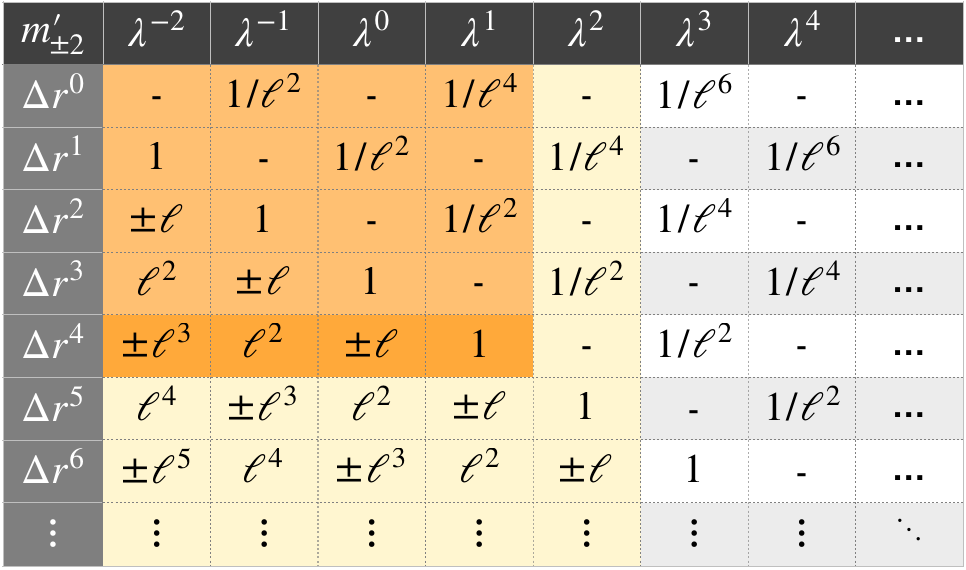}
    \caption{Large-$\ell$ behaviour of the modes of a singular field with \emph{odd} powers of $\rho$ for $|m'|=|s|$ (top) and $|m'|=|s|+2$ (bottom). Some cases, denoted by a $\pm$, have a directionally dependent value as $\Delta r \to 0$. Light coloured regions correspond to pieces required for the checks in Sec.~\ref{sec:validation} while dark coloured regions correspond to extra pieces included in the expressions we provide online \cite{PuncturesRepository}. Yellow coloured regions correspond to pieces included in the first order puncture that is used to compute the second order Ricci tensor (see Sec.~\ref{sec:d2R}). As $|m'|$ increases from $|s|$ the columns shift leftwards. The $|m'|=1, 3, 4$ cases are not shown but are similarly included for the `singular times regular' punctures, while the `singular times singular' puncture includes the $|m'|=1$ case.
} 
    \label{fig:SingularFieldStructure}
\end{figure}
\begin{figure*}[!htp]
    \centering
    \includegraphics[width=0.9\linewidth]{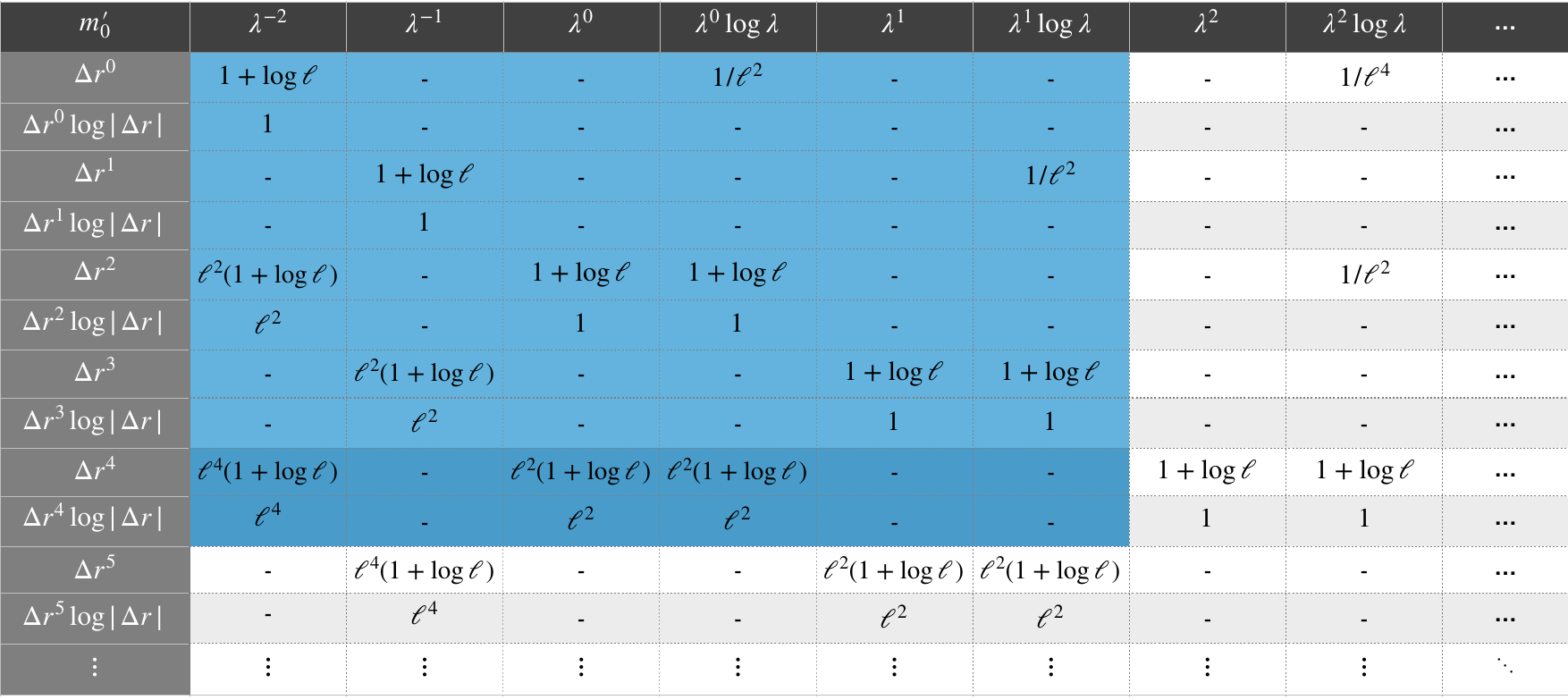}
    \caption{Structure of the modes of a singular field with \emph{even} powers of $\rho$ for $|m'|=|s|$. In this case we get additional logarithmic (in $\lambda$, $\Delta r$ and $\ell$) contributions compared to the odd-power case. Cases with a dash indicate that there is no large-$\ell$ contribution, although in some case there is a contribution for a specific $\ell$.}
    \label{fig:SingularFieldStructure-even}
\end{figure*}
After performing the integral over $\alpha$ in Eq.~\eqref{Rnilm} for the $\mathring{h}^n_{\mu\nu}$'s given in Eqs.~\eqref{eq:hP1coord}--\eqref{eq:hmscoord}, we find expressions of the form~\cite{Pound:2021qin}
\begin{equation}
    (\delta^2+2)^{(n+2)/2}\sum_{i}r_i\delta^{2i} + |\delta|\delta^{n+1}\sum_{i}s_{i}\delta^{2i}, \label{eq:rhooddint}
\end{equation}
for integrals of odd powers of $\rho$ and
\begin{equation}
    \delta^{n+2}(\delta^{2}+2)^{(n+2)/2}\sum_{i}p_{i}\delta^{2i} 
        + \log\biggl(\frac{\delta^{2}+2}{\delta^2}\biggr)\sum_{i}q_{i}\delta^{2i}, \label{eq:rhoevenint}
\end{equation}
for integrals of even powers of $\rho$ and $\log\rho$.
The sum ranges and coefficient values depend on the specific power of $\rho$ and the values of $\ell$ and $m'$ being considered.
When integrated over $\beta$, Eq.~\eqref{eq:rhooddint} leads to sums of elliptic integrals of the first, second, and third kinds.
The direct integration of Eq.~\eqref{eq:rhoevenint} is less straightforward, but is fortunately not necessary as it suffices to work with series expansions valid through some power of $\Delta r$. In that case, the resulting $\beta$ integrals simplify dramatically for both Eqs.~\eqref{eq:rhooddint} and \eqref{eq:rhoevenint}. As discussed later in Sec.~\ref{sec:beta_int}, the final result is a sum of polynomials, elliptic integrals and derivatives of hypergeometric functions.

The specific order in $\Delta r$ that the series expansion must be taken to depends on the quantity being computed and the degree of regularity required---see Figs.~\ref{fig:SingularFieldStructure} and \ref{fig:SingularFieldStructure-even}, which display the large-$\ell$ and smoothness-in-$\Delta r$ behaviour of the modes when re-expanded in powers of $\Delta r$. In this work we retain terms through $\order(\Delta r^4)$, one order higher than is strictly necessary for the checks in Sec.~\ref{sec:validation}. We include the additional order in anticipation that it may be useful---particularly if combined with a calculation of the order-$\lambda^2$ piece of the puncture---in calculations involving the Teukolsky formalism, where additional derivatives need to be taken~\cite{Leather:2025InPrep}.

For a similar reason, the number of $m'$ modes that we must calculate is also constrained.
As discussed in Ref.~\cite{Pound:2021qin}, the highest mode we need is $|m'|=|s|+n$, where $n$ is the number of derivatives that we wish to take. This is illustrated in the lower panel in Fig.~\ref{fig:SingularFieldStructure}, which illustrates how the behaviour of the modes becomes more smooth as $|m'|$ moves away from $|s|$.

\subsubsection{Integration over \texorpdfstring{$\alpha$}{α}}\label{sec:int_alpha}

The strategy for integration over $\alpha$ follows that laid out in Sec.~IV~B~1 of Ref.~\cite{Miller:2016hjv}.
However, that reference only dealt with scalar quantities, even $m'$ (due to the powers of $\sin\beta$ and $\cos\beta$ that appear; see the preceding section) and odd powers of $\rho$.
Here we recap the method of Ref.~\cite{Miller:2016hjv} and extend it to cover odd $m'$, tensorial quantities, even powers of $\rho$ and terms featuring $\log\rho$.

We first recall that since $\mathcal{W}^n_{\mathbf{m}}$ is a polynomial in $\frac{1-\cos\alpha}{2}$, we can rewrite it in terms of even powers of $\rho$.
Thus, all integrals in the scalar sector with even $m'$ can be written in the form
\begin{equation}
    \int_{-1}^1\rho^n P_{\ell}^{m'}(x)\odif{x}, \label{eq:rhonint}
\end{equation}
where $x=\cos\alpha$.
As presented in Eq.~(44) of Ref.~\cite{Miller:2016hjv}, for $m'=0$, we see that for odd $n$,
\begin{align}
    \int_{-1}^{1}(\delta^2+1-x)^{n/2}P^0_{\ell}\odif{x} ={}& I(n,\ell,0), \label{eq:rhonintm0}
\end{align}
where
\begin{widetext}
\begin{align}
    I(n,\ell,0) = \Biggl[\frac{(-1)^{\frac{n+1}{2}}(\delta^{2}+2)^{\frac{n}{2}+1}\bigl[\bigl(\frac{1}{2}\bigr)_{\frac{n+1}{2}}\bigr]^2}{(\ell-n/2)_{n+2}}\sum_{k=0}^{\ell}\frac{(-1)^k\delta^{2k}(\ell-k+1)_{2k}}{2^{k}k!(n/2-k+1)_k} - |\delta|\delta^{n+1}\sum_{k=0}^{\ell}\frac{\delta^{2k}(\ell-k+1)_{2k}}{2^{k}k!(n/2+1)_{k+1}}\Biggr]. \label{eq:Inl0}
\end{align}
\end{widetext}
Following Ref.~\cite{Miller:2016hjv}, the $m'>0$ modes can then be found by taking advantage of the definition of the associated Legendre polynomials in terms of the ordinary Legendre polynomials,
\begin{equation}
    P_{\ell}^m(x) = (-1)^m(1-x^2)^{m/2}\frac{d^m}{dx^m}P_{\ell},
\end{equation}
and integrating Eq.~\eqref{eq:rhonint} by parts $m'$ times to get an expression of the form of Eq.~\eqref{eq:rhonintm0} plus boundary terms.
Concretely, we find that
\begin{align}
    I(n,\ell,m') ={}& \int_{-1}^1R^{n/2}(x) P_{\ell}^{m'}(x)dx \nonumber \\
    ={}& (-1)^{m'}\sum_{j=0}^{m'}a_{jm'n}\Bigl(1+j-m'+\frac{n}{2}\Bigr)_{m'} \nonumber \\
        & \times I(n+2(j-m'),\ell,0) \nonumber \\
    & + \Biggl[\sum_{r=0}^{m'-1}\sum_{j=0}^{m'}a_{jm'n}\Bigl(1+j+\frac{n}{2}-r\Bigr)_{r} \nonumber \\
    & \times R^{j+\frac{n}{2}-r}(x)\frac{d^{m'-r-1}}{dx^{m'-r-1}}P_\ell(x)\Biggr]_{x=-1}^{x=1},
\end{align}
where
\begin{equation}
    R(x) = \delta^2+1-x, \label{eq:Rdeltadef}
\end{equation}
and
\begin{align}
    a_{jm'n} ={}& \sum_{i=0}^{j}(-1)^{\frac{m'}{2}-n+i}\binom{\frac{m'}{2}}{i}\binom{\frac{m'}{2}}{j-i} \nonumber \\
        & \times (\delta^2)^{\frac{m'}{2}-i}(2+\delta^{2})^{\frac{m'}{2}-j+i}. \label{eq:ajmndef}
\end{align}
Finally, as noted before, the window function can be written in terms of powers of $\rho$ (and thus $R$).
Therefore, we introduce the function $\mathcal{A}^{(p)}_{(n,\ell,m')}$ as shorthand for the integral of $\rho^n$ multiplied by a window function of order $|m'|+p$.
$\mathcal{A}^{(p)}_{(n,\ell,m')}$ can be written as a linear combination of $I(n,\ell,m')$ with different $n$ and is explicitly given by
\begin{align}
    \mathcal{A}^{(p)}_{(n,\ell,m')} \coloneqq{}& B^{n/2}\int_{-1}^1\mathcal{W}^4_{|m'|+p}R^{n/2}P_{\ell}^{m'}\odif{x} \nonumber \\
        ={}& B^{n/2}\sum_{k=0}^{m'+p+2}b_{km'p}I(n+k,\ell,m'), \label{eq:Aintdef}
\end{align}
where $m'$ and $k$ are even, $b_{km'p}$ is a function of $\delta^{2}$ and
\begin{equation}
    B\coloneq 2r_0^2 f_{0} U_0^2\chi, \label{eq:Bdef}
\end{equation}
so that $\rho^2=R(x)B$.
In Eq.~\eqref{eq:Aintdef}, $p$ provides us with the option to increase the regularisation of the integral using a smoother window function if needed.
We will use this in the vector and tensor sectors ($|s|=1$ and $|s|=2$), where a stronger smoothing by the window function is required to maintain convergence.

Returning now to the case of $m'$ odd, we leverage the various recurrence relations for the associated Legendre polynomials to write the integrals in terms of $m'$ even, which we can calculate using the above method.
As an example, in the scalar sector, we have integrals of the form
\begin{equation}
    \int_{-1}^{1}\rho^nP_{\ell}^{m'}(x)\sqrt{1-x^2}\odif{x}.
\end{equation}
This can be written in terms of even-$m'$ integrals using the relation
\begin{align}
    \sqrt{1-x^2}P_{\ell}^{m} ={}& \frac{1}{2\ell+1}[(\ell-m+1)(\ell-m+2)P_{\ell+1}^{m-1} \nonumber \\
        & - (\ell+m-1)(\ell+m)P_{\ell-1}^{m-1}].
\end{align}
Therefore,
\begin{align}
    \int_{-1}^{1}\rho^nP_{\ell}^{m'}(x)\sqrt{1-x^2}\odif{x} \nonumber \\
    \MoveEqLeft[10] \begin{multlined}= \frac{1}{2\ell+1}\Bigl[(\ell-m'+1)(\ell-m'+2)\mathcal{A}^{(0)}_{(n,\ell+1,m'-1)} \\
        - (\ell+m'-1)(\ell+m')\mathcal{A}^{(0)}_{(n,\ell-1,m'-1)}\Bigr]. \end{multlined}
\end{align}

In the vector sector, for $i=4,8$, due to the integral over $\beta$, we only encounter $m'$ odd.
Two different kinds of integrals appear and are given by
\begin{align}
    \int_{-1}^{1}\frac{\rho^n}{\sqrt{1-x^2}}\frac{(\ell+1)(\ell+m')P_{\ell-1}^{m'}-\ell(\ell-m'+1)P_{\ell+1}^{m'}}{2\ell+1} \odif{x} \nonumber \\
    \MoveEqLeft[24] = \frac{1}{2}\Bigl[(\ell-m'+1)(\ell+m)\mathcal{A}^{(2)}_{(n,\ell,m'-1)} - \mathcal{A}^{(0)}_{(n,\ell,m'+1)}\Bigr],
\end{align}
and
\begin{align}
    \int_{-1}^{1}\frac{\rho^nm'}{\sqrt{1-x^{2}}}P_{\ell}^{m'}\odif{x} \nonumber \\
        \MoveEqLeft[6] \begin{multlined}= -\frac{1}{2}\Bigl[(\ell+m'-1)(\ell+m')\mathcal{A}^{(2)}_{(n,\ell-1,m'-1)} \\
            + \mathcal{A}^{(0)}_{(n,\ell-1,m'+1)}\Bigr].\end{multlined}
\end{align}
For $i=5,9$, we only have $m'$ even non-vanishing.
Again, we have two kinds of integrals which are given by
\begin{align}
    \int_{-1}^{1}\frac{\rho^{n}}{2\ell+1}[(\ell+1)(\ell+m')P_{\ell-1}^{m'}-\ell(\ell-m+1)P_{\ell+1}^{m'}]\odif{x} \nonumber \\
        \MoveEqLeft[20] \begin{multlined}=\frac{1}{2\ell+1}\Bigl[(\ell+1)(\ell+m')\mathcal{A}^{(0)}_{(n,\ell-1,m')} \\
         -\ell(\ell-m'+1)\mathcal{A}^{(0)}_{(n,\ell+1,m')}\Bigr],\end{multlined}
\end{align}
and
\begin{align}
    \int_{-1}^{1}\rho^{n}m'P_{\ell}^{m'}\odif{x} = m'\mathcal{A}^{(0)}_{(n,\ell,m')}.
\end{align}

Finally, in the tensor sector, we encounter four different types of integrals, two for even $m'$ and two for odd $m'$.
For $m'$ even they are
\begin{widetext}
\begin{align}
    \begin{multlined}\int_{-1}^{1}\frac{\rho^n}{(1-x^{2})(2\ell-1)(2\ell+1)(2\ell+3)}\Bigl[(\ell+1)(\ell+2)(2\ell+3)(\ell+m'-1)(\ell+m)P_{\ell-2}^{m'} \\
        - 2(\ell-1)(\ell+2)(2\ell+1)(\ell^2+\ell-3(m')^2)P_{\ell}^{m'} + (\ell-1)\ell(2\ell-1)(\ell-m'+1)(\ell-m'+2)P_{\ell+2}^{m'}\Bigr]\odif{x}\end{multlined} \nonumber \\
        \MoveEqLeft[40] = \frac{1}{2}(\ell-m'+1)(\ell-m'+2)(\ell+m'-1)\mathcal{A}^{(4)}_{(n,\ell,m'-2)} + (m')^2\mathcal{A}^{(2)}_{(n,\ell,m')} + \frac{1}{2}\mathcal{A}^{(0)}_{(n,\ell,m'+2)},
\end{align}
and
\begin{align}
    \int_{-1}^{1}\frac{\rho^nm'}{(1-x^2)(2\ell+1)}\Bigl[(\ell+2)(\ell+m')P_{\ell-1}^{m'} - (\ell-1)(\ell-m'+1)P_{\ell+1}^{m'}\Bigr] \nonumber \\
        \MoveEqLeft[28] = -\frac{1}{4}(\ell+m')(\ell-m'+1)(\ell-m'+2)(\ell-m'+3)\mathcal{A}^{(4)}_{(n,\ell+1,m'-2)} \nonumber \\
            \MoveEqLeft[27] - \frac{1}{2}m'(\ell-m'+1)\mathcal{A}^{(2)}_{(n,\ell+1,m')} + \frac{1}{4}\mathcal{A}^{(0)}_{(n,\ell+1,m'+2)},
\end{align}
while for $m'$ odd, they are
\begin{align}
    \int_{-1}^{1}\frac{\rho^n}{\sqrt{1-x^2}}\Bigl[\Bigl(1+\ell+(m')^2 +(\ell+1)(\ell+2)x\Bigr)P_{\ell}^{m'} + (\ell-m'+1)\Bigl((\ell-m'+2)P_{\ell+2}^{m'} - (2\ell+5)xP_{\ell+1}^{m'}\Bigr)\Bigr]\odif{x} \nonumber \\
        \MoveEqLeft[46] = \frac{1}{2m'(2\ell-1)(2\ell+1)(2\ell+3)}\Bigl[\bigl(2\ell-2\ell^3-3m'+\ell m'+8\ell^2m'+4\ell^3m'-2\ell(m')^2-8\ell^2(m')^2\bigr)(\ell+2)(\ell+m') \nonumber \\
        \MoveEqLeft[45]\times(\ell+m'-1)\mathcal{A}^{(2)}_{(n,\ell-1,m'-1)} + \bigl(2\ell-2\ell^3+3m'-\ell m'-8\ell^2m'-4\ell^3m'-2\ell(m')^2-8\ell^2(m')^2\bigr)(\ell+2)\mathcal{A}^{(0)}_{(n,\ell-1,m'+1)} \nonumber \\
        \MoveEqLeft[45] + (\ell-1)(\ell-m'+1)(\ell-m'+2)\Bigl(2\ell(\ell+1)(\ell+2)+\ell m'(3-4\ell(\ell+1))-2(m')^{2}(\ell+1)(4\ell+3)\Bigr)\mathcal{A}^{(2)}_{(n,\ell+1,m'-1)} \nonumber \\
        \MoveEqLeft[45] + (\ell-1)\Bigl(2\ell(\ell+1)(\ell+2)+\ell m'(2\ell-1)(2\ell+3)-2(m')^{2}(\ell+1)(4\ell+3)\Bigr)\mathcal{A}^{(0)}_{(n,\ell+1,m'+1)}\Bigr],
\end{align}
\end{widetext}
and
\begin{align}
    \int_{-1}^{1}\frac{\rho^n}{\sqrt{1-x^2}}\Bigl[(\ell-m'+1)P_{\ell+1}^{m'}-(\ell+2)xP_{\ell}^{m'}\Bigr]\odif{x} \nonumber \\
        \MoveEqLeft[20] = \frac{1}{2m'}\Bigl[(m'-1)(m'-\ell-1)(\ell+m')\mathcal{A}^{(2)}_{(n,\ell,m'+1)} \nonumber \\
        \MoveEqLeft[19] + (m'+1)\mathcal{A}^{(0)}_{(n,\ell,m'+1)}\Bigr].
\end{align}
After performing these integrals, we then perform a series expansion in powers of $\delta$, as described earlier.

The strategy we use for even powers of $\rho^n$ and for $\log\rho$ is slightly different.
While we know that these cases lead to expressions of the form of Eq.~\eqref{eq:rhoevenint} we have not been able to determine the specific $n$-, $\ell$- and $m'$-dependence in $p_{i}$ and $q_{i}$.
This results in us not having an equivalent expression to Eq.~\eqref{eq:Inl0} for these cases.
Instead, we explicitly calculate expressions for the required $n$ and $m'$ values
over a range of values of $\ell$ in the scalar, vector and tensor sectors, before performing a series expansion in $\delta$ to the appropriate order.
From there, we can determine the dependence on $\ell$ for each order in $\delta$ through the use of \textsc{Mathematica}'s \texttt{FindSequenceFunction}.
As an example, we find that
\begin{align}
    \int_{-1}^{1}\rho^{-10}P_{\ell}^0\odif{x} ={}& \frac{1}{192B^{5}}\biggl[\frac{48}{\delta^8} - \frac{8\ell(\ell+1)}{\delta^6} \nonumber \\
        & + \frac{(\ell-1)\ell(\ell+1)(\ell+2)}{\delta^{4}} + \order(\delta^{-2})\biggr].
\end{align}
This is then repeated for all values of $n$ and $m'$ required, and for the different types of integrals that appear in the scalar, tensor and vector sectors.

\subsubsection{Integration over \texorpdfstring{$\beta$}{β}}\label{sec:beta_int}

After performing the integration in $\alpha$ and series expansion in $\delta$, we rewrite all the remaining dependence in $\beta$ (from powers of $\sin\beta$ and $\cos\beta$, and from $e^{-im'\beta}$) in terms of powers of $\chi$ (given in Eq.~\eqref{eq:chi_def}).
Unlike Ref.~\cite{Wardell:2015ada}, we also find terms of the form $\sim\chi^{n}\log\chi$ appearing from the integrals of even powers of $\rho$.
Both cases can be readily integrated using
\begin{equation}
    \int_0^{2\pi} \chi^n\odif{\beta} = 2\pi \mathcal{F}_{-n}\biggl(\frac{M}{r_0f_0}\biggr), \label{eq:chi_int}
\end{equation}
where $\mathcal{F}_{p}(k) \coloneqq {}_{2}F_{1}\bigl(p,\tfrac{1}{2},1;k\bigr)$, and
\begin{equation}
    \int_{0}^{2\pi} \chi^n\log\chi\odif{\beta} = -2\pi \mathcal{F}'_{-n}\biggl(\frac{M}{r_0f_0}\biggr),
\end{equation}
where $\mathcal{F}'_{p}(k) \coloneqq \partial_{a}\bigl({}_{2}F_{1}(a,\tfrac{1}{2},1;\tfrac{M}{r_0f_0})\bigr)\bigr|_{a=p}$.
In the case that $n=-1/2$, $n=1/2$ or $n$ is an integer, Eq.~\eqref{eq:chi_int} reduces to an elliptic integral of the first kind, $\mathcal{K}\bigl(\frac{M}{r_0f_0}\bigr)$, elliptic integral of the second kind, $\mathcal{E}\bigl(\frac{M}{r_0f_0}\bigr)$, or a polynomial in $\frac{M}{r_0f_0}$, respectively.

We can then use the standard recurrence relations for the hypergeometric function to express integrals of all other half-integer powers in terms of $\mathcal{K}\bigl(\frac{M}{r_0f_0}\bigr)$ and $\mathcal{E}\bigl(\frac{M}{r_0f_0}\bigr)$,
\begin{equation}
    \mathcal{F}_{p+1}(k) = \frac{(2p-1)(k-2)}{2p(k-1)}\mathcal{F}_{p}(k) + \frac{p-1}{p(k-1)}\mathcal{F}_{p-1}(k).
\end{equation}
Additionally, by taking the appropriate derivative of the recurrence relation, we can find a similar expression for $\mathcal{F}'_{p}(k)$.
This is given by
\begin{align}
    \mathcal{F}'_{p+1}(k) ={}& \frac{(2p-1)(k-2)}{2p(k-1)}\mathcal{F}'_{p}(k) + 
    \frac{p-1}{p(k-1)}\mathcal{F}'_{p-1}(k) \nonumber \\
        & + \frac{k-2}{2p^{2}(k-1)}\mathcal{F}_{p}(k) + \frac{1}{p^{2}(k-1)}\mathcal{F}_{p-1}(k).
\end{align}

With the integral of $\beta$ completed, we have the decomposition of the second-order singular field in $(\ell m')$-modes.
These can then straightforwardly be converted to $(\ell m)$-modes by using the expression involving the Wigner-D matrix from Eq.~\eqref{eq:flm_lmp}.

\subsection{Regular field}\label{sec:res_field}

When evaluating certain parts of the second-order singular field, we require the value of the first-order regular field $h^{\R1}_{\mu\nu}$ and its derivatives evaluated on the worldline. We compute these by evaluating the sum of residual-field BLS modes, $h^{\calR1}_{i\ell m}$, as in Eq.~\eqref{eq:h modes}. To calculate the residual field modes, we  calculate the retarded field modes, $h^1_{i\ell m}$ and then subtract the first-order puncture modes, $h^{\calP1}_{i\ell m}$, from them.
The retarded field modes are calculated numerically using the \textsc{h1Lorenz} package~\cite{h1Lorenz} from the Black Hole Perturbation Toolkit; for more details, see Ref.~\cite{Akcay:2013wfa}.
We then use the mode-decomposed expressions for the punctures and subtract them from the respective retarded field modes so that we are left with the modes of the first-order residual field.

We next sum over modes numerically to obtain the full regular field and its derivatives on the worldline (note that the residual and regular fields agree on the worldline). We first sum over $m$ to obtain summands that depend only on $\ell$.
Instead of only summing over computed $\ell$-mode data, we use knowledge of the large-$\ell$ behaviour of the modes to accelerate the convergence of the sum.
The first-order puncture field that we use contains four total orders in distance (i.e. $\lambda^{-1}$, $\lambda^{0}$, $\lambda^{1}$ and $\lambda^{2}$).
As illustrated in Fig.~\ref{fig:SingularFieldStructure}, this means that the $\ell$ modes of the residual field fall off as $h^{\calR1}_{\mu\nu} \sim \ell^{-4}$; the $\ell$ modes of its derivatives, also as $h^{\calR1}_{\mu\nu,\alpha} \sim \ell^{-4}$; and those of its second derivatives, as $h^{\calR1}_{\mu\nu,\alpha\beta} \sim \ell^{-2}$.

As discussed in Ref.~\cite{Wardell:2015ada}, each order in $\ell^{n}$ in the residual field has the form\footnote{The coefficients for $n$ odd and negative vanish, so we only need concern ourselves with the behaviour for $n$ even and negative.}
\begin{equation}
    \Lambda_{\ell,s,n} \coloneq \frac{2^{n-2s}(2\ell+1)(\ell-s+1)_{2s}}{(2\ell-2s+n+1)(\ell-s+\tfrac{n}{2}+\tfrac{3}{2})_{2s-n}}, \label{eq:largel}
\end{equation}
where $s\in\mathbb{N}_0$, $n\in\mathbb{Z}$ and $(a)_n$ denotes the Pochhammer symbol. The normalisation factor $2^{n-2s}$ is not essential, but is included so that $\Lambda_{\ell,s,n} = \ell^n+ \mathcal{O}(\ell^{n-1})$.
An important feature of this behaviour is that for any even $n\leq-2$ the sum over $\ell$ vanishes:
\begin{equation}
    \sum_{\ell=0}^{\infty}\Lambda_{\ell,s,n} = 0 \quad \text{for}\quad n\le -2,\quad \frac{n}{2}\in\mathbb{Z}. \label{eq:largelsum}
\end{equation}
We also note that $\Lambda_{\ell,s,n} = 0$ for $\ell<s$ and $n$ even, so that the sum also vanishes when starting at $\ell=s$.
Thus, we can accelerate convergence of the sum over modes by fitting the large-$\ell$ behaviour of the modes to Eq.~\eqref{eq:largel} (with a sum over even negative values of $n$ to some $n_{\text{max}}$) and subtracting this fit mode-by-mode from our calculated $\ell$ modes \cite{Detweiler:2002gi}.
It is important to emphasise that because the sum over $\ell$ is zero, we are essentially `subtracting zero' from the final result.
Thus, we never change the final numerical answer; we only speed up the convergence of the sum over modes.

For our calculations, we perform the series acceleration with $n_{\text{max}}=-4$ and $s=0$, fitting the large-$\ell$ behaviour of each term to \begin{equation}
    c_{-2}\Lambda_{\ell,0,-2} + c_{-4}\Lambda_{\ell,0,-4}, \label{eq:large_l_behaviour}
\end{equation}
This speeds the convergence up a sufficient amount while still being numerically stable to fit to.
After fitting and subtracting mode by mode, we then use $\sum_{\ell=\ell_{\text{max}}}^\infty c_{-4}\Lambda_{\ell,m,-4}$ as a (conservative) estimate of the mode-sum truncation error.

There are two final considerations that simplify things: (i) due to up-down symmetry, parts of the regular field (and its derivatives) with an odd number of \(\theta\) components are identically zero; (ii) terms where the number of $t$ components plus the number of $\phi$ components is odd do not require the ``convergence acceleration'' procedure described above as they already converge exponentially (i.e., the singular field does not contribute to them).

\section{Second-order Ricci tensor}
\label{sec:d2R}

\subsection{Calculation far from the worldline}
\label{sec:RicciOutside}

We now turn to the calculation of the second-order Ricci tensor, $\delta^2 R_{\alpha \beta}[h^1, h^1]$. In most of the spacetime, we evaluate this using the mode-coupling scheme developed in Ref.~\cite{Spiers:2023mor}. In particular, we compute the modes of the first-order retarded metric perturbation, $h^1_{i\ell m}$, up to $\ell=50$ on a grid of radial points using the \textsc{h1Lorenz} code \cite{h1Lorenz}. We then pass the output to the \textsc{SecondOrderRicci} code \cite{SecondOrderRicci}, which evaluates the modes $\delta^2 R_{i\ell m}[h^1,h^1]$ of the second-order Ricci tensor using the mode coupling formula
\begin{equation}
    \delta^{2}R_{i\ell m}[h^1,h^1] = \sum_{\substack{i_{1}\ell_{1}m_{1} \\ i_{2}\ell_{2}m_{2}}}\mathscr{D}_{i\ell m}^{i_1\ell_1m_1i_2\ell_2m_2}[h^1_{i_1\ell_1m_1},h^1_{i_2\ell_2m_2}].\label{eq:d2Rilm}
\end{equation}
The coefficients $\mathscr{D}_{i\ell m}^{i_1\ell_1m_1i_2\ell_2m_2}$ are derived in Ref.~\cite{Spiers:2023mor} and are given explicitly in the \textsc{PerturbationEquations} package \cite{PerturbationEquations}. The sums over $i_1$, $i_2$, $m_1$ and $m_2$ run over a finite range ($1\leq i_n\leq 10$, $-\ell_n\leq m_n\leq \ell_n$ for a given $\ell_1$ and $\ell_2$), and the sums over $\ell_1$ and $\ell_2$ run from 0 to $\infty$.  The coefficients enforce the constraints $m_1+m_2=m$ and $|\ell_1-\ell_2|\leq \ell\leq \ell_1+\ell_2$.

\begin{figure*}[htb!]
    \centering
    \includegraphics[width=0.48\linewidth]{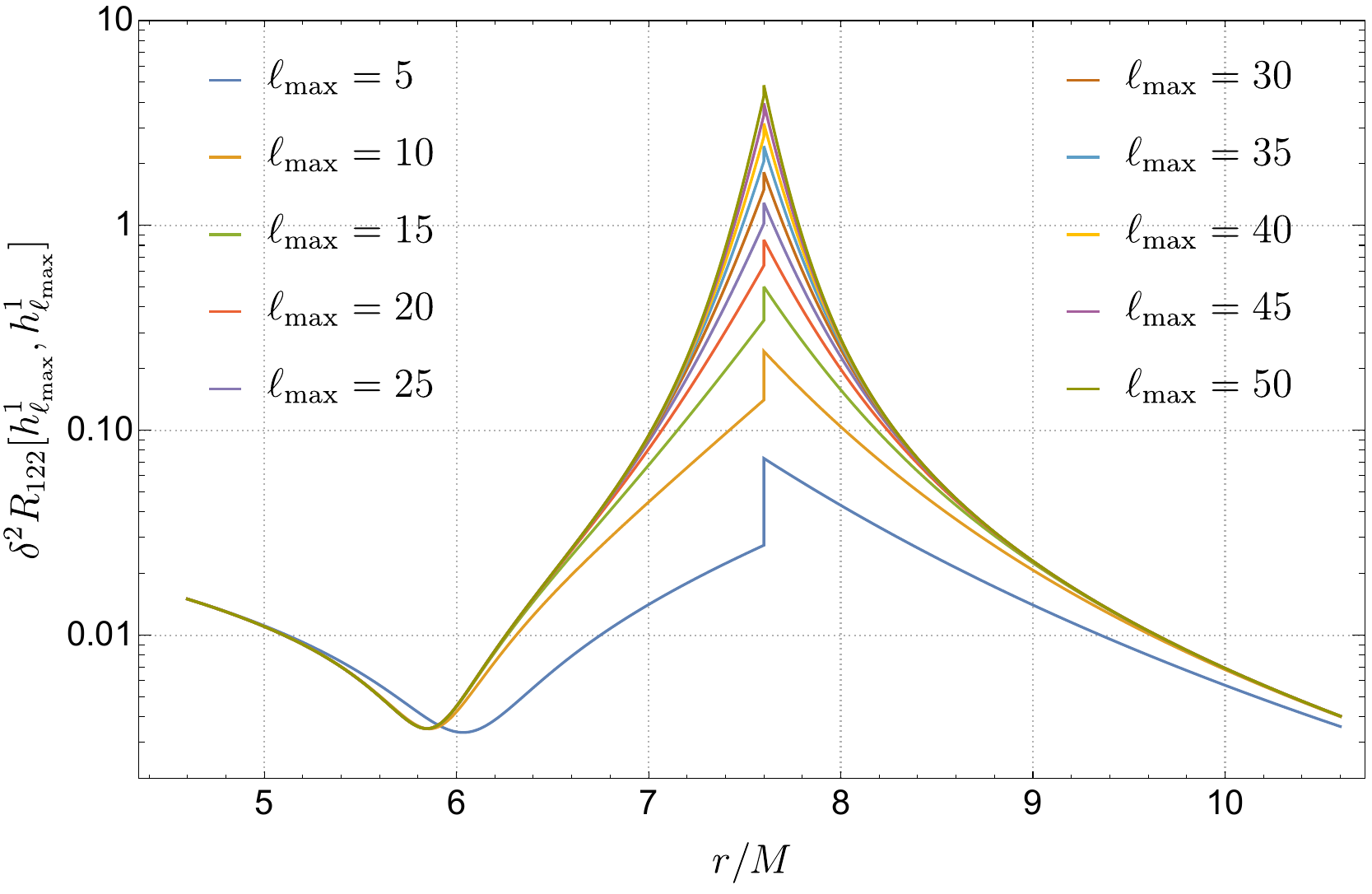}
    \hspace{5mm}
    \includegraphics[width=0.48\linewidth]{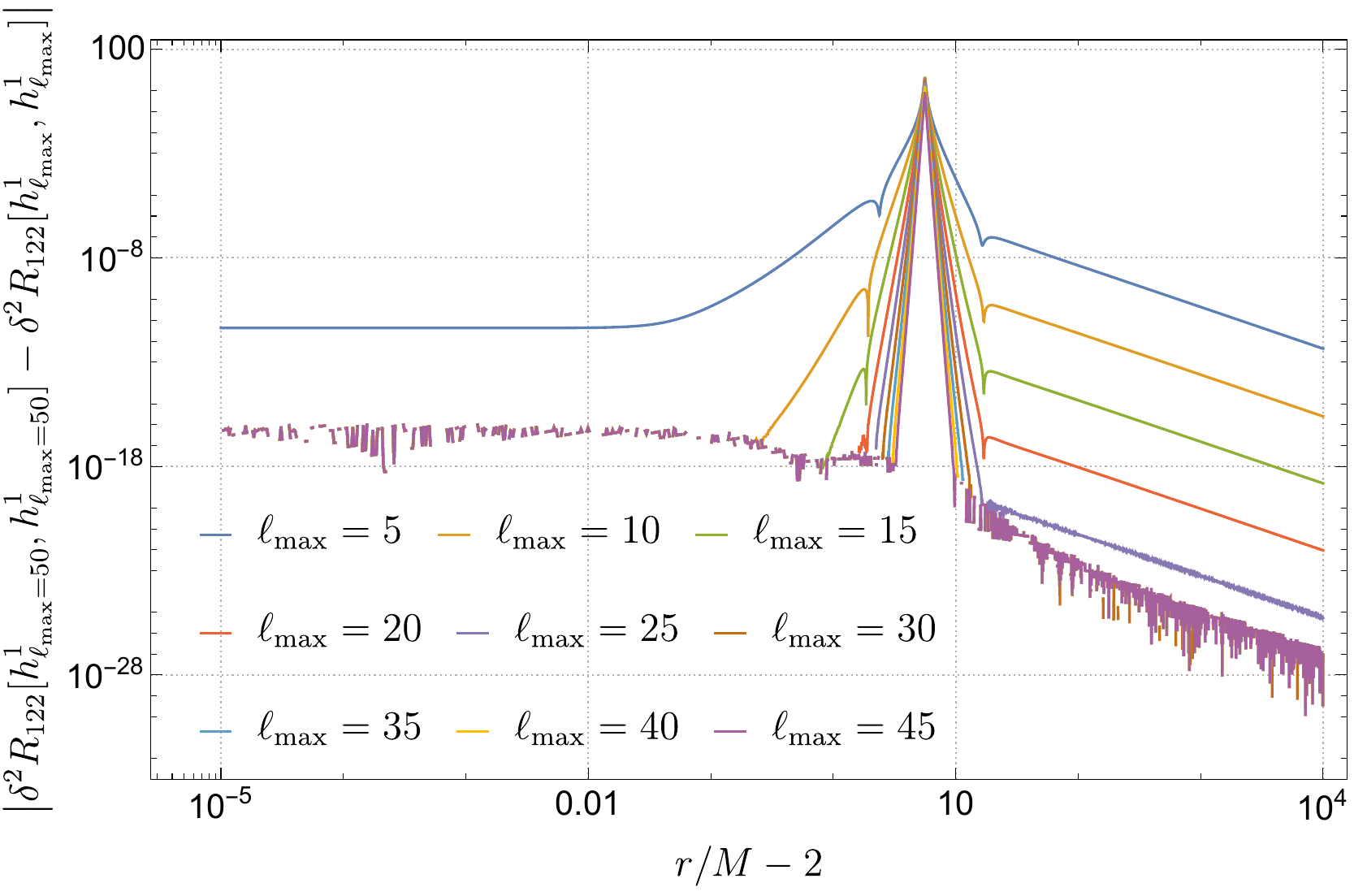}
    \caption{(\textit{Left panel}) The second-order Ricci tensor $\delta^2R_{122}[h^1_{\ell_{\rm max}},h^1_{\ell_{\rm max}}]$ for a variety of $\ell_{\rm max}$ values for a particle at $r_0 = 7.6M$. Curves of increasing $\ell_{\rm max}$ are stacked from bottom to top. 
    (\textit{Right panel}) Convergence of $\left|\delta^2R_{122}[h^1_{\ell_{\rm max}=50},h^1_{\ell_{\rm max}=50}] - \delta^2R_{122}[h^1_{\ell_{\rm max}},h^1_{\ell_{\rm max}}]\right|$ with $\ell_{\rm max}$ for a particle at $r_0 = 7.6M$. Here curves of increasing $\ell_{\rm max}$ are stacked from top to bottom. Note the convergence is very rapid far from the worldline, but it becomes unacceptably slow (while remaining formally exponential) close to the worldline.
    }
    \label{fig:retret}
\end{figure*}
Since the second-order Ricci tensor is a smooth function everywhere except at the particle, the sums over $\ell_1$ and $\ell_2$ converge exponentially everywhere except on the sphere intersecting the worldline (i.e., at the single radial grid point $r=r_0$). We therefore obtain a highly accurate approximation by truncating the sums at $\ell_1 = \ell_{\rm max} = \ell_2$ (in practice we choose $\ell_{\rm max} = 50$). The convergence with $\ell_{\rm max}$ is shown in Fig.~\ref{fig:retret}. We see that the sum converges rapidly for radial points sufficiently far away from the worldline, justifying our choice $\ell_{\rm max}=50$ as striking a good balance between computational cost and accuracy.

\subsection{Calculation close to the worldline}

\label{sec:RicciInside}

Figure \ref{fig:retret} highlights a practical problem: close to $r=r_0$ the convergence is slow and the choice $\ell_{\rm max}=50$ is clearly inadequate. This can be understood from the behavior of the second-order Ricci tensor close to the worldline. It diverges as the fourth power of inverse distance from the worldline, its modes correspondingly diverge as $(r-r_0)^{-2}$ and the mode-sum also formally diverges on the worldline, yet it is a perfectly smooth function off the worldline. This non-uniform behaviour leads to a very slow (but still formally exponential) convergence close to $r=r_0$. 

\begin{figure*}[htb!]
    \centering
    \includegraphics[width=0.48\linewidth]{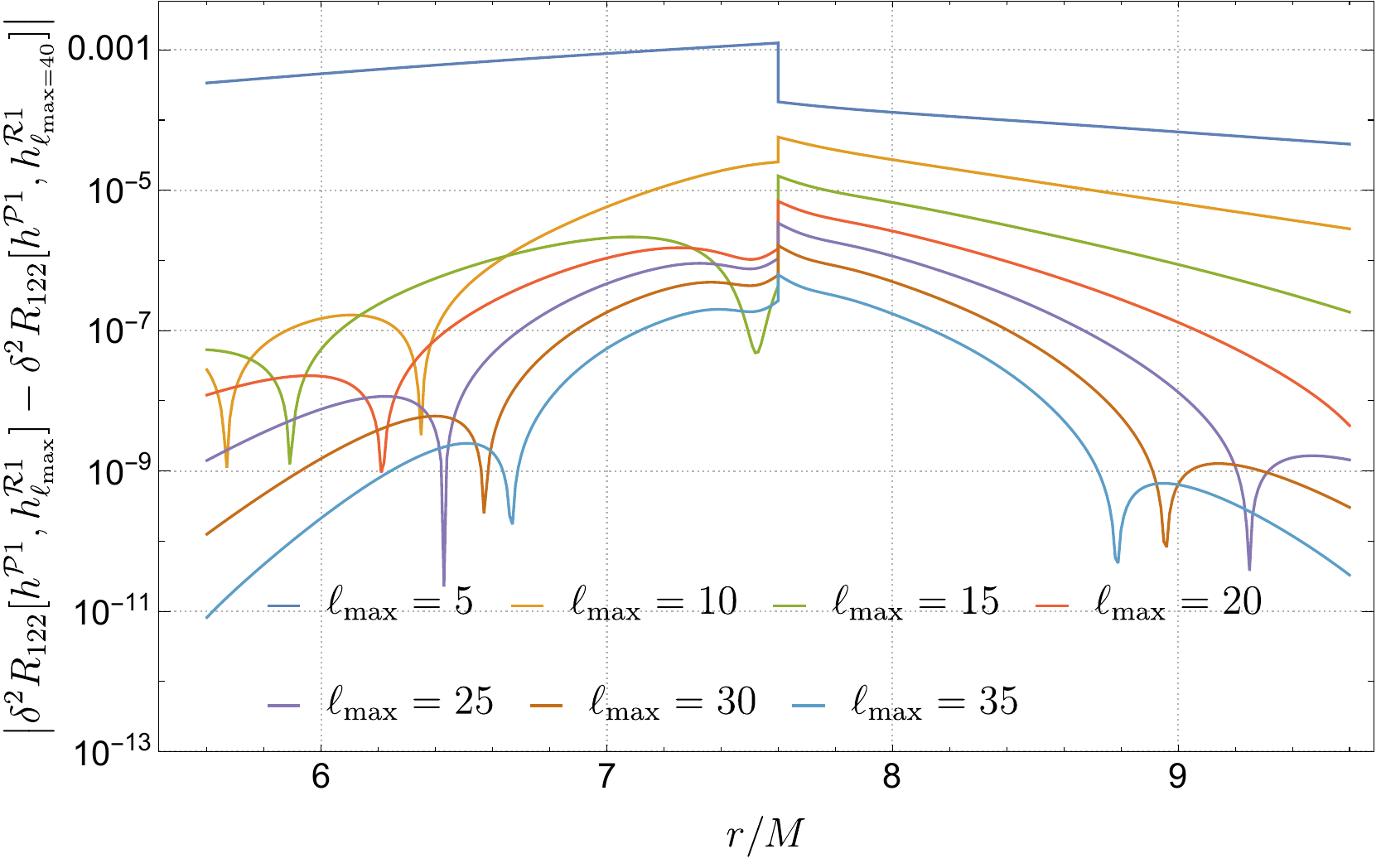}\hspace{5mm}
    \includegraphics[width=0.48\linewidth]{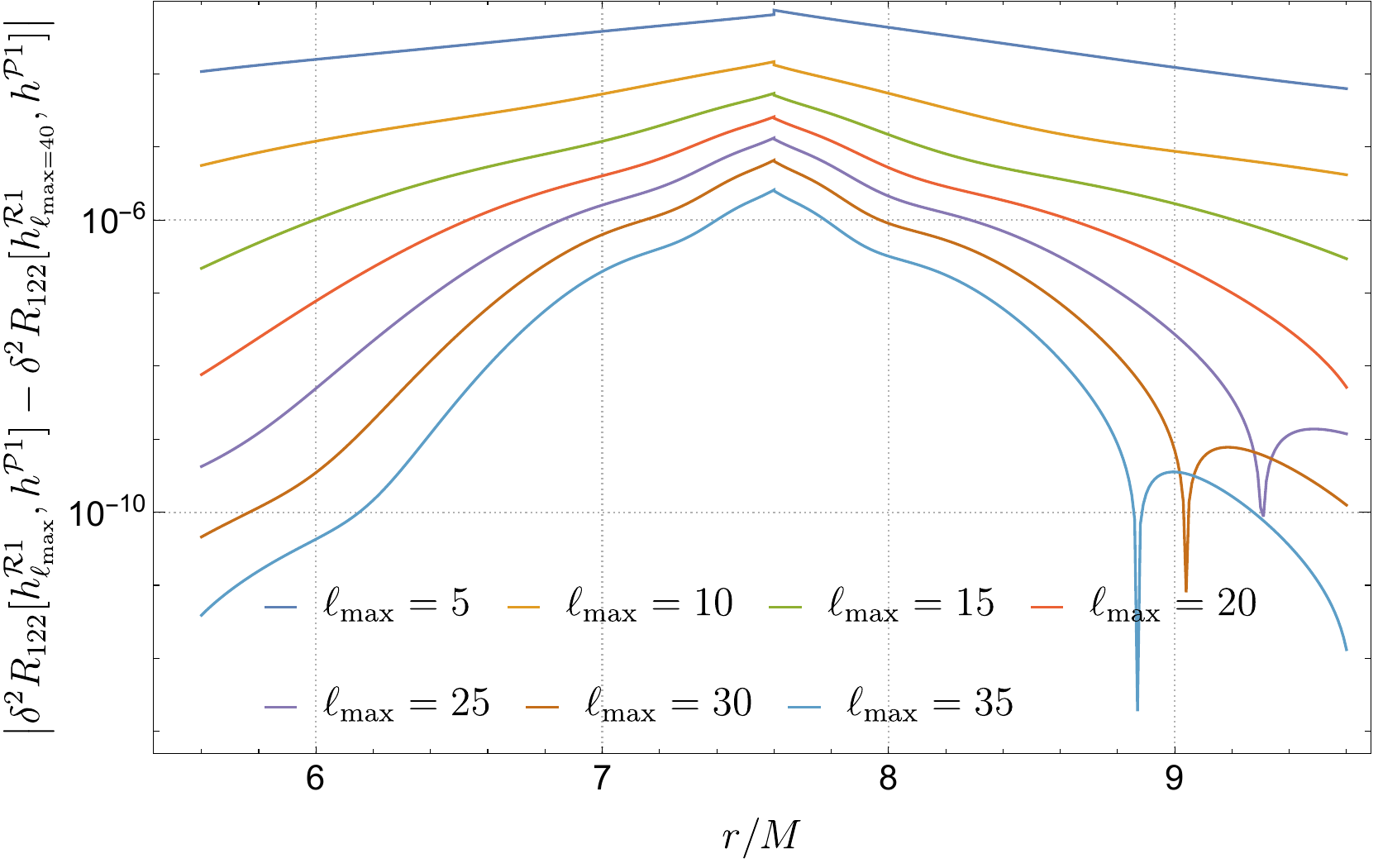}
    \caption{(\textit{Left panel}) Convergence of $\left|\delta^2R_{122}[h^{\calP1},h^{\calR1}_{\ell_{\rm max}}] - \delta^2R_{122}[h^{\calP1},h^{\calR1}_{\ell_{\rm max=40}}]\right|$ with $\ell_{\rm max}$ for a particle at $r_0 = 7.6M$.
    (\textit{Right panel}) Convergence of $\left|\delta^2R_{122}[h^{\calR1}_{\ell_{\rm max}},h^{\calP1}] - \delta^2R_{122}[h^{\calR1}_{\ell_{\rm max=40}},h^{\calP1}]\right|$ with $\ell_{\rm max}$ for a particle at $r_0 = 7.6M$. In both cases, curves of increasing $\ell_{\rm max}$ are stacked from top to bottom. Note the contrast to Fig.~\ref{fig:retret}: even though the convergence is now polynomial rather than exponential, in practice this polynomial convergence produces a much more accurate result close to the worldline.
    }
    \label{fig:SR+RS}
\end{figure*}

\begin{figure}[htb!]
    \centering
    \includegraphics[width=\linewidth]{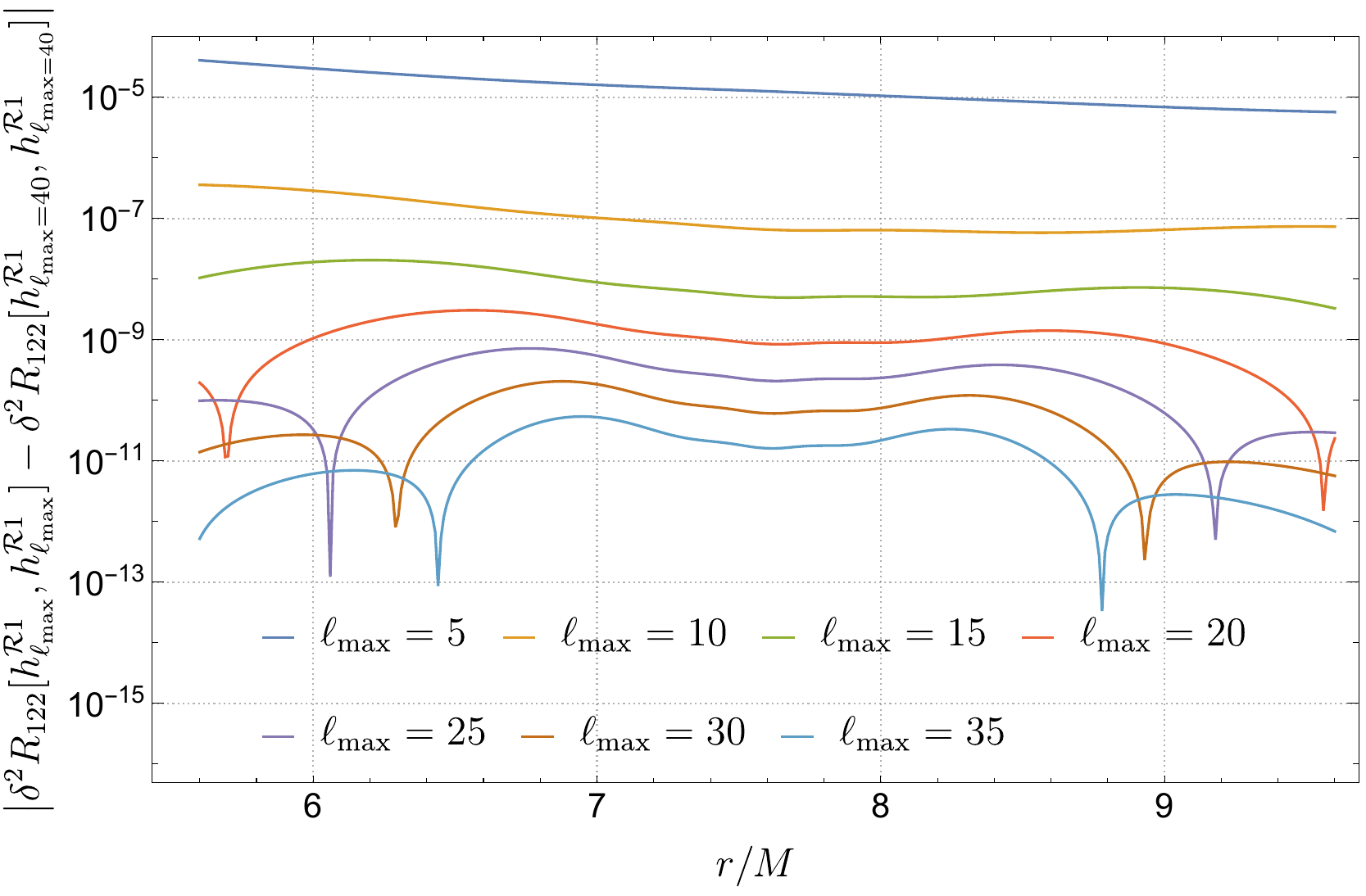}
    \caption{
    Convergence of $\left|\delta^2R_{122}[h^{\calR1}_{\ell_{\rm max}},h^{\calR1}_{\ell_{\rm max}}]\right.$ $- \left.\delta^2R_{122}[h^{\calR1}_{\ell_{\rm max}=40},h^{\calR1}_{\ell_{\rm max=40}}]\right|$ with $\ell_{\rm max}$ for a particle at $r_0 = 7.6M$. Curves of increasing $\ell_{\rm max}$ are stacked from top to bottom. 
    }
    \label{fig:RR}
\end{figure}

To address this problem, we follow the method developed in Ref.~\cite{Miller:2016hjv} and introduce a worldtube around $r=r_0$. Inside the worldtube we use the split of the first-order metric perturbation into puncture and residual pieces, $h^1 = h^{\calP1} + h^{\calR1}$, to split the second-order Ricci tensor into a sum of four terms:
\begin{align}
    \delta^{2}R_{\ilm}[h^1,h^1] &= \delta^{2}R_{i\ell m}[h^{\calR1},h^{\calR1}] + \delta^{2}R_{i\ell m}[h^{\calP1},h^{\calR1}]\nonumber \\
    & + \delta^{2}R_{\ilm}[h^{\calR1},h^{\calP1}] + \delta^{2}R_{\ilm}[h^{\calP1},h^{\calP1}]. \label{eq:d2R-split}
\end{align}
Note that this is an exact equality, not an approximation. Here, we mimic the naming convention in the puncture, Eq.~\eqref{eq:hS2_sc}, by referring to the terms on the right as the RR (regular times regular), SR (singular times regular), RS (regular times singular), and SS (singular times singular) terms, respectively. Also note that while the second-order Ricci tensor was defined to be symmetric in its two arguments in some papers (e.g., ~\cite{Pound:2015fma}), the concrete expressions used in \textsc{SecondOrderRicci} have not been symmetrized, leading to different expressions for the RS and SR pieces.

\subsubsection{Calculation of SR, RS and RR Ricci tensor via mode coupling}
\label{sec:SR-RS-RR-Ricci}

The first three terms on the right hand side of Eq.~\eqref{eq:d2R-split} are obtained using the mode-coupling formula~\eqref{eq:d2Rilm}, as was done outside the worldtube. The only caveat we need to be careful of is that, for Eq.~\eqref{eq:d2R-split} to be an equality, the \emph{exact same} puncture and residual field must be used in all four terms. In particular, our mode decomposition of the puncture must be done exactly and cannot, for example, be obtained as a series expansion in $\Delta r$.

To achieve this, we start from the first-order puncture given schematically in Eq.~\eqref{eq:hP1coord} and multiply it by a window function $\mathcal{W}_{10}^4 $ such that all modes up to $m'=10$ are not contaminated by the non-smoothness at the south pole, while simultaneously ensuring that the puncture is not affected through the first four orders in distance from the worldline.

We next use the closed-form expressions for the integration over $\alpha$ described in Sec.~\ref{sec:int_alpha} \emph{without} series expanding in $\delta$. This makes the integration over $\beta$ more difficult. In principle it can still be written as a sum of elliptic integrals (of the first, second and third kinds). However, due to the complexity of the expressions, in practice we evaluate the $\beta$ integrals numerically using \textsc{Mathematica}'s \texttt{NIntegrate} on a grid of $r$ values inside the worldtube. We use extended precision and set the tolerance of the integration such that the resulting integral is accurately computed to approximately 16 decimal places.

Much of the advantage of working with a rotated coordinate system --- including being able to do the $\alpha$ and $\beta$ integrals analytically, obtaining closed-form expressions for arbitrary $\ell$, and needing to include only a small number of $m'$ modes -- is lost as a result of needing to perform the mode decomposition exactly. However, one advantage remains: as demonstrated in Fig.~4 of Ref.~\cite{Miller:2016hjv}, the contribution from higher $m'$ modes to a given $m$ mode falls off exponentially. In practice, this means that we only need to compute up to $m'=10$ in order to determine $h^{\calP1}_{i\ell m}$ to within our error tolerance of approximately 16 digits.

Next, we obtain the residual field modes by subtracting the puncture field modes from retarded field modes produced using the \textsc{h1Lorenz} code,
\begin{equation}
    h^{\calR1}_{i\ell m} =h^{1}_{i\ell m} - h^{\calP1}_{i\ell m},
\end{equation}
and we then compute the $\R\R$, $\R\S$ and $\S\R$ contributions to the Ricci tensor using the \textsc{SecondOrderRicci} code.

The modes $\delta^{2}R_{i\ell m}[h^{\calP1},h^{\calR1}]$, $\delta^{2}R_{\ilm}[h^{\calR1},h^{\calP1}]$, and $\delta^{2}R_{\ilm}[h^{\calP1},h^{\calP1}]$ are finite everywhere, but the use of a puncture for the first-order metric perturbation causes the convergence with $\ell_{\rm max}$ of the mode-coupling sum to be polynomial rather than exponential, and this happens not just on the worldline but everywhere the puncture is used. However, it turns out that, for moderate $\ell_{\rm max}$, the result is more accurate than the equivalent (exponentially convergent) sum over retarded-field modes. This is illustrated in Figs.~\ref{fig:retret}, ~\ref{fig:SR+RS} and \ref{fig:RR}. 

\subsubsection{Calculation of the SS Ricci tensor}
\label{sec:SS-Ricci}

We now turn to the calculation of the $\S\S$ contribution to the second-order Ricci tensor.
We compute the modes $\delta^2 R_{i\ell m}[h^{\calP1},h^{\calP1}]$ in much the same way as we compute the modes $h_{i\ell m}^{\calP1}$ of the first-order metric perturbation:
we evaluate $\delta^2 R_{\mu\nu}[h^{\calP1},h^{\calP1}]$ as a coordinate expression in the rotated $(t,r,\alpha,\beta)$ coordinate system described in Sec.~\ref{sec:rotatedcoordinates} and then integrate against the BLS tensor harmonics.
The coordinate expression for the second-order Ricci tensor involves the coordinate components of $h^{\calP1}_{\mu\nu}$; for these we use the \emph{exact same} puncture (including the window function) as in Sec.~\ref{sec:SR-RS-RR-Ricci}.

To obtain our coordinate expression for $\delta^2R_{\mu\nu}$ in $(t,r,\alpha,\beta)$ coordinates, we start from the expressions in ``2+2D'' form given in Appendix~B of Ref.~\cite{Spiers:2023mor}. These expressions are valid in any coordinate system $(x^a,\theta^A)$, where $x^a$ are coordinates on the $t$-$r$ plane and $\theta^A$ are coordinates on the two-sphere of constant $x^a$. However, there is a subtlety in utilizing these expressions: they are not valid for a time-dependent angular coordinate system. 
We account for this issue, and derive the appropriate alteration of the 2+2D expressions, using an extension of the method developed in Ref.~\cite{Miller:2016hjv}. Specifically, we adopt the ``2D'' method described in Appendix~A of that reference, suitably adapted to tensor fields as opposed to scalars. In this approach, at each instant $t$ we define new angular coordinates in which the particle is only instantaneously at the north pole; this contrasts with a 4D method that would utilize 4D coordinates in which the particle would be permanently at rest at the north pole.
 
We first introduce the notation $\alpha^{A'} = (\alpha,\beta)$ and use indices $A', B', \ldots$ for tensors in the tangent or cotangent space of the 2D spherical submanifold charted by $(\alpha, \beta)$. Following Ref.~\cite{Miller:2016hjv}, we define
 \begin{equation}
     \dot{\alpha}^{A'} = \dfrac{\partial \alpha^{A'}}{\partial t} = \Omega(-\cos\beta,\cot\alpha\sin\beta),
 \end{equation} the Jacobian $\Omega_A^{\ A'}\coloneqq \partial \alpha^{A'}/\partial\theta^A$, and its inverse $\Omega^A_{\ A'}\coloneqq \partial \theta^{A}/\partial\alpha^{A'}$. Now consider a term $\partial_t T_{A}$ in the original 2+2D expressions in a time-independent angular coordinate system. This is a component of the tensor $(\partial_t T_{A})d\theta^A$. We can rewrite it in the time-dependent coordinates using $T_A = \Omega_A^{\ A'}T_{A'}$, $d\theta^A=\Omega^A_{\ B'}d\alpha^{B'}$, and the chain rule
\beq
 \partial_t\bigr|_{\theta^C} T_A = \left(\partial_t\bigr|_{\alpha^{C'}}+\dot\alpha^{C'}D_{C'}\right)T_A\;,
\eeq
where $D_{C'}$ is the covariant derivative compatible with the unit-sphere metric $\Omega_{A'B'}=\diag(1, \sin^2\alpha)$.
(We are free to use $D_{C'}$ in place of $\partial_{C'}$ since it is acting on a component, which is a scalar.) This gives  
\begin{multline}
(\partial_t T_{A})d\theta^A \\= \left(\partial_t\bigr|_{\alpha^{C'}}T_{B'}+\dot\alpha^{C'}D_{C'}T_{B'} +C^{A'}_{\ B'}T_{A'}\right)d\alpha^{B'},
\end{multline}
where we have defined
\begin{equation}
     C^{A'}_{\ B'} \coloneqq \dot{\alpha}^{C'} C^{A'}_{B'C'} = 
     \Omega\begin{bmatrix}
        0 & \sin^2\alpha\sin \beta \\
        -\sin \beta & 0
     \end{bmatrix}
 \end{equation}
 with $C^{A'}_{B'C'}\coloneqq  \Omega^A_{\ B'} D_{C'}\Omega_A^{\ A'}$. Generalizing to higher-rank tensors, we see that the 2+2D expressions in Ref.~\cite{Spiers:2023mor} remain valid if we simply replace unprimed indices with primed ones and replace time derivatives with
 \begin{align}
     \partial_t T_{A_1\cdots A_n}&\to \partial_t T_{A_1'\cdots A_n'} +\dot{\alpha}^{C'} D_{C'} T_{A_1'\cdots A_n'} \nonumber\\
     & + C_{A_1'}{}^{B'}T_{B'A_2'\cdots A_n'} + \cdots + C_{A_n'}{}^{B'}T_{A_1'\cdots A_{n-1}'B'}.
 \end{align}
The $\partial_t$ on the right-hand side is taken at fixed $(\alpha,\beta)$; we note that such a partial derivative vanishes when acting on the puncture in the case of a quasicircular orbit~\cite{Miller:2016hjv}.

The resulting 2+2D expression can readily be expanded in explicit coordinate form. We then substitute the expression for the puncture into this coordinate expression for the second-order Ricci tensor, in the process making sure to avoid making any approximations that would lead to Eq.~\eqref{eq:d2R-split} no longer being an equality. This unfortunately leads to very large, unwieldy expressions that are not readily amenable to analytic integration. Instead, we resort to numerical 2D integration over $\alpha$ and $\beta$ using \textsc{Mathematica}'s \texttt{NIntegrate}. We set the tolerance of the integration such that the resulting integral is accurately computed to approximately 8 decimal places.
 
The calculation of the $\S\S$ Ricci tensor is implemented in the \textsc{SecondOrderRicciSS} \cite{SecondOrderRicciSS} code. For a single $(\ell,m')$ at a single radial point this calculation is slow but tolerable, taking on the order of a few seconds for a low-$(\ell,m')$ mode on a typical CPU (more challenging high-$(\ell,m')$ are significantly slower by a factor of approximately 10--100). However, with a significant number of modes (we have computed up to $\ell=10$), $10$ BLS fields (for the $10$ components of the metric), a set of radial grid points (we typically choose a uniform grid of $401$ points with a spacing of $0.01M$ inside a worldtube of radius $2M$), and a grid of different $r_0$ values (we chose $141$ equally spaced points between $r_0=6M$ and $r_0 = 20M$, plus a handful of higher-radius cases) the computation time quickly multiplies up so that our total computational cost for the $\S\S$ Ricci calculation was on the order of $3$ million CPU hours. This totally dominates the overall cost of the second-order calculation, and should be the first target for future optimisations. Some possible optimisations that are currently being explored include:
\begin{itemize}
    \item Performing the subtraction $2\delta^2 R_{\alpha \beta}[\mathring{h}^S,\mathring{h}^S] - E^0_{\alpha \beta}[\mathring{h}^{SS}]$ first at the 4D level and then decomposing into modes. This would make the resulting 2D integrand smoother and more amenable to numerical integration.
    \item Performing at least one of the two integrations analytically \cite{Bourg:2024cgh}.
    \item Switching to an $m$-mode scheme in which the calculation of the second-order Ricci tensor is dramatically simpler \cite{PanossoMacedo:2024pox}.
\end{itemize}

\subsubsection{Total second-order Ricci tensor}

\begin{figure*}[!htp]
    \centering
    \includegraphics[width=0.495\linewidth]{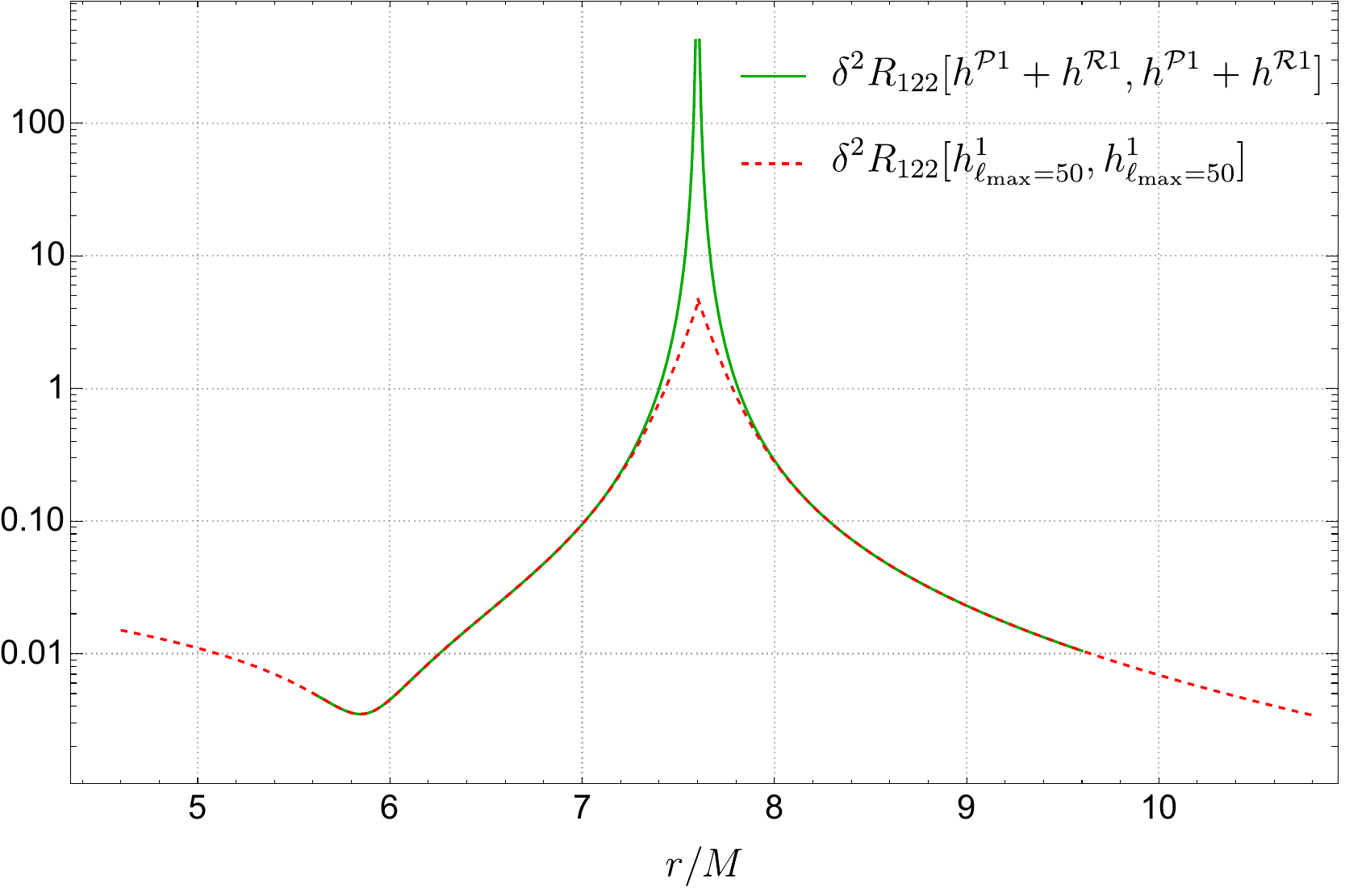}
    \includegraphics[width=0.495\linewidth]{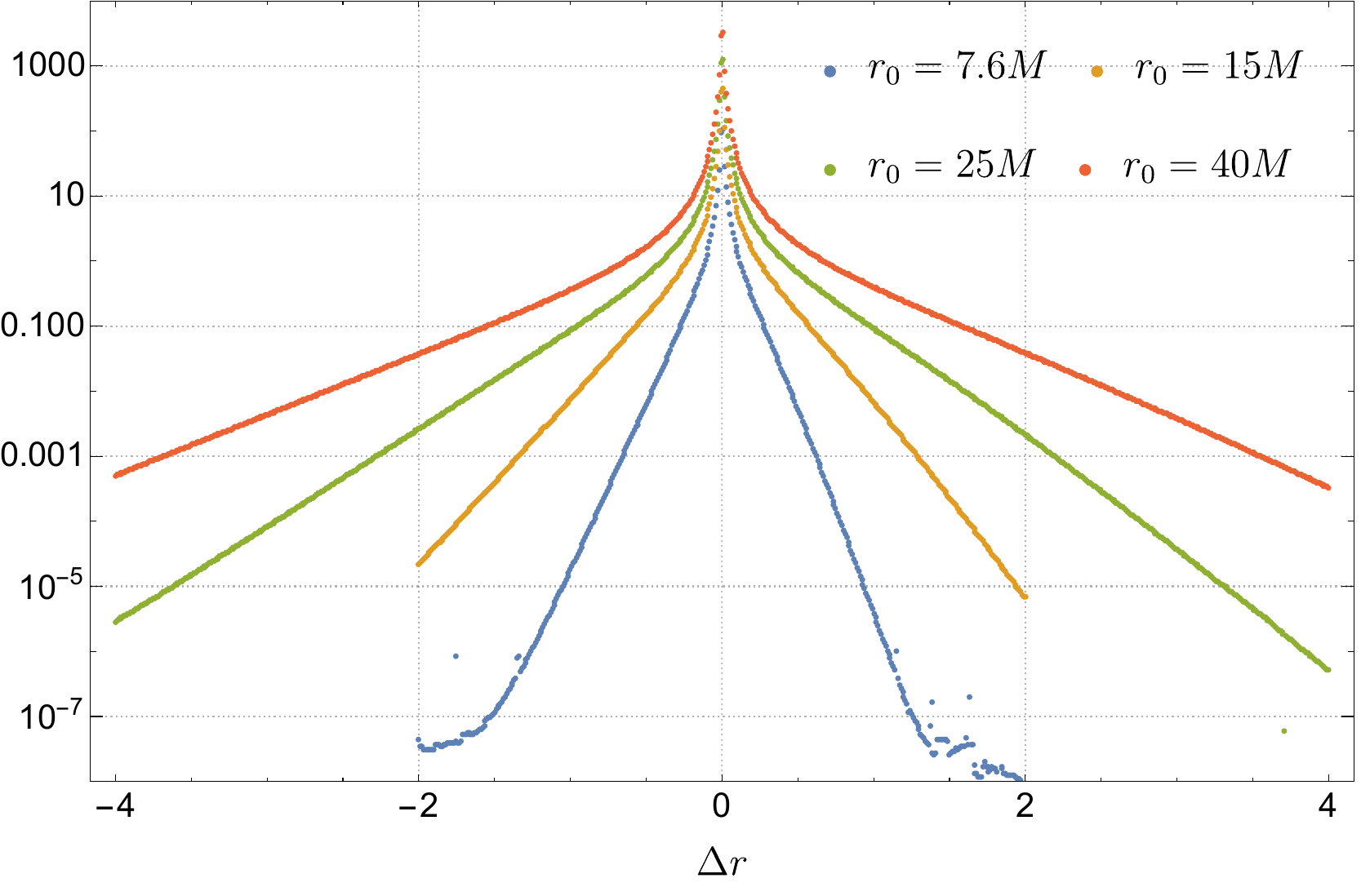}
    \caption{
    Comparison of the two different methods to compute $\delta^2R_{i\ell m}[h^1,h^1]$ near the worldline for the $i=1$, $l=2$, $m=2$ mode.
    (\emph{Left panel}) Results for $r_0=7.6M$. 
    Here the dashed (red) curve shows $\delta^2R_{122}[h^1,h^1]$ computed via mode coupling using $h^1$ modes up to $\ell_{\rm max} = 50$. The solid green curve shows the Ricci tensor computed using the $\delta^2R_{122}[h^{\calP1}+h^{\calR1}, h^{\calP1}+h^{\calR1}]$ decomposition using $\ell_{\rm max} = 40$. (\emph{Right panel}) Relative difference between the two ways of computing $\delta^2R_{122}[h^1,h^1]$ for a variety of orbital radii as a function of $\Delta r = r - r_0$.}
    \label{fig:RetRetRicci}
\end{figure*}

In Fig.~\ref{fig:RetRetRicci} we see the benefit of our approach to computing the second-order Ricci tensor as compared to mode coupling with the retarded field only. The finite number of $h^1_{i\ell m}$ modes means that mode coupling cannot capture the logarithmic divergence in $\delta^2R_{i\ell m}[h^1,h^1]$ at $r=r_0$.
Far from the worldline the two approaches agree, as expected.

\begin{figure*}[htb!]
    \centering
    \includegraphics[width=0.495\linewidth]{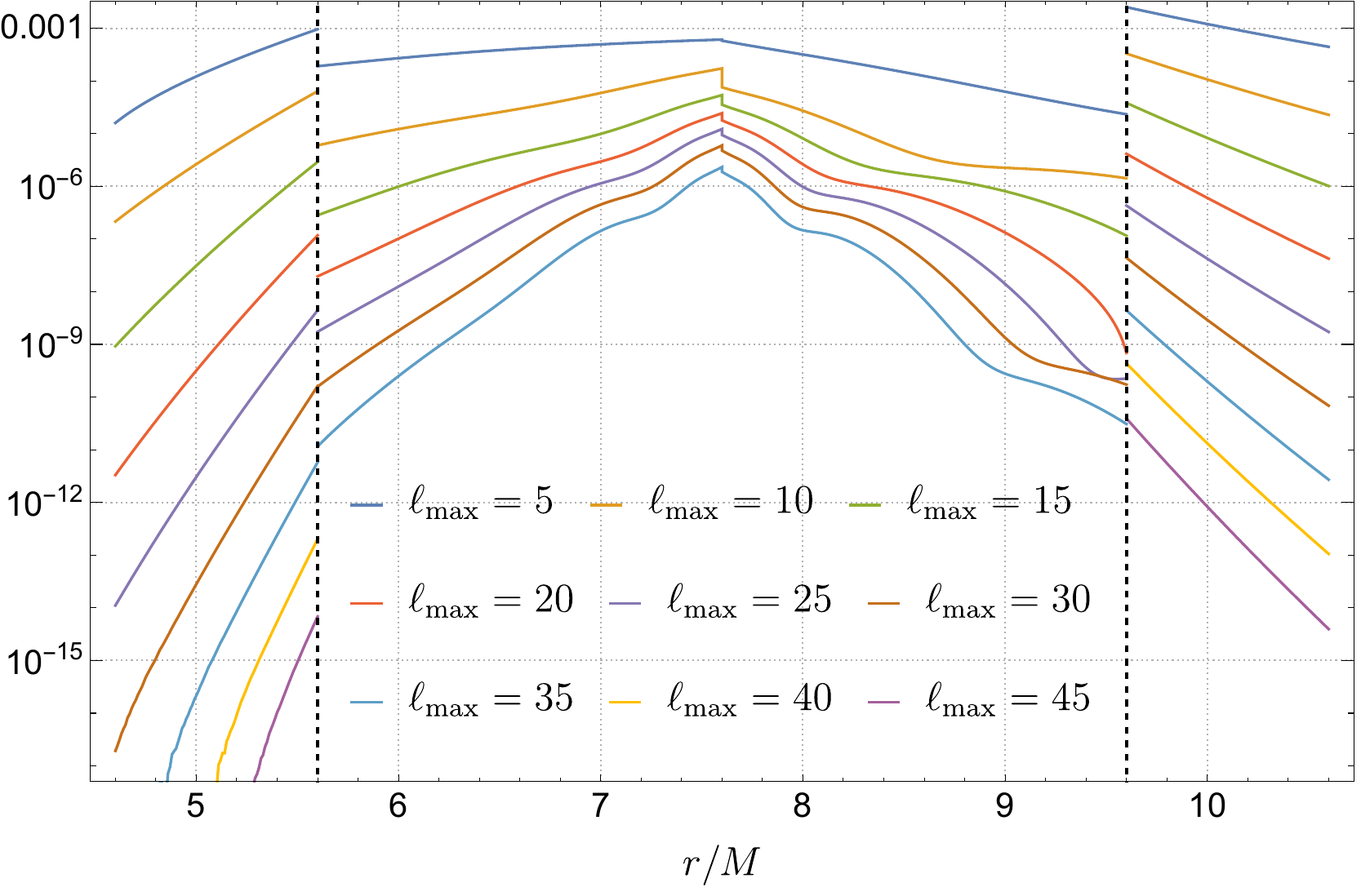}
    \includegraphics[width=0.495\linewidth]{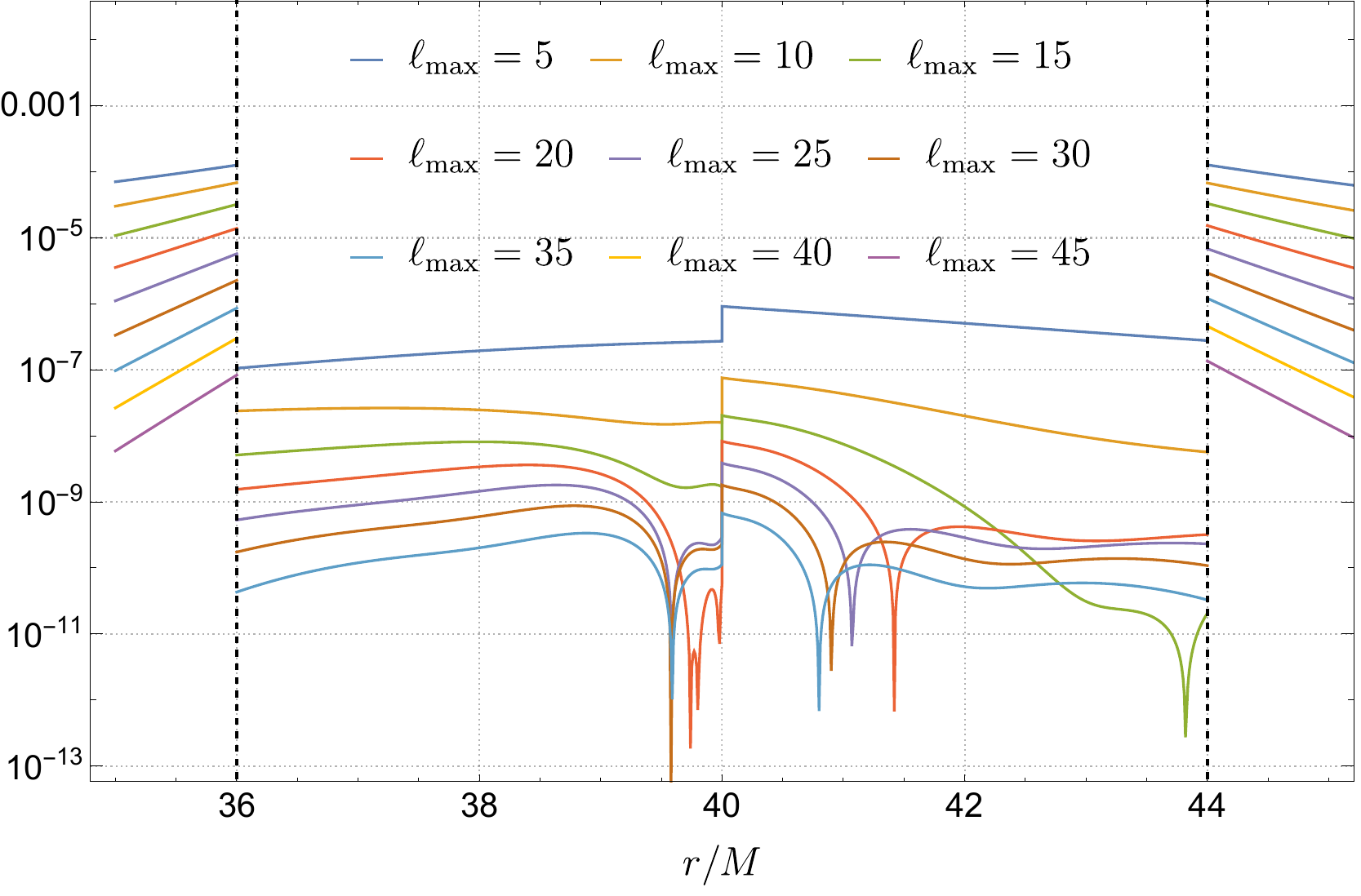}
    \caption{
The difference $\left|\delta^2R_{122}[h^{1}_{\ell_{\text{max}}},h^{1}_{\ell_{\text{max}}}]-\delta^2R_{122}[h^{1}_{\ell_{\text{max}=50}},h^{1}_{\ell_{\text{max}=50}}]\right|$ for a sequence of $\ell_{\rm max}$ values, computed using the method described in Sec.~\ref{sec:RicciOutside} outside the worldtube and $\bigl|\delta^2R_{122}[(h^{\calP1}+h^{\calR1})_{\ell_{\text{max}}},(h^{\calP1}+h^{\calR1})_{\ell_{\text{max}}}]-\delta^2R_{122}[(h^{\calP1}+h^{\calR1})_{\ell_{\text{max}=40}},(h^{\calP1}+h^{\calR1})_{\ell_{\text{max}=40}}]\bigr|$ computed using the method described in Sec.~\ref{sec:RicciInside}  inside the worldtube. In both panels the worldtube radius is marked by the vertical, dashed lines, and curves of increasing $\ell_{\rm max}$ are stacked from top to bottom. (\textit{Left panel}) Data for $r_0 = 7.6M$ and a worldtube boundary at $|\Delta r| = 2M$. (\textit{Right panel}) Data for $r_0 = 40M$ and a worldtube boundary at $|\Delta r| = 4M$. Note the clear need for a larger worldtube for larger $r_0$. Indeed, we see on the right panel that even a worldtube twice as wide does not make the accuracy comparable inside and outside the worldtube.
    }
    \label{fig:d2G-hS-plus-hR}
\end{figure*}

There is a balance between the width of the worldtube and the number of $\ell_1$ and $\ell_2$ modes included in the sum on the inside and the outside of the worldtube. This is highlighted in the left panel of Fig.~\ref{fig:d2G-hS-plus-hR}. We see that for moderate orbital radii $r_0 \lesssim 20M$, a worldtube size of $|\Delta r| = 2M$ is appropriate when using $\ell_{\rm max} = 50$ outside the worldtube and $\ell_{\rm max} = 40$ inside the worldtube.\footnote{These specific values of $\ell_{\rm max}$ were chosen as the number of modes that could be computed in a reasonable time; the calculation of the puncture modes is significantly slower than the retarded field modes and becomes increasingly challenging for larger $\ell$, hence the lower value of $\ell_{\rm max}$ chosen inside the worldtube.}

At fixed $\Delta r$ we find the agreement worsens as $r_0$ increases.
This is expected, as for weak-field orbits $\delta^2R_{i\ell m}[h^{\calP1},h^{\calP1}]$ dominates and this piece is poorly captured by $\delta^2R_{i\ell m}[h^1,h^1]$ computed via mode coupling.
The requirement for comparable accuracy either side of the worldtube boundary between the two ways of computing $\delta^2R_{i\ell m}[h^1,h^1]$ motivates setting the worldtube boundary at $|\Delta r| = 2M$ for $r_0 \lesssim 20M$ and at $|\Delta r| = 4M$ for $r_0 \gtrsim 20M$.
As shown in the right panel of Fig.~\ref{fig:d2G-hS-plus-hR}, for $r_0 \gg 20M$ it is clear that there would be a benefit in using an even larger worldtube (or increasing $\ell_{\text{max}}$ outside the worldtube). However, we opted not to do so due to the significant computational cost associated with evaluating the SS Ricci at additional points inside an enlarged worldtube.

\section{Slow-evolution source terms}
\label{sec:slow_time}

The remaining piece of the second-order source is the term $\breve{E}^1_{ij \ell m}h^{1}_{j\ell m}$, which accounts for the slow evolution of the first-order metric perturbation due to the gradual inspiral of the compact object and absorption of radiation by the black hole. An explicit expression for this is given in Eq.~(65) of Ref.~\cite{Miller:2023ers}. Translated into the notation of this paper, it reads
\beq\label{breveE1ijlm}
\breve E^1_{ij\ell m} h_{j\ell m} := \Box^1_{\ell m} h_{i\ell m} + \breve{\cal M}_{ij}^1 h_{j\ell m},
\eeq
with
\begin{multline}
 \Box^1_{\ell m} = \frac{1}{4}\Big[\left(2 H\partial_{r^*}+ H'\right)\vec\partial_{\cal V}\\
			 -\left(1-H^2\right)\left(2i\omega_m\vec\partial_{\cal V}+i m F_0\right)\Big]\label{tildeBox1}
\end{multline}
and
\begin{subequations}\label{Mij1}
\begin{align}
\breve{\mathcal{M}}^1_{1j} h_{j\ell m} &= -\frac{1}{2}ff'H\vec\partial_{\cal V} h_{6\ell m}, \label{M1-1}\\
\breve{\mathcal{M}}^1_{2j} h_{j\ell m} &= -\frac{f'}{2}\Bigl[ f H \vec\partial_{\cal V} h_{6\ell m} \nonumber\\
&\qquad - \left(1-H\right)\vec\partial_{\cal V}\left( h_{2\ell m}- h_{1\ell m}\right) \Bigr],\label{M1-2}\\
\breve{\mathcal{M}}^1_{4j} h_{j\ell m} &= \frac{f'}{4}\left(1-H\right)\vec\partial_{\cal V}\left( h_{4\ell m}- h_{5\ell m}\right),\label{M1-4}\\
\breve{\mathcal{M}}^1_{8j} h_{j\ell m} &= \frac{f'}{4}\left(1-H\right)\vec\partial_{\cal V}\left( h_{8\ell m}- h_{9\ell m}\right),\label{M1-8}\\
\breve{\mathcal{M}}^1_{ij} h_{j\ell m} &= 0\,\,\,\,\text{for}\,\,\,\,\,i=3,5,6,9,10.
\end{align}
\end{subequations}
Here we recall Eq.~\eqref{d/dV} for the parametric directional derivate $\vec{\partial}_{\cal V}$.

The derivatives $\partial_{\delta M}h^1_{i\ell m}$ and $\partial_{\delta S}h^1_{i\ell m}$ can be calculated analytically from the Lorenz-gauge mass and spin perturbations in Appendix~D of Ref.~\cite{Miller:2020bft}. Moreover, these derivatives only enter for $m=0$ and $\ell=0,1$, since linear mass and spin perturbations vanish for other modes. 

The main computational task is thus to obtain the parametric derivative $\partial_{\Omega} h^1_{i\ell m}$ of the first-order metric perturbation. A method for computing this is detailed in Ref.~\cite{Durkan:2022fvm} for the case of interest here, namely quasicircular inspirals in Schwarzschild spacetime within the Lorenz gauge (see also Ref.~\cite{Miller:2023ers}, which computed the same results using a different method, as well as Ref.~\cite{PanossoMacedo:2022fdi}, which computed analogous results for a scalar field using yet another method). The essential idea is to formulate field equations for $\partial_\Omega h^1_{i\ell m}$ by differentiating the field equation for $h^1_{i\ell m}$ with respect to $\Omega$. This leads to a field equation in which $h^1_{i\ell m}$ acts as a noncompact source for $\partial_\Omega h^1_{i\ell m}$.

Rather than directly solving the coupled Lorenz gauge field equations with an unbounded source---an approach that is numerically challenging---the authors of Ref.~\cite{Durkan:2022fvm} employ a more efficient method using a gauge transformation from Regge-Wheeler gauge to Lorenz gauge. They compute the $\Omega$ derivative of the Regge-Wheeler-Zerilli master functions and use Berndtson's gauge transformation \cite{Berndtson:2007gsc} to construct the Lorenz-gauge metric perturbation and its parametric derivative.

To handle the non-compactness of the source in the parametric-derivative equations, Ref.~\cite{Durkan:2022fvm} applies the \emph{method of partial annihilators}. This involves applying the Regge-Wheeler operator twice, converting the equation into a higher-order differential equation with a compact distributional source. This allows the problem to be solved using variation of parameters with purely compact support, bypassing the difficulties of unbounded integrals.

Reference~\cite{Durkan:2022fvm} systematically implements the method for both the Regge-Wheeler and Zerilli master functions, including the handling of additional gauge fields required for a complete Lorenz-gauge representation. The final output is a set of high-accuracy inhomogeneous solutions for $\partial_{\Omega} h^1_{i\ell m}$. Substituting this into Eq.~\eqref{breveE1ijlm} then yields the required slow-evolution piece of the source.

A key feature of the source term $\breve{E}^1_{ij\ell m}h^1_{j\ell m}$ is its strong dependence on slicing. This feature has been stressed in Refs.~\cite{Miller:2020bft,Miller:2023ers}. For example, we can see from Eq.~\eqref{tildeBox1} that in $t$ slicing ($H=0$), the source near the boundaries is dominated by the term $\propto\omega_m\vec\partial_{\cal V}$; this is because, in $t$ slicing, the first-order field behaves as $h^1_{i\ell m}\propto e^{\pm im\Omega r^*}$ near the boundaries, and the derivative $\partial_\Omega$ acts on the oscillatory exponential $e^{\pm im\Omega r_*}$. In $u$ or $v$ slicing ($H=\pm1$), this contribution to Eq.~\eqref{tildeBox1} vanishes, and $\breve{E}^1_{ij\ell m}h^1_{j\ell m}$ becomes well behaved at the boundaries. 

\begin{figure}
    \centering
    \includegraphics[width=\columnwidth]{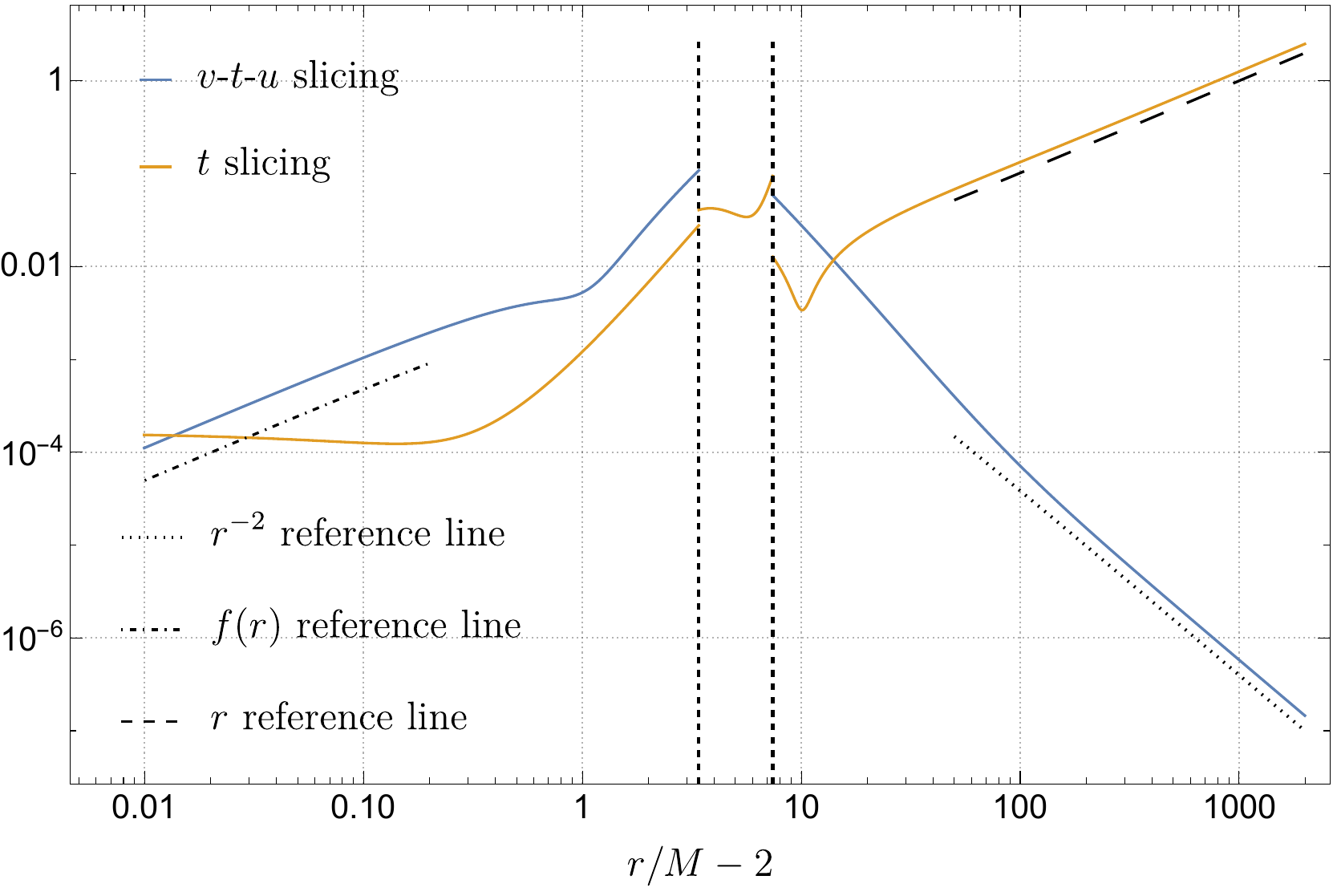}
    \caption{Absolute value of the $i=1$, $\ell=2$, $m=2$ component of the slow-evolution source, $|\breve E^1_{ij\ell m}h^1_{j\ell m}|$ (outside the worldtube) or $|\breve E^1_{ij\ell m}h^1_{j\ell m}+\breve E^0_{ij\ell m}h^{\ms}_{j\ell m}|$ (inside the worldtube). The orbital radius is  $r_0=7.4M$, and the worldtube boundaries, indicated by vertical dotted lines, are at $r_0\pm2M$. Of note are: (i) the difference in asymptotic behaviour depending on the choice of slicing; (ii) the  smoothness of the source arising from the use of the multiscale puncture inside the worldtube.}
    \label{fig:E1h1}
\end{figure}

Figure~\ref{fig:E1h1} displays the absolute value of $\breve E^1_{ij\ell m}h^1_{j\ell m}$ (outside the worldtube) and $\breve E^1_{ij\ell m}h^1_{j\ell m}+\breve E^0_{ij\ell m}h^{\ms}_{j\ell m}$ (inside the worldtube) in both $t$ slicing and $v$-$t$-$u$ slicing. We observe very poor behavior toward the boundaries in $t$ slicing and much improved behavior there in $v$-$t$-$u$ slicing. The rates of falloff in either direction, indicated by the reference lines, are consistent with the predictions in Refs.~\cite{Miller:2020bft,Miller:2023ers}. 

In the figure, we also observe the cancellation between $\breve E^1_{ij\ell m}h^1_{j\ell m}$ and $\breve E^0_{ij\ell m}h^{\ms}_{j\ell m}$ inside the worldtube, leaving a regular effective source. We discuss this in detail in the next section. There we check the cancellation using the undamped quantities $ E^1_{ij\ell m}h^1_{j\ell m}$ and $ E^0_{ij\ell m}h^{\ms}_{j\ell m}$, for consistency with our other checks, while here we plot the damped quantities for completeness, as they are the versions that were used in the numerical implementation in Refs.~\cite{Pound:2019lzj, Warburton:2021kwk}.

\section{Validation of the effective source}\label{sec:validation}

To check the punctures and the second-order Ricci tensor, we substitute them into the appropriate field equations and gauge conditions and verify that they have the correct level of regularity at the particle; see Table~\ref{tab:puncture_orders} for the expected regularity levels. We perform our tests at both the covariant level and at the level of modes. Throughout these checks we use punctures obtained by truncating the covariant singular fields at fourth relative order in $\lambda$, discarding terms of order $\lambda^3$ and above for $h^{\S 1}_{\mu\nu}$ and of order $\lambda^2$ and above for $h^{\S 2}_{\mu\nu}$. For coordinate singular fields, we obtain punctures by keeping only the orders given explicitly in Eqs.~\eqref{eq:hP1coord}--\eqref{eq:hmscoord} and ignoring all higher-order terms. At the level of modes, in cases where we further expand in powers of $\Delta r$ we keep terms through order~$\Delta r^4$.

\begin{table*}[t!]
    \centering
    \begin{ruledtabular}
    \begin{tabular}{ccl}
        Field & Action of operator & Singularity structure  \\
        \midrule
        $h^{\calP2}_{\mu\nu}$ & $\cdots$ & $\displaystyle\frac{1}{\lambda^2}+\frac{1}{\lambda} +\lambda^0+\log\lambda+\lambda+\lambda\log\lambda$  \\[.8em]
        $Z^0_{\mu}[\mathring{\bar{h}}^{\calP2}] + Z^1_{\mu}[\mathring{\bar{h}}^{\calP 1}]$ & $\sim\nabla$ & $\displaystyle\frac{0}{\lambda^3}+\frac{0}{\lambda^2}+\frac{0}{\lambda}+\frac{0}{\lambda}\log\lambda+0\!\cdot\!\lambda^0+0\!\cdot\!\log\lambda+\order(\lambda\log\lambda)$ \\[.8em]
        $S^{2, \rm eff}_{\mu\nu}$ (Eq.~\eqref{eq:EFE2 tt}) & $\sim \nabla^2$ & $\displaystyle \frac{0}{\lambda^4}+\frac{0}{\lambda^3}+\frac{0}{\lambda^2}+\frac{0}{\lambda^2}\log\lambda+\frac{0}{\lambda}+\frac{0}{\lambda}\log\lambda + \order(\log\lambda)$ \\[1.2em]
        $h^{\calP2}_{i\ell m} $ & $\sim \int_{S^2}\odif{\Omega}$ & $P^{(1)}_{ilm}(\Delta r)\log|\Delta r| + P^{(2)}_{ilm}(\Delta r)\,{\rm sgn}(\Delta r)+P^{(3)}_{ilm}(\Delta r)$  \\[0.8em]
        $Z^0_{ij}h^{\calP2}_{j\ell m} + Z^1_{ij}h^{\calP1}_{j\ell m}$ & $\sim \partial_{r}\int_{S^2}\odif{\Omega}$ & $\;\;\displaystyle\frac{0}{\Delta r} \;\;\hspace{1pt} + 0\cdot \delta(\Delta r) \; + \cdots + \order\bigl[(\Delta r)^3\log|\Delta r|\bigr]+\order\bigl[(\Delta r)^3\,\text{sgn}(\Delta r)\bigr] + \text{smooth terms}$ \\[0.8em]
        $S^{2, \rm eff}_{i\ell m}$ (Eq.~\eqref{S2effilm}) & $\sim \partial^2_r\int_{S^2}\odif{\Omega}$ & $\displaystyle\frac{0}{(\Delta r)^2} + 0\cdot \delta'(\Delta r) + \cdots + \order\bigl[(\Delta r)^2\log|\Delta r|\bigr]+\order\bigl[(\Delta r)^2\,\text{sgn}(\Delta r)\bigr] + \text{smooth terms}$ \\[-.9em]&&
    \end{tabular}
    \end{ruledtabular}
    \caption{Singularity structure of the second-order puncture field and quantities constructed from it, where $P^{(n)}_{i\ell m}$ are smooth cubic polynomials in $\Delta r$. 
    We include four total orders in distance in the puncture field, omitting terms of order $\lambda^2$ and higher. Since the Lorenz vector $Z_\mu[\bar{h}]$ and effective source involve one and two derivatives of the puncture, respectively, they involve terms that are one and two orders more singular than the puncture itself. However, the puncture's construction ensures that the first four orders in $\lambda$ cancel in each case. By performing the $(\ell m)$ decomposition, we increase the regularity of each quantity by two orders, with $1/\lambda^2$ terms becoming either $\log(\Delta r)$ (if arising from $h^{\S\S}_{\mu\nu}$) or ${\rm sgn}(\Delta r)$ (if arising from $h^{\delta z}_{\mu\nu}$). The end result, at the level of modes, is that the Lorenz vector is $C^2$ at $r=r_0$, and the effective source is $C^1$. In the punctures we provide online~\cite{PuncturesRepository}, the smooth polynomials $P^{(n)}_{i\ell m}$ are quartic in $\Delta r$, but we stress that the $(\Delta r)^4$ terms are incomplete because $\lambda^2$ terms in $h^{\calP2}_{\mu\nu}$ (which we omit) would also contribute $(\Delta r)^4$ terms in $h^{\calP2}_{i\ell m}$.}
    \label{tab:puncture_orders}
\end{table*}

\subsection{Covariant punctures}

The covariant punctures for $h^{\S\S}_{\mu\nu}$, $h^{\delta m}_{\mu\nu}$ and $h^{\delta z}_{\mu\nu}$ were already confirmed to satisfy the respective field equations to all orders derived in Ref.~\cite{Pound:2014xva}.
The leading two orders of the $h^{\S\R}_{\mu\nu}$ puncture were also successfully checked in Ref.~\cite{Pound:2014xva}, but the authors noted that the  complexity of the expressions made it unfeasible to confirm whether the highest-order term in $h^{\S\R}_{\mu\nu}$ (of order $\lambda^1$) satisfied the correct field equation.

We revisited the calculation as part of our checks for this paper, and in the process discovered an incorrectly symmetrised term in our expression for the order-$\lambda$ piece of $h^{\S\R}_{\mu\nu}$.
This was corrected, and we are now able to show that all orders derived in $h^{\S\R}_{\mu\nu}$ satisfy the wave equation~\eqref{eq:E_hSR}.
We have also confirmed that, for quasicircular orbits, $h^{\ms}_{\mu\nu}$ satisfies Eq.~\eqref{eq:E_hms} to the correct order in $\lambda$ and that the gauge conditions from Eqs.~\eqref{eq:Z_hSS} and~\eqref{eq:Z_hSR} are satisfied.
As such, we are confident that the covariant expressions for the second-order punctures are correct through the appropriate order in $\lambda$ (displayed in Table~\ref{tab:puncture_orders}). 

\subsection{Mode-decomposed punctures}

At the level of modes, we begin by checking that the punctures satisfy the mode-decomposed gauge conditions to the correct degree of regularity. We then check that they correctly cancel the singularities in the field equations. In both types of checks, we restrict to $t$ slicing as that is the slicing we adopt inside the worldtube.

Many of our checks on the mode-decomposed punctures involve asymptotic equalities, which in the present context refers to quantities that are equal up to smooth terms and non-smoothness at some order in $\Delta r$. We denote these using $\simeq$ so that, for example, $f(h) \simeq0$ implies $f(h) = \text{ smooth pieces } + \order[\Delta r^{i} \log(\Delta r)] + \order[\Delta r^{j} \text{sgn}(\Delta r)]$ for some integers $i$ and $j$ that are determined by the order of the puncture. Note that smooth, nonvanishing terms arise as a result of the mode decomposition; these terms would vanish when summed over modes and evaluated on the worldline.

\subsubsection{Checks of the gauge condition}

To check that the mode-decomposed punctures satisfy the gauge conditions~\eqref{eq:Z_cond multiscale 1} and \eqref{eq:Z_cond multiscale 2}, we work with the mode-decomposed versions of those conditions.
These are explicitly given in Eqs.~(154)--(158) of Ref.~\cite{Miller:2020bft}, which we reproduce here as
\begin{align}
    Z^0_{kj}h^{\calP1}_{j\ell m} &\simeq0,\\ Z^0_{kj}h^{\calP2}_{j\ell m} + Z^1_{kj}h^{\calP1}_{j\ell m} &\simeq 0, \label{eq:gauge_condition_modes}
\end{align}  
where
\begin{subequations}\label{eq:gauge_condition_Z0}
\begin{align}
    Z^0_{1j}h^n_{j\ell m} ={}& i \omega_{m}(h^n_{1\ell m} + f h^n_{3\ell m}) \nonumber \\
        & + \frac{f}{r}(rh^n_{2\ell m,r} + h^n_{2\ell m} - h_{4\ell m}), \label{eq:gauge_condition_Z0_1} \\
    Z^0_{2j}h^n_{j\ell m} ={}& i\omega_{m}h^n_{2\ell m} + f\Bigl[h^n_{1\ell m,r} - fh^n_{3\ell m,r} \nonumber \\
        & +\frac{1}{r}\bigl(h^n_{1\ell m} - h^n_{5\ell m} - fh^n_{3\ell m} - 2fh^n_{6\ell m}\bigr)\Bigr], \label{eq:gauge_condition_Z0_2} \\
    Z^0_{3j}h^n_{j\ell m} ={}& i\omega_{m}h^n_{4\ell m} + \frac{f}{r}\bigl[rh^n_{5\ell m,r} + 2h^n_{5\ell m} \nonumber \\
        & + \ell(\ell+1)h^n_{6\ell m} - h^n_{7\ell m}\bigr], \label{eq:gauge_condition_Z0_3} \\
    Z^0_{4j}h^n_{j\ell m} ={}& i\omega_{m}h^n_{8\ell m} + \frac{f}{r}\bigl(rh^n_{9\ell m,r} + 2h^n_{9\ell m} - h^n_{10\ell m}\bigr), \label{eq:gauge_condition_Z0_4}
\end{align}
\end{subequations}
and
\begin{subequations}\label{eq:gauge_condition_Z1}
\begin{align}
    Z^1_{1j}h^{\calP n}_{j\ell m} \simeq{}& -(\vec{\partial}_{\cal V}h^{\calP n}_{1\ell m} + f \vec{\partial}_{\cal V}h^{\calP n}_{3\ell m}), \label{eq:gauge_condition_Z1_1} \\
    Z^1_{2j}h^{\calP n}_{j\ell m} \simeq{}& -\vec{\partial}_{\cal V}h^{\calP n}_{2\ell m}, \label{eq:gauge_condition_Z1_2} \\
    Z^1_{3j}h^{\calP n}_{j\ell m} \simeq{}& -\vec{\partial}_{\cal V}h^{\calP n}_{4\ell m}, \label{eq:gauge_condition_Z1_3} \\
    Z^1_{4j}h^{\calP n}_{j\ell m} \simeq{}& -\vec{\partial}_{\cal V}h^{\calP n}_{8\ell m}, \label{eq:gauge_condition_Z1_4}
\end{align}
\end{subequations}
with $\vec{\partial}_{\cal V}$ given by Eq.~\eqref{d/dV}.
Note that we have set \(H=0\) in the original expressions as we are using \(t\) slicing; see Ref.~\cite{Miller:2020bft} for more details.

Equation~\eqref{eq:gauge_condition_modes} can be further subdivided into a condition on $h^{\S\S}_{i\ell m}$ and a condition on all other puncture fields, following from the mode decomposition of Eqs.~\eqref{eq:Z_hSS} and~\eqref{eq:Z_hSR}. We start with the `singular times singular' piece of the puncture as it can be checked independently of the other fields.
To do so, we substitute the appropriate $i$ modes into Eq.~\eqref{eq:gauge_condition_Z0} and perform a series expansion in $\Delta r$.
This is done for modes satisfying $0\leq \ell\leq 10$ and $0\leq m\leq \ell$.
For all tested modes, we find that
\begin{equation}
    Z^0_{kj}h^{\S\S}_{j\ell m} = \text{ smooth pieces } + \order[\Delta r^{3} \log(\Delta r)],
\end{equation}
agreeing with the expected regularity displayed in Table~\ref{tab:puncture_orders}.

We now move to the `singular times regular' pieces.
As discussed in Sec.~\ref{sec:int_lm}, the modes of the `singular times regular' piece of the second-order singular field have a structurally different form to those of the `singular times singular' piece.
Instead of introducing logarithms, the mode decomposition instead introduces terms $\sim |\Delta r|$.
Terms of this form can then introduce jumps when derivatives are taken unless they are multiplied by a sufficiently high power of $\Delta r$.
To check that we have the regularity that is expected from Table~\ref{tab:puncture_orders}, we therefore need to show that there are no jumps in the gauge condition after applying up to two radial derivatives.
That is, we need to show that
\begin{align}
    \calJ^{(0)}[Z^0_{kj}h^2_{j\ell m} + Z^1_{kj}h^1_{j\ell m}] ={}& 0, \label{eq:jump_con} \\
    \calJ^{(1)}[Z^0_{kj}h^2_{j\ell m} + Z^1_{kj}h^1_{j\ell m}] ={}& 0, \label{eq:jump2_con} \\
    \calJ^{(2)}[Z^0_{kj}h^2_{j\ell m} + Z^1_{kj}h^1_{j\ell m}] ={}& 0\label{eq:jump3_con}
\end{align}
is satisfied, where
\begin{equation}
    \calJ^{(n)}[g(r)] \coloneqq \lim_{r\to r_0^+}\frac{d^ng(r)}{dr^n} - \lim_{r\to r_0^-}\frac{d^ng(r)}{dr^n}\label{eq:jump_func}
\end{equation}
gives the jump at the worldline in the $n$th derivative of $g(r)$, and $h^2_{j\ell m}$ refers to the `singular times regular' piece of the second-order singular field.

We do these checks analytically by substituting the mode decompositions into Eq.~\eqref{eq:jump_con} to check there are no jumps in the gauge conditions.
On doing so, we find that we do not need to combine all of the `singular times regular' pieces for this to be satisfied; only certain combinations are needed.
These are
\begin{align}
    \calJ^{(0)}[Z^0_{kj}h^{\delta z}_{j\ell m}] ={}& 0, \label{eq:J0_dz} \\
    \calJ^{(0)}[Z^0_{kj}(h^{\S\R}_{j\ell m} + h^{\delta m}_{j\ell m})] ={}& 0, \label{eq:J0_SRdm} \\
    \calJ^{(0)}[Z^0_{kj}h^{\ms}_{j\ell m} + Z^1_{kj}h^{\calP1}_{j\ell m}] ={}& 0. \label{eq:J0_ms}
\end{align}%

However, to show that Eqs.~\eqref{eq:jump2_con} and~\eqref{eq:jump3_con} are satisfied, we need to combine all of the fields, as in
\begin{align}
    \calJ^{(1)}[Z^0_{kj}(h^{\S\R}_{j\ell m} + h^{\delta m}_{j\ell m} + h^{\delta z}_{j\ell m} + h^{\ms}_{j\ell m}) + Z^1_{kj}h^{\calP1}_{j\ell m}] ={}& 0, \label{eq:J1_SR} \\
    \calJ^{(2)}[Z^0_{kj}(h^{\S\R}_{j\ell m} + h^{\delta m}_{j\ell m} + h^{\delta z}_{j\ell m} + h^{\ms}_{j\ell m}) + Z^1_{kj}h^{\calP1}_{j\ell m}] ={}& 0, \label{eq:J2_SR}
\end{align}
as well as using the analytical expressions~\eqref{eq:deltar} for \(\deltar\) and~\eqref{eq:r0dot} for \(\rodot\).
Additionally, demonstrating Eq.~\eqref{eq:J1_SR} is satisfied requires the use of the gauge conditions for $h^{\R1}_{\mu\nu}|_{\gamma}$, and for Eq.~\eqref{eq:J2_SR} we had to substitute actual numerical values for $h^{\R1}_{\mu\nu}|_{\gamma}$ and its derivatives due to the complexity of the resulting expression.

\begin{figure*}[!htp]
    \centering
    \includegraphics[width=.985\linewidth, trim={25 22 10 15}]{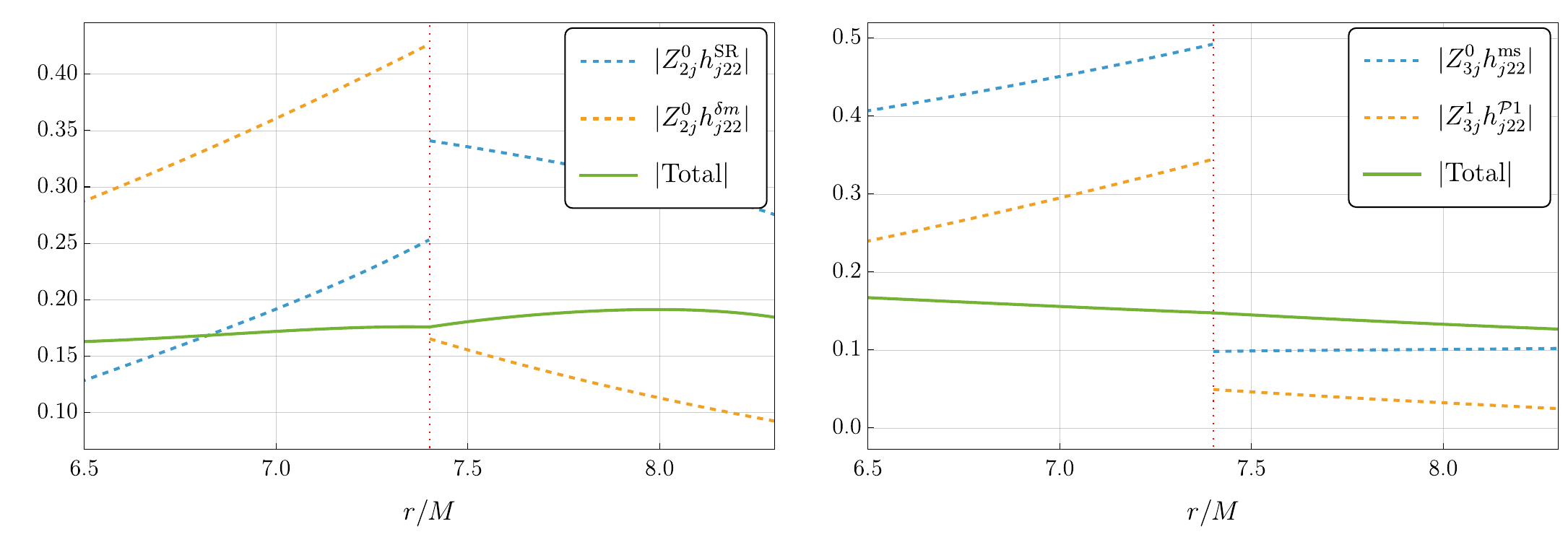}
    \caption{Modes of the Lorenz vector $Z_\mu[h]$ constructed from the puncture fields. While the individual fields do not satisfy the gauge condition $Z_\mu=0$, certain combinations of the puncture fields do. Both plots have the particle at $r_0=7.4M$, which is indicated by a vertical, dotted red line, and are for $\ell=m=2$. (\emph{Left panel}) Plots of $Z^0_{2j}$, given by Eq.~\eqref{eq:gauge_condition_Z0_2}, applied to $h^{\S\R}_{i\ell m}$ (dashed blue) and $h^{\delta m}_{i\ell m}$ (dashed orange) as a function of $r$. While both pieces feature a jump discontinuity at the particle, their sum (solid green) is continuous there; that is, $\calJ^{(0)}[Z^0_{2j}(h^{\S\R}_{j22}+h^{\delta m}_{j22})]=0$, as in Eq.~\eqref{eq:J0_SRdm}. (\emph{Right panel}) Similar to the left panel but for $Z^0_{3j}h^{\ms}_{j22}$ (dashed blue) and $Z^1_{3j}h^{\calP1}_{j22}$ (dashed orange), where $Z^0_{3j}$ and $Z^1_{3j}$ are given by Eqs.~\eqref{eq:gauge_condition_Z0_3} and~\eqref{eq:gauge_condition_Z1_3}, respectively. The individual fields are discontinuous but their sum, as expected from Eq.~\eqref{eq:J0_ms}, is continuous, $\calJ^{(0)}[Z^0_{3j}h^{\ms}_{j22} + Z^1_{3j}h^{\calP1}_{j22}] = 0$.}
    \label{fig:GaugeConditionJumpsJ0}%
\end{figure*}%

\begin{figure*}[!htp]
    \centering
    \includegraphics[width=.985\linewidth, trim={25 22 10 15}]{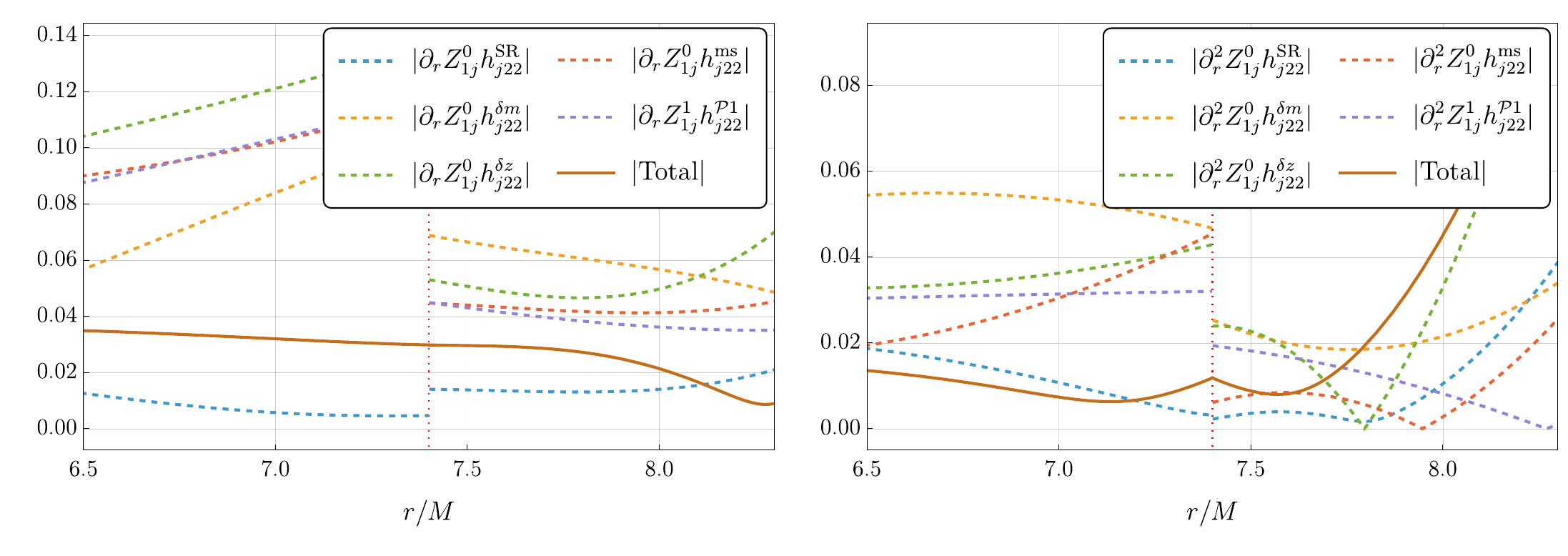}
    \caption{First (\emph{left panel}) and second (\emph{right panel}) radial derivatives of the second-order Lorenz vector, $Z^0_{1j}h^2_{j22} + Z^1_{1j}h^1_{j22}$, where the individual terms are given by Eqs.~\eqref{eq:gauge_condition_Z0_1} and~\eqref{eq:gauge_condition_Z1_1}, respectively. The (absolute value of the) derivatives of the Lorenz vector constructed from the individual fields are given by dashed lines, while their sum is given by a solid brown line. As in Fig.~\ref{fig:GaugeConditionJumpsJ0}, the individual fields do not satisfy the gauge condition, but their sums do. However, unlike Fig.~\ref{fig:GaugeConditionJumpsJ0}, we now require contributions from all of the puncture fields to ensure that the gauge condition is continuous at the particle's orbital radius. This is true for both first and second radial derivatives as found in Eqs.~\eqref{eq:J1_SR} and~\eqref{eq:J2_SR}. Both plots are for $\ell=m=2$ and for a particle at $r_0=7.4M$ (indicated with a vertical, dotted red line).}
    \label{fig:GaugeConditionJumpsJ1J2}
\end{figure*}

Our checks of Eqs.~\eqref{eq:J0_dz}--\eqref{eq:J2_SR} covered the cases $0\leq\ell\leq 5$ and $0\leq m \leq \ell$. In our check of Eq.~\eqref{eq:J2_SR}, we used numerical data for the case $r_0=7.4M$.
Representative plots demonstrating our check of Eqs.~\eqref{eq:J0_SRdm} and~\eqref{eq:J0_ms} are provided in Fig.~\ref{fig:GaugeConditionJumpsJ0}, and plots demonstrating our check of Eqs.~\eqref{eq:J1_SR} and \eqref{eq:J2_SR} are provided in Fig.~\ref{fig:GaugeConditionJumpsJ1J2}.

\subsubsection{Checks of the field equation}

With the mode-decomposed punctures successfully satisfying the Lorenz gauge condition, we move on to showing that they satisfy the Lorenz gauge field equations.
Individually, the fields should satisfy the mode-decomposed versions of Eqs.~\eqref{eq:E_hSS_T}--\eqref{eq:E_hms_T}:
\begin{align}
    E^0_{ij\ell m}h^{\S\S}_{j\ell m} \simeq{}& -\frac{rf}{2a_{i\ell}}\delta^{2}R_{i\ell m}[h^{\calP1},h^{\calP1}], \label{eq:E0_RSS_modes} \\
    E^0_{ij\ell m}h^{\S\R}_{j\ell m} \simeq{}& \frac{rf}{a_{i\ell}}(4\pi\TeffTR^{\Ric}_{i\ell m} - Q^{\Ric,0}_{i\ell m}[h^{\calP1}]), \label{eq:E0_RSR_modes} \\
    E^0_{ij\ell m}h^{\delta m}_{j\ell m} \simeq{}& \frac{4\pi rf}{a_{i\ell}}\TeffTR^{\delta m}_{i\ell m}, \label{eq:E0_Rdm_modes} \\
    E^0_{ij\ell m}h^{\delta z}_{j\ell m} \simeq{}& \frac{4\pi rf}{a_{i\ell}}\TeffTR^{\delta z}_{i\ell m}, \label{eq:E0_Rdz_modes} \\
    E^0_{ij\ell m}h^{\ms}_{j\ell m} \simeq{}& \frac{rf}{4a_{i\ell}}(16\pi\TeffTR^{\ms}_{i\ell m} + E^1_{ij\ell m}h^{\calP1}_{j\ell m}). \label{eq:E0_Rms_modes}
\end{align}
The main difference between the field equation and gauge condition checks is the appearance of distributions on the particle's worldline through the Detweiler stress-energy tensor.
To extract the distributional content of the non-stress-energy pieces of the field equations, we interpret the radial derivatives in $E^n_{ij\ell m}$ and $Q^{\Ric,0}_{i\ell m}$ as distributional derivatives so that $\frac{d|\Delta r|}{dr} = 2\Theta(\Delta r) - 1$ and $\frac{d^2|\Delta r|}{dr^2} = 2\delta(\Delta r)$.
The modes of $\barT_{i\ell m}^2$ are calculated by integrating Eq.~\eqref{eq:TDet_ms} against the appropriate BLS modes and rescaling according to Eq.~\eqref{ERres=Seff}.

We begin by examining Eqs.~\eqref{eq:E0_Rdm_modes}--\eqref{eq:E0_Rms_modes}, that is, the $\delta m$, $\delta z$ and multiscale punctures.
It is straightforward to show that the distributional content on the RHS is the same as that found when acting on the mode-decomposed metric perturbations with $E^0_{ij\ell m}$.
We note also that $E^1_{ij\ell m}h^{\calP1}_{j\ell m}$ does not contain any delta functions as the operator only features one derivative and thus can never produce a delta function; see App.~\ref{sec:E1Form} for the explicit form of $E^1_{ij\ell m}$.

\begin{figure*}[!htp]
    \centering
    \includegraphics[width=.985\linewidth, trim={10 25 10 20}]{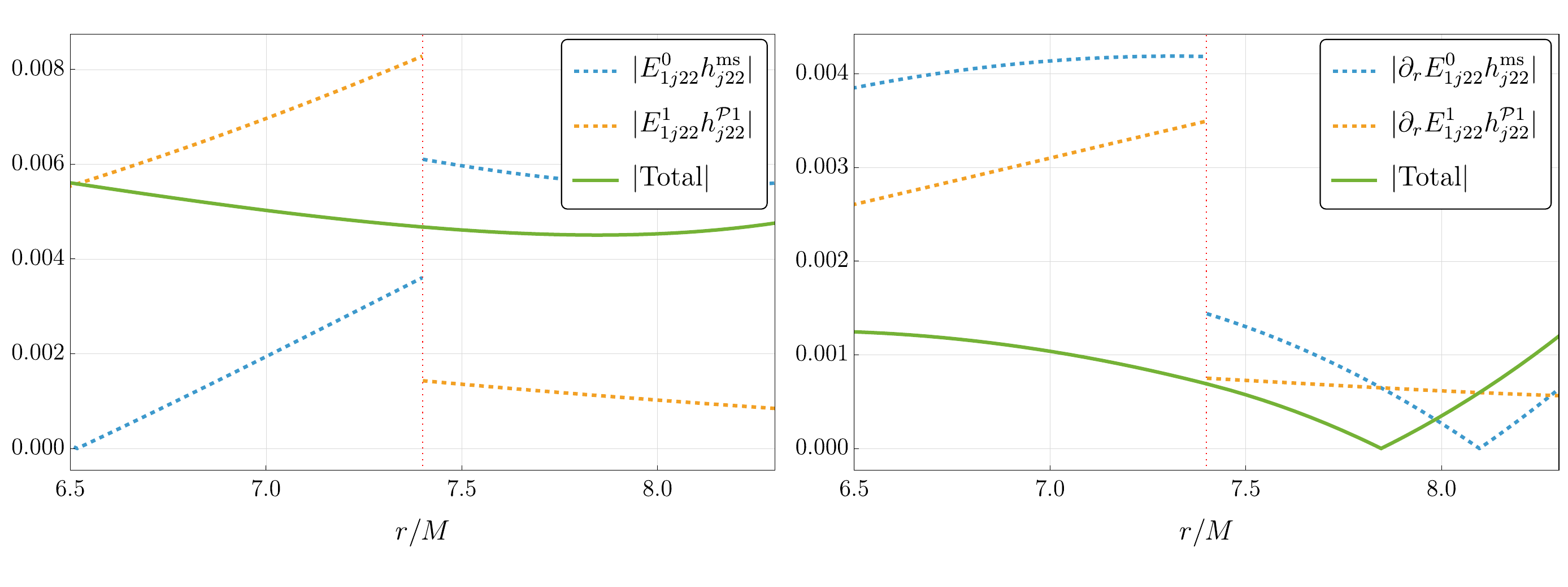}
    \caption{Demonstration that the multiscale parts of the effective source are (at least) $C^1$. Both plots are for the $i=2$, $\ell=m=2$ mode of the source with the particle at $r_0=7.4M$, which is indicated by the vertical, dotted red line. (\emph{Left panel}) Absolute value of $E^0_{1j}h^{\ms}_{j22}$ (dashed blue) and $E^1_{1j}h^{\calP1}_{j22}$ (dashed orange), with their sum in solid green. (\emph{Right panel}) Radial derivatives of these terms. In both panels, the individual terms feature a jump discontinuity at the worldline, but their sums are continuous, as expected from Eqs.~\eqref{eq:J0_Ems} and~\eqref{eq:J1_Ems}.}
    \label{fig:E0msE1S1}
\end{figure*}

Checking the smoothness of the fields, we find that, individually,
\begin{align}
    \calJ^{(0)}[E^0_{ij\ell m}h^{\delta m}_{j\ell m}] ={}& \calJ^{(1)}[E^0_{ij\ell m}h^{\delta m}_{j\ell m}] = 0, \label{eq:J0_Edm} \\
    \calJ^{(0)}[E^0_{ij\ell m}h^{\delta z}_{j\ell m}] ={}& \calJ^{(1)}[E^0_{ij\ell m}h^{\delta z}_{j\ell m}] = 0, \label{eq:J0_Edz}
\end{align}
demonstrating that the field equations for these pieces are $C^1$ on the worldline, as expected.
The field equation for the multiscale puncture is also smooth to the correct order, as
\begin{align}
    \calJ^{(0)}[E^0_{ij\ell m}h^{\ms}_{j\ell m} + E^1_{ij\ell m}h^{\calP1}_{j\ell m}] ={}& 0, \label{eq:J0_Ems} \\
    \calJ^{(1)}[E^0_{ij\ell m}h^{\ms}_{j\ell m} + E^1_{ij\ell m}h^{\calP1}_{j\ell m}] ={}& 0. \label{eq:J1_Ems}
\end{align}
Equations~\eqref{eq:J0_Edm}--\eqref{eq:J1_Ems} have been checked analytically for generic $r_0$ (and thus, without specific values for the residual field and its derivatives), and for $0 \leq \ell \leq 5$ and $0 \leq m \leq \ell$.
Figure~\ref{fig:E0msE1S1} provides an example of the field equation for the multiscale piece of the puncture.
It demonstrates that while both $E^0_{ij\ell m}h^{\ms}_{j\ell m}$ and $E^1_{ij\ell m}h^{\calP1}_{j\ell m}$ (and their derivatives) feature jump discontinuities, their sum does not.

We move next to confirm Eq.~\eqref{eq:E0_RSR_modes} is satisfied to the correct order.
Using the same methods as when verifying Eqs.~\eqref{eq:E0_Rdm_modes}--\eqref{eq:E0_Rms_modes}, it is straightforward to extract the delta content and jumps from $E^0_{ij\ell m}h^{\S\R}_{j\ell m}$ and $\TeffTR^{\text{Ric}}_{i\ell m}$.
However, extracting the delta content and jumps from $Q^{\text{Ric},0}_{i\ell m}$ is slightly more complicated.
This is because $Q^{\text{Ric},0}_{i\ell m}$ is calculated via a sum of products of $h^{\calR1}_{i\ell m}$ and $h^{\calP1}_{i\ell m}$.
While the infinite sum is formally convergent, in practice it is only calculated up to a certain finite $\ell_{\text{max}}$, chosen to ensure convergence up to a certain level of accuracy.

In previous work~\cite{Miller:2016hjv}, the terms in the sum were determined to fall off as $\sim \ell_{\text{max}}^{1-k}$ where $k$ is the order of the puncture used.
As we are using a first-order puncture with four total orders ($k=4$), we would then expect the terms in the sum to fall off as $\sim \ell_{\text{max}}^{-3}$, and for the sum itself to converge as $\sim\ell_{\text{max}}^{-2}$.
By the same arguments, we expect the jump in the derivative of the sum to converge as $\sim \ell_{\text{max}}^{-1}$, because we can think of taking a radial derivative as reducing the degree of smoothness by one order.

Note that this expected rate of convergence for the delta content and the jump in the field at $r_0$ is a worst-case scenario. In many instances the convergence is faster, a fact that can ultimately be traced back to either parity (i.e., the large-$\ell$ behaviour of $h^{\R1}_{i\ell m}$ does not contain any \emph{odd} inverse powers of $\ell$) or to the fact that some components of the circular-orbit metric perturbation are more regular than others.

To test this, we calculate $Q^{\text{Ric},0}_{i\ell m}$ at $r_0=7.4M$ for all ten $i$ modes and for $0\leq\ell\leq 2$ and $0\leq m \leq \ell$ using expressions for $\delta^{2}R_{i\ell m}$ from the \textsc{PerturbationEquations}~\cite{PerturbationEquations} package. We sum modes in the range $\ell_1 \le \ell_{\text{max}}$ and $\ell_2 \le \ell_{\text{max}}+\ell$.
We then extract the delta content, jumps and jumps in the derivatives at the worldline and compare to those from $E^0_{ij\ell m}h^{\S\R}_{j\ell m}$ and $\TeffTR^{\text{Ric}}_{i\ell m}$.
For all $(i,\ell,m)$ modes tested, we find that Eq.~\eqref{eq:E0_RSR_modes} tends to $0$ at (at least) the expected rate when increasing $\ell_{\text{max}}$.
\begin{figure*}[!htp]
    \centering
    \includegraphics[width=.95\linewidth]{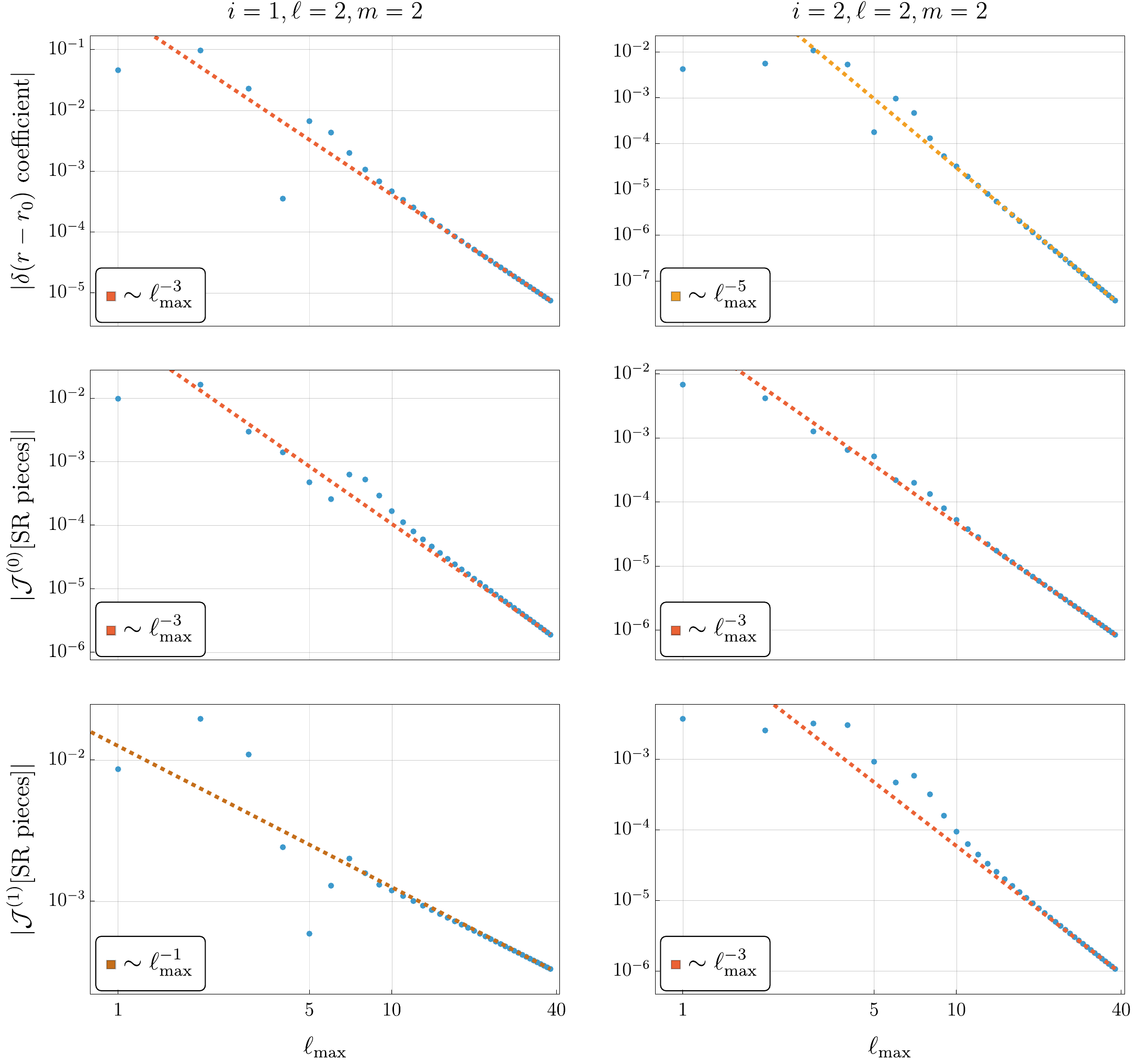}
    \caption{Convergence to zero of singular pieces of the SR source term~\eqref{eq:E0_RSR_modes} at the worldline as a function of the number of first-order modes (denoted by $\ell_{\text{max}}$) included in the calculation of  $Q^{\Ric,0}_{i\ell m}$. Data is displayed for $r_0=7.4M$ and $\ell=m=2$, with $i=1$ (\emph{left column}) or $i=2$ (\emph{right column}). From top to bottom, the rows show the (absolute value of the) coefficient of the radial delta function in the source, the source's jump at the worldline, and the jump in its radial derivative at the worldline. In all plots we see that the total converges to zero with $\ell_{\text{max}}$ at least as rapidly as the expected convergence rate. As mentioned in the body of the paper, certain modes converge faster due to parity or to being constructed from more regular components of the metric perturbation. This is seen in the right-hand column, where the top and bottom plots demonstrate more rapid convergence than the conservative estimate.
}
    \label{fig:SRDeltaCheck}
\end{figure*}
As a representative example, in Fig.~\ref{fig:SRDeltaCheck} we provide plots of the delta content, jumps and jumps in the derivative for Eq.~\eqref{eq:E0_RSR_modes} for $i=1 \text{ and } 2$ and $\ell=m=2$ with $\ell_{\text{max}}$ ranging from $1$ to $38$.

\begin{figure*}[htb!]
    \centering
    \includegraphics[width=0.495\linewidth]{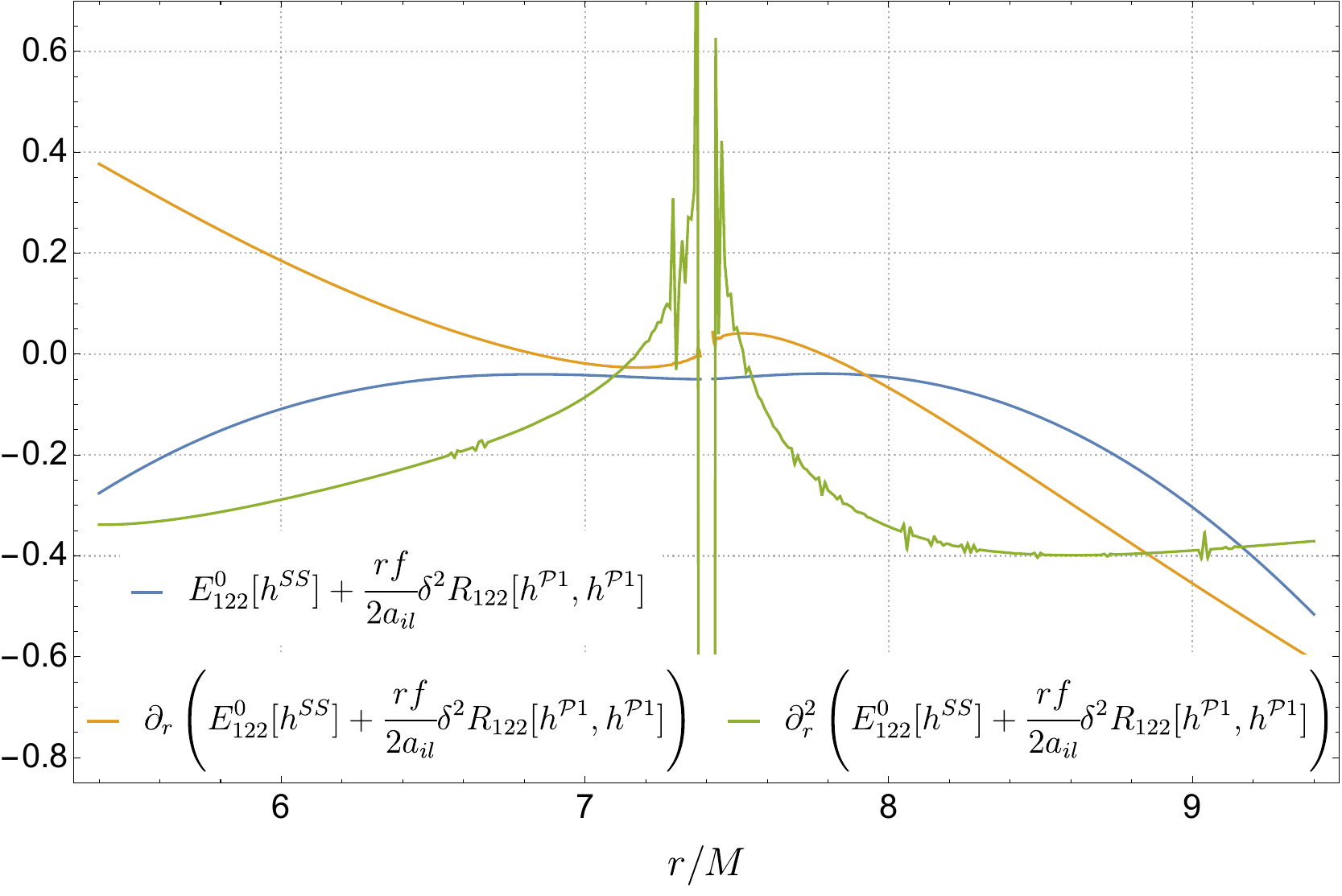}
    \includegraphics[width=0.495\linewidth]{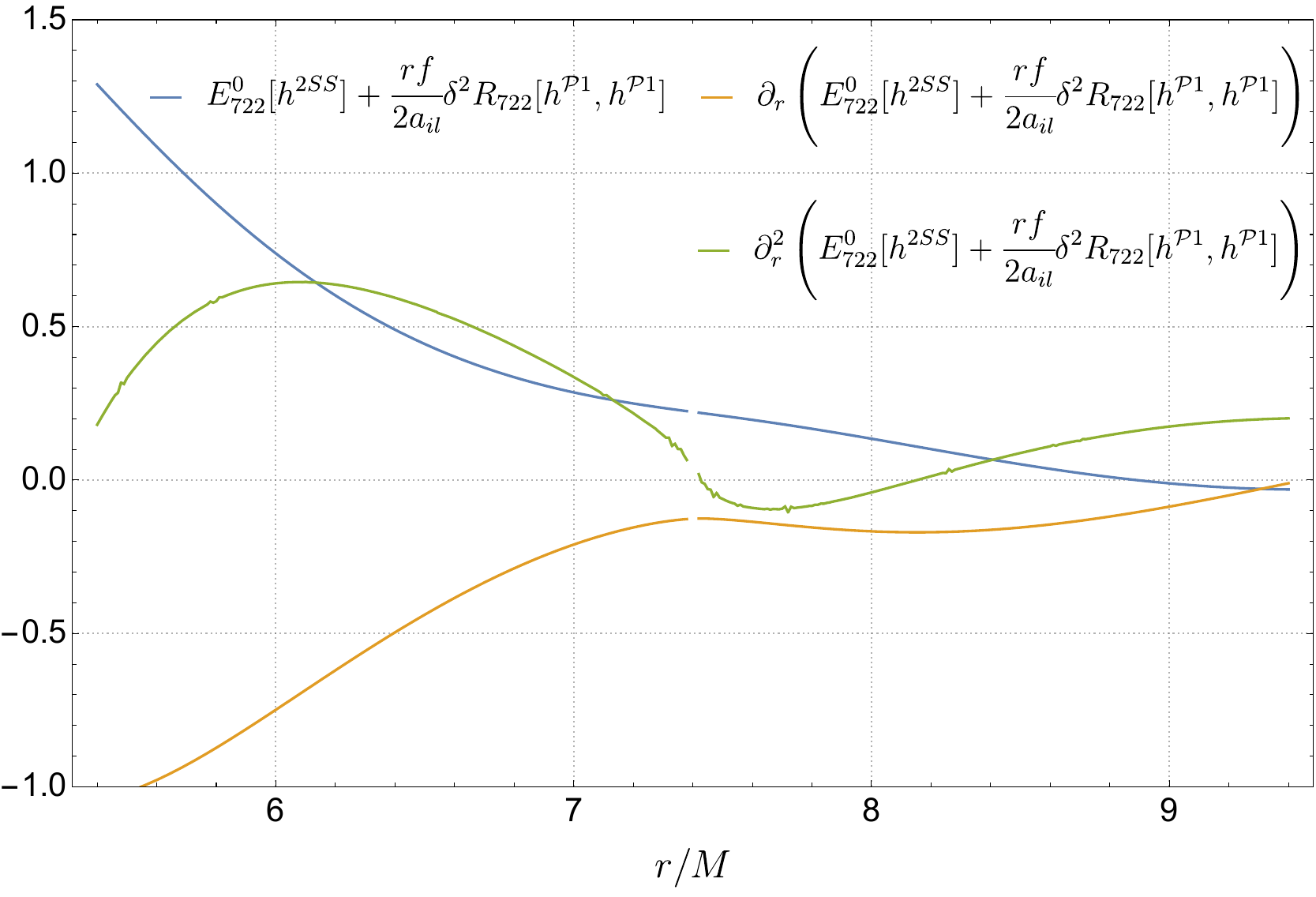}
    \caption{Check of the SS part of the effective source for $r_0=7.4M$. We see that $E^0_{ij\ell m}h^{\S\S}_{j\ell m} + \frac{rf}{2a_{i\ell}}\delta^{2}R_{i\ell m}[h^{\calP1},h^{\calP1}]$ exhibits the expected $C^1$ regularity displayed in Table~\ref{tab:puncture_orders} for the case $i=1$, $\ell=2$, $m=2$ (\emph{left panel}). The corresponding case for $i=7$ (\emph{right panel}) is smoother ($C^2$), suggesting that this component of the source is more regular than others.
    }
    \label{fig:d2GSS-E0hSS_l2m2}
\end{figure*}
Finally, we turn to the check of the SS part, Eq.~\eqref{eq:E0_RSS_modes}. Checking the smoothness of the fields, we find that
\begin{align}
    E^0_{ij\ell m}h^{\S\S}_{j\ell m}& +\frac{rf}{2a_{i\ell}}\delta^{2}R_{i\ell m}[h^{\calP1},h^{\calP1}]= \nonumber \\
    & \text{ smooth pieces } + \order[\Delta r^{2} \log(\Delta r)],
\end{align}
agreeing with the expected regularity displayed in Table~\ref{tab:puncture_orders}.

In this case, since the SS Ricci is only available numerically our check is entirely numerical. Figure~\ref{fig:d2GSS-E0hSS_l2m2} provides a representative example. The left panel shows the case $r_0=7.4M$, $i=1$, $\ell=2$, $m=2$. We see that Eq.~\eqref{eq:E0_RSS_modes} and its first derivative are continuous at the worldline. There is some numerical noise in the second derivative, but despite this we see a behaviour consistent with the expected $\log | \Delta r|$ divergence. The right panel shows the corresponding $i=7$ field. In that case, the second derivative also appears to be continuous, suggesting this component is more smooth than the others.

\section{Complete effective source}
\label{sec:effsource}

\begin{figure*}[htb!]
    \centering
    \includegraphics[width=0.495\linewidth]{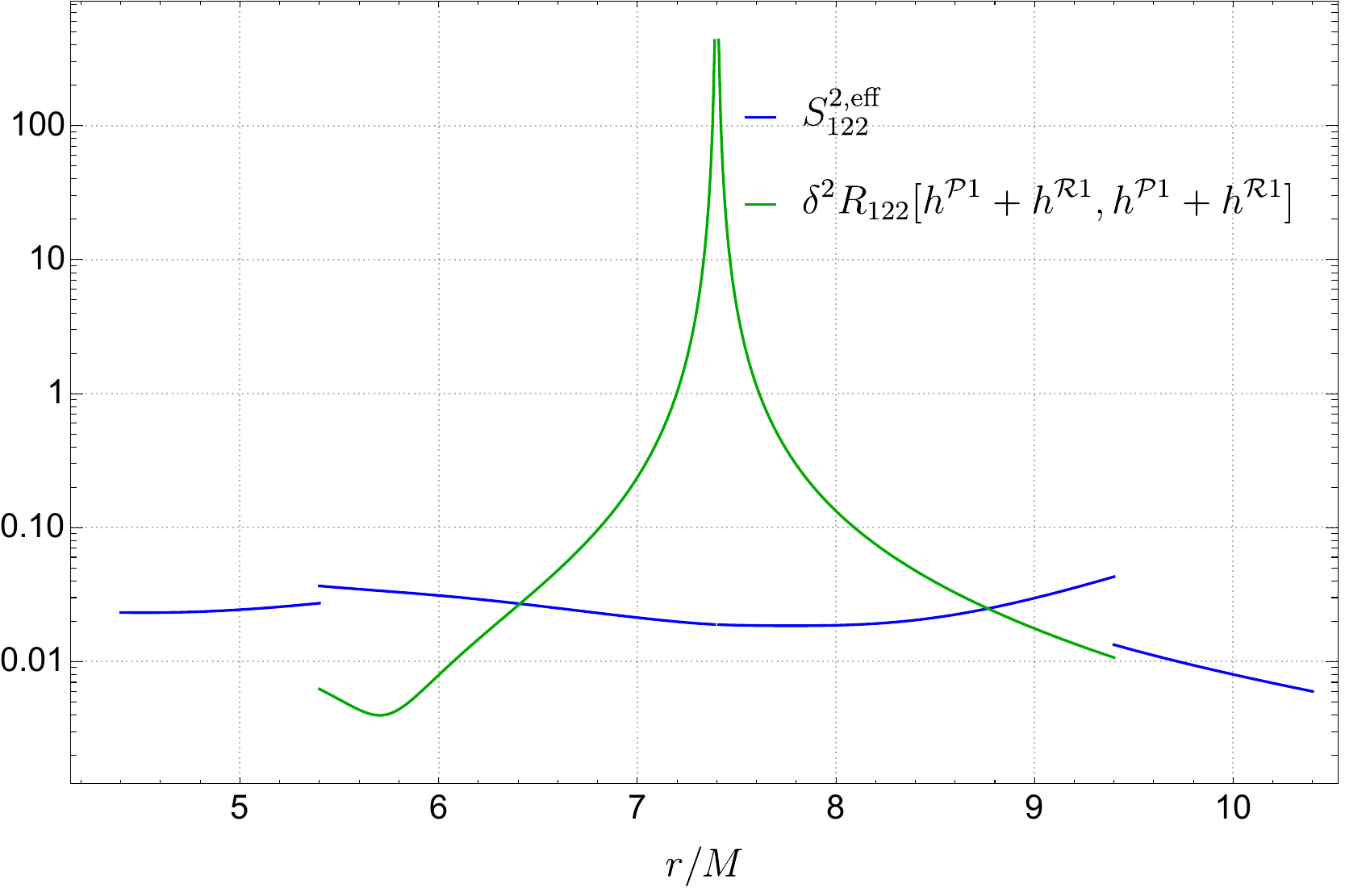}
    \includegraphics[width=0.495\linewidth]{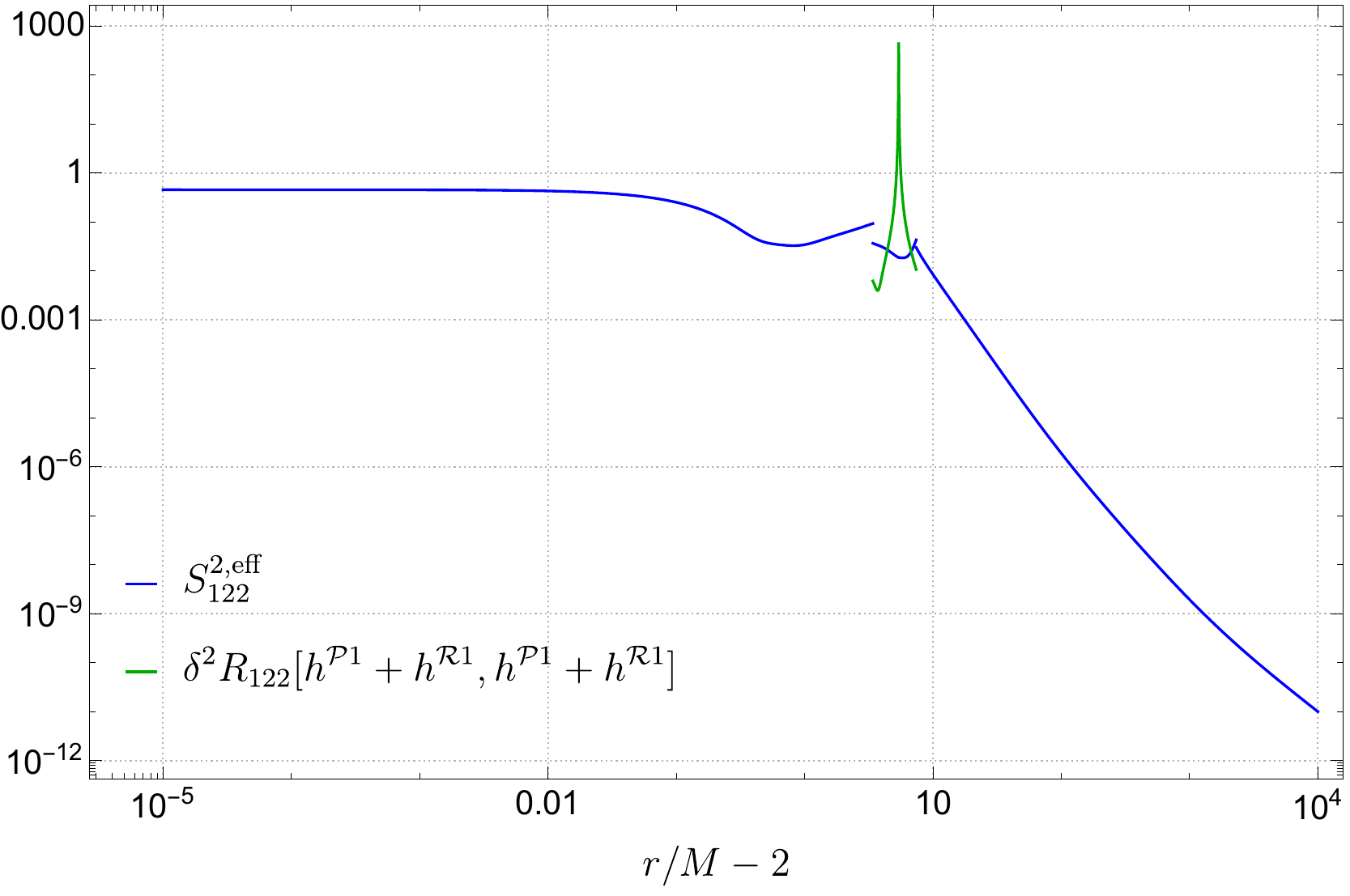}
    \caption{Full effective source near the worldline on $t$-slicing (\emph{left panel}) and throughout our computational domain on $v$-$t$-$u$ slicing (\emph{right panel}) for the case $r_0=7.4M$, $i=1$, $\ell=2$, $m=2$. In both panels, the blue curve shows the effective source, which is finite everywhere. In contrast, the green curve shows that the source without the punctures (i.e. the second order Ricci tensor) diverges towards the worldline. We see similar results for other $(i,\ell,m)$ modes.}
    \label{fig:Seff}
\end{figure*}

Having validated the effective source's various contributions, we now sum them to obtain the complete effective source across the spacetime. This is shown in Fig.~\ref{fig:Seff} in $v$-$t$-$u$ slicing. The left panel focuses on the region around the particle, while the right panel zooms out to show our full numerical domain.

In practice, the effective source used in Refs.~\cite{Pound:2019lzj,Warburton:2021kwk} differs somewhat from the one displayed here. As described in Ref.~\cite{Miller:2023ers}, our concrete calculations involve additional puncture fields in a near-horizon region and a large-$r$ region. These punctures enforce asymptotic (physical, retarded) boundary conditions. At large $r$, the physical boundary conditions on the Lorenz-gauge metric perturbations are derived using an asymptotic post-Minkowskian expansion explained in Ref.~\cite{Cunningham:2024dog}. The near-horizon boundary conditions are obtained using an analogous method in a vicinity of the horizon (inspired by Ref.~\cite{Barack:1998bw}). Since the derivation and implementation of these conditions is quite involved, we defer presentation of it to future papers.

\section{Conclusion}
\label{sec:conclusion}

In this paper we have provided the complete effective source for second-order gravitational self-force calculations in Schwarzschild spacetime, restricted to quasicircular inspirals in a multiscale formulation of the field equations.

The source contains several ingredients, each of which we have stringently tested:
\begin{enumerate}
    \item A puncture field expanded in tensor spherical harmonics. In Sec.~\ref{sec:punctures}, we have developed a multiscale expansion of the puncture field (valid for generic orbits in Kerr) and the tensor spherical-harmonic decomposition (restricted to quasicircular orbits in Schwarzschild spacetime). The puncture field is made up of several pieces, which satisfy specific field equations and gauge conditions. We have verified that all conditions on the puncture fields are satisfied. This includes establishing consistency, at the level of the final mode decomposition, with the distributional form of the field equations developed by two of us in Ref.~\cite{Upton:2021oxf}.
    \item A quadratic source term, $\delta^2 R_{i\ell m}$, constructed from products of first-order fields and their first and second derivatives. In Sec.~\ref{sec:d2R}, we have described our calculation of this source, following the strategy in Ref.~\cite{Miller:2016hjv}. Inside a worldtube around the particle, this involves (i) constructing the more regular parts of $\delta^2 R_{i\ell m}$ from sums of products of first-order (puncture and residual field) modes, using coupling formulas from Ref.~\cite{Spiers:2023mor}, and (ii) constructing the most singular part of $\delta^2 R_{i\ell m}$ by numerically evaluating two-dimensional angular integrals of the four-dimensional $\delta^2 R_{\mu\nu}[h^{{\cal P}1},h^{{\cal P}1}]$. Outside the worldtube, we use mode-coupling alone, calculating $\delta^2 R_{i\ell m}$ from first-order retarded field modes. We have verified consistency between these two calculations at the worldtube boundary, and we have verified that the various pieces of this source term correctly cancel pieces of the second-order puncture field near the particle. 
    \item A piece arising from the slow evolution of the first-order field, displayed in Sec.~\ref{sec:slow_time}. This piece of the source is highly dependent on the choice of time slicing in the multiscale expansion. In addition to verifying that this source correctly cancels a certain piece of the puncture field, we have also verified that it has the correct behavior at the horizon and infinity in our $v$-$t$-$u$ slicing.
\end{enumerate}

In these calculations and checks, we have expanded the second-order puncture to two orders higher (in powers of distance $\lambda$ from the particle) than necessary. To obtain the correct asymptotic fluxes, we only need to include enough terms to (i) yield an integrable effective source (i.e., a source that diverges no more strongly than $\log\Delta r$ at the level of modes), (ii) be consistent with $C^1$ residual field modes $h^{{\cal R}2}_{i\ell m}$ at $r=r_0$ (since we would otherwise need to impose jump conditions for the residual field modes). This implies we only need to control terms of order $1/\lambda^2$ and $1/\lambda$ in the second-order puncture. We have instead consistently included terms through linear order in $\lambda$. There are several reasons for this. First, it leads to a smoother effective source. Second, it allows us to more robustly test every piece of the effective-source calculations. Third, it will ultimately be needed in a calculation of the local self-force on the particle (which involves a derivative of the residual field). Fourth, it is required in second-order Teukolsky calculations, which involve two additional derivatives of the fields; we will present those calculations in a followup paper~\cite{Leather:2025InPrep}.

The work we have presented here, alongside Refs.~\cite{Miller:2020bft,Bonetto:2021exn,Miller:2023ers,Spiers:2023mor,Cunningham:2024dog}, provides a nearly complete description of the methods underlying the second-order calculations and waveform generation in Refs.~\cite{Pound:2019lzj,Warburton:2021kwk,Wardell:2021fyy}. We have only left out details on the following:
\begin{enumerate}
    \item The calculation of large-$r$ boundary conditions (following the method explained in Ref.~\cite{Cunningham:2024dog}) and near-horizon boundary conditions. These boundary conditions are enforced using the puncture method explained in Ref.~\cite{Miller:2023ers} and will be fully detailed in future papers.
    \item The $\delta M$ and $\delta S$ contributions to the source arising from their contributions to $h^1_{\mu\nu}$. As explained in Ref.~\cite{Mathews:2025nyb}, we separate these out, compute them using the same methods described in this paper, and appeal to linearity to add them in at the end.
\end{enumerate}

Even in the restricted context of quasicircular, nonspinning binaries, numerous natural followup calculations present themselves:
\begin{enumerate}
    \item A complete 1PA waveform model requires calculation of second-order ($\varepsilon^3$) horizon fluxes, while Ref.~\cite{Warburton:2021kwk} only calculated fluxes to future null infinity. 
    \item The fluxes at future null infinity in Ref.~\cite{Warburton:2021kwk} were incomplete as they omitted ``memory distortion'' terms recently discovered in Ref.~\cite{Cunningham:2024dog}. These should be calculated.
    \item The 1PA waveform generation in Ref.~\cite{Wardell:2021fyy} (see also Ref.~\cite{Albertini:2022rfe}) relied on an assumed energy balance law, in which the particle's mechanical energy decreases at the rate of emission of asymptotic fluxes. Recent and forthcoming work~\cite{Trestini:2025nzr, Grant:2025InPrep} shows that this law needs to be corrected.
    \item We aim to also calculate the local second-order self-force. The dissipative piece of the force can be calculated from the oscillatory, $m\neq0$ modes we have focused on here. The work done by this dissipative force should be related to the asymptotic energy fluxes through the corrected balance law mentioned above.
    \item The conservative pieces of the self-force will additionally require the calculation of stationary, $m=0$ modes. These conservative forces will have a direct contribution from gravitational memory, as shown in Ref.~\cite{Cunningham:2024dog} (see also~\cite{Porto:2024cwd}).
\end{enumerate}

As alluded to above, other work~\cite{Leather:2025InPrep} will present the calculation of fluxes from the second-order Teukolsky formalism of Refs.~\cite{Spiers:2023cip,Spiers:2023mor}, rather than the Lorenz-gauge method of this paper, utilizing the spectral method of Refs.~\cite{PanossoMacedo:2022fdi,Leather:2024mls}. Calculations of the effective source in a Kerr background are also ongoing, building on methods developed here and in Refs.~\cite{Osburn:2022bby,Dolan:2023enf,Bourg:2024cgh,PanossoMacedo:2024pox}.

\begin{acknowledgments}
SDU and AP acknowledge the support of a Royal Society University Research Fellowship and the ERC Consolidator/UKRI Frontier Research Grant GWModels (selected by the ERC and funded by UKRI [grant number EP/Y008251/1]). NW acknowledges support from a Royal Society -- Research Ireland University Research Fellowship. This publication has emanated from research conducted with the financial support of Research Ireland under grant number 22/RS-URF-R/3825.
This work is supported by ERC grant EMRIWaveforms (\href{https://doi.org/10.3030/101200625}{https://doi.org/10.3030/101200625}).
LB acknowledges support from the STFC via grant No.~ST/B001170/1.

This work makes use of the Black Hole Perturbation Toolkit.
\end{acknowledgments}

\appendix

\section{Derivation of the trace-reversed Detweiler stress-energy tensor}\label{sec:T_eff_TR_deriv}

In this appendix we derive the trace-reversed stress-energy source $\bar T_{\mu\nu}$ given in Eq.~\eqref{eq:T_eff_TR_intro}.

In Ref.~\cite{Upton:2021oxf}'s original derivation of the Detweiler stress-energy tensor~\eqref{eq:T_def} in the Lorenz gauge, the field equations were written in self-consistent form in terms of perturbations of the Einstein tensor,
\begin{equation}
    \delta G_{\mu\nu}[\e h^1 + \e^{2}h^2] + \e^2\delta^{2}G_{\mu\nu}[h^1,h^1] = 8\pi T_{\mu\nu} + \order(\e^3). \label{eq:EFE_sc_T}
\end{equation}
This equation was made well-defined, despite its quadratic singularities, by adopting the canonical Detweiler definition of the second-order Einstein tensor,
\begin{equation}
    \delta^{2}G_{\mu\nu}[h^1,h^1] \coloneqq \lim_{s\to 0^+}\delta^{2}G_{\mu\nu}^s[h^1,h^1], \label{eq:d2G_net}
\end{equation}
where
\begin{align}
    \delta^{2}G_{\mu\nu}^s[h^1,h^1] &\coloneqq \bigl(-\delta G_{\mu\nu}[h^{\S\S}]+ \delta^{2}G_{\mu\nu}[h^{\R1},h^{\S1}] \nonumber \\
        &\quad + \delta^{2}G_{\mu\nu}[h^{\S1},h^{\R1}] + \delta^{2}G_{\mu\nu}[h^{\R1},h^{\R1}]\bigr)\theta^-_s \nonumber \\
        &\quad + \delta^{2}G_{\mu\nu}[h^1,h^1]\theta^+_s.
\end{align}
Here $\theta^-_s$ is a smooth window function equal to 1 in an open neighbourhood of the worldline and vanishing outside a tube $\rho<s$, with $\rho$ the proper distance from the worldline; similarly, $\theta^+_s\coloneqq 1-\theta^-_s$ is equal to 1 outside the tube and vanishes in a neighbourhood of the worldline. The core idea here is that we have replaced the quadratic quantity $\delta^2G_{\mu\nu}[h^{\S1},h^{\S1}]$ with the linear quantity $-\delta G_{\mu\nu}[h^{\S\S}]$ in a neighbourhood of the worldline. 

Using the definition~\eqref{eq:d2G_net}, in Ref.~\cite{Upton:2021oxf} two of us were able to show that $T_{\mu\nu}$ in Eq.~\eqref{eq:EFE_sc_T} is uniquely given by the Detweiler stress-energy tensor~\eqref{eq:T_def}. The derivation establishes that any solution to Eq.~\eqref{eq:EFE_sc_T}, with Eqs.~\eqref{eq:d2G_net} and~\eqref{eq:T_def}, locally agrees with the metric outside a small compact object (which is obtained from first principles using matched asymptotic expansions).

When performing calculations in the Lorenz gauge, it is often useful to work with the trace-reversed Einstein field equations instead of Eq.~\eqref{eq:EFE_sc_T}.  If the stress-energy tensor and metric were smooth fields, the trace-reversed field equations would read $R_{\mu\nu}[{\sf g}]=8\pi\left(\delta^\alpha_\mu \delta^\beta_\nu-\frac{1}{2}{\sf g}_{\mu\nu}{\sf g}^{\alpha\beta}\right)T_{\alpha\beta}$, and one might naively try to start from this equation with the Detweiler stress-energy tensor on the right-hand side. However, the stress-energy source would then be manifestly ill defined as it would involve products of the (singular) metric ${\sf g}_{\mu\nu}$ with the Dirac delta function in $T_{\mu\nu}$. A correct approach is instead to start from the vacuum field equations away from the worldline, which (in the Lorenz gauge) take the form
\begin{equation}
    E_{\mu\nu}[\e h^1 + \e^2h^2] - 2\e^{2}\delta^{2}R_{\mu\nu}[h^1,h^1] = \order(\e^3). \label{eq:dR_EFE_sc}
\end{equation}
If we promote this equation to a domain including the worldline, we can find the correct point-particle source term by demanding consistency with the local form $h^{\S n}_{\mu\nu}+h^{\R n}_{\mu\nu}$ of the metric perturbations derived from matched expansions.
Taking this approach and appealing to methods in Ref.~\cite{Upton:2021oxf}, we will demonstrate that, when promoted to the full spacetime, Eq.~\eqref{eq:dR_EFE_sc} becomes
\begin{equation}
    E_{\mu\nu}[\e h^1 + \e^2h^2] - 2\e^{2}\delta^{2}R_{\mu\nu}[h^1,h^1] = -16\pi\TeffTR_{\mu\nu} + \order(\e^3), \label{eq:dR_EFE_sc_T}
\end{equation}
where $\TeffTR_{\mu\nu}$ is the trace reversal of the Detweiler stress-energy tensor with respect to the effective metric,
\begin{equation}
    \TeffTR_{\mu\nu} = \left(\delta^\alpha_\mu\delta^\beta_\nu - \frac{1}{2}\geff_{\mu\nu}\geff^{\alpha\beta}\right)\Teff_{\alpha\beta} + \order(\e^3),\label{eq:T_eff_TR}
\end{equation}
rather than with respect to ${\sf g}_{\mu\nu}$ or $g_{\mu\nu}$. 

This result is not surprising, given the properties of the Detweiler stress-energy tensor derived in Ref.~\cite{Upton:2021oxf}.
It was shown in that reference that the indices of $T_{\mu\nu}$ need to be raised and lowered with the effective metric, not the physical metric (nor the background metric). In short, when manipulating the indices of the Detweiler stress-energy tensor, we must always use the effective metric to obtain the result that is compatible with a distributional treatment of the perturbed Einstein field equations.

\subsection{Distributional analysis}\label{sec:Teff_dist}

In performing our distributional analysis, we skip over many steps as they are functionally equivalent to those performed in Ref.~\cite{Upton:2021oxf}; see Sec.~V of that reference for details of the full calculation.
We also adopt Detweiler's canonical definition from Eq.~\eqref{eq:d2G_net} but apply it to the second-order Ricci tensor.

To begin, we write the Einstein equations with an as-yet-undetermined stress-energy source,
\begin{align}
    -16\pi\TeffTR_{\mu\nu} ={}& \lim_{s\to0^+}\Bigl\{\e E_{\mu\nu}[h^{\S1}] + \e^2 \Bigl(E_{\mu\nu}[h^{\S\R}] + E_{\mu\nu}[h^{\delta m}] \nonumber \\
        & - 4 Q^{\Ric}_{\mu\nu}[h^{\S1}]\Bigr)\Bigr\}\theta^-_s + \order(\e^3). \label{eq:TeffTR_def}
\end{align}
Here $
Q^{\Ric}_{\mu\nu}[h] = \frac{1}{2}\delta^{2}R_{\mu\nu}[h^{\R1},h]+\frac{1}{2}\delta^{2}R_{\mu\nu}[h,h^{\R1}]$
is the smooth linear operator defined in Eq.~\eqref{eq:Qdef}, and we have used (i) $E_{\mu\nu}[h^{n}]=\lim_{s\to0^+}(\theta^-_s+\theta^+_s)E_{\mu\nu}[h^{n}]$, (ii) the vacuum field equation~\eqref{eq:dR_EFE_sc} in the region $\rho>s$, and (iii) the fact that the regular fields satisfy the vacuum field equations also for $\rho<s$. 

We now determine $\bar T_{\mu\nu}$ by integrating Eq.~\eqref{eq:TeffTR_def} against a test field $\phi^{\mu\nu}$:
\begin{align}
    \MoveEqLeft[4] -16\pi\int\phi^{\mu\nu}\TeffTR_{\mu\nu}\odif{V} \nonumber \\
        ={}& \lim_{s\to0}\int\phi^{\mu\nu}_s\Bigl\{\e E_{\mu\nu}[h^{\S1}] + \e^2\bigl(E_{\mu\nu}[h^{\S\R}] \nonumber \\
            & + E_{\mu\nu}[h^{\delta m}] - 4 Q^{\Ric}_{\mu\nu}[h^{\S1}]\bigr)\Bigr\}\odif{V},\label{eq:int phi Tbar}
\end{align}
where we defined the new test field $\phi^{\mu\nu}_s\coloneqq \phi^{\mu\nu}\theta^-_s$. We next move the linear operators onto the test field via the definition $\int \phi L[\psi]dV\coloneqq \int \psi L^\dagger[\phi]dV$ (for a test field $\phi$ and integrable function $\psi$). Using the fact that $\int F dV = \lim_{R\to0^+}\int_{\rho>R}F dV $ for any integrable function $F$, integrating by parts, and discarding the remaining volume integral as its integrand satisfies the vacuum Einstein field equations (in the vacuum region $\rho>R$), we reduce the right-hand side of Eq.~\eqref{eq:int phi Tbar} to
\begin{align}
     \MoveEqLeft[4] \int\bigl(E_{\mu\nu}[\phi_s]\{\e h_{\S1}^{\mu\nu} + \e^2(h_{\S\R}^{\mu\nu} + h_{\delta m}^{\mu\nu})\} - 4\e^{2}Q^{\Ric\dagger}_{\mu\nu}[\phi]h^{\mu\nu}_{\S1}\bigr)\odif{V} \nonumber \\
        ={}& -\lim_{R\to 0^+}\int_{\rho=R}\Bigl\{K^E_\alpha[\e h^{\S1} + \e^{2}(h^{\S\R} + h^{\delta m})] \nonumber \\
            & - 4\e^{2}K^{Q_{\Ric}}_\alpha[h^{\S1}]\Bigr\}\odif{\Sigma^\alpha},
\end{align}
before taking the limit $s\to0$.
In obtaining this result we have used the fact that the linearised Einstein operator is self-adjoint. We can write the surface element as
\begin{equation}
    \odif{\Sigma^\alpha} = -R^2 n^\alpha\odif{\tau}\odif{\Omega} + \order(R^3)
\end{equation}
with outward-directed unit normal $n^\alpha$.
The operators $K^{E/Q_{\rm Ric}}_\alpha[h]$ appearing in the integrand are given by
\begin{equation}
    K^E_\alpha[h] = \phi^{\mu\nu}_sh_{\mu\nu;\alpha} - h_{\mu\nu}\phi^{\mu\nu}_{s\ \; ;\alpha}, \label{eq:KE}
\end{equation}
and
\begin{align}
    K^{Q_{\Ric}}_\alpha[h] ={}& -\frac{\phi_s^{\mu\nu}}{8}\Bigl(4h^{\R1}_{\alpha\beta}\delta\Gamma^\beta_{\mu\nu}[h] + 4\barh_{\alpha\beta}\delta\Gamma^\beta_{\mu\nu}[h^{\R1}] \nonumber \\
        & + h_{\mu\nu}h^{\R1}_{\beta}{}^{\beta}{}_{;\alpha} + 4h_{\mu}{}^{\beta}(\barh^{\R1}_{\nu\alpha;\beta} - h^{\R1}_{\nu\beta;\alpha}) \nonumber \\
        & - 2g_{\alpha\mu}h_{\beta\gamma;\mu}h_{\R1}^{\beta\gamma}\Bigr) - \frac{\phi^{\mu\nu}_s{}_{;\rho}}{4}\Bigl(h_{\mu\nu}h^{\R1}_{\alpha}{}^{\rho} \nonumber \\
        & + g_{\alpha\mu}\delta_{\nu}^{\rho}h^{\beta\gamma}h^{\R1}_{\beta\gamma} - 2g_{\alpha\mu}h_{\nu\beta}h_{\R1}^{\beta\rho}\Bigr), \label{eq:KQRic}
\end{align}
where
\begin{equation}
    \delta\Gamma^\alpha_{\mu\nu}[h] = \frac{1}{2}g^{\alpha\beta}(2h_{\beta(\mu;\nu)}-h_{\mu\nu;\beta})
\end{equation}
is the linear perturbation of the Christoffel symbol.

Working in Fermi-Walker coordinates $(\tau,x^i)$, with $x^i=\rho n^i$ and $n^\alpha=(0,n^i)$, we can substitute the known Fermi-Walker expressions for the metric perturbations~\cite{Pound:2014xva} into Eqs.~\eqref{eq:KE}--\eqref{eq:KQRic}, perform the angular integral, and take the limit. We find
\begin{equation}
    \lim_{R\to 0^+}\int_{\rho=R}K^E_\alpha[h^{\S1}]\odif{\Sigma}^\alpha = 8\pi m\int\left(g_{\alpha\beta} + 2u_\alpha u_\beta\right)\phi^{\alpha\beta}\odif{\tau} \label{eq:KEhS1}
\end{equation}
at first order in $\e$, and
\begin{align}
    \MoveEqLeft[4] \lim_{R\to 0^+}\int_{\rho=R}K^E_\alpha[h^{\S\R}]\odif{\Sigma}^\alpha = 0, \label{eq:KEhSR} \\
    \MoveEqLeft[4] \lim_{R\to 0^+}\int_{\rho=R}K^E_\alpha[h^{\delta m}]\odif{\Sigma}^\alpha \nonumber \\
        ={}& \frac{4\pi m}{3}\int(2h^{\R1}_{\alpha\beta} + g_{\alpha\beta}h^{\R1}_{\gamma}{}^{\gamma} + 12h^{\R1}_{\alpha\gamma}u_{\beta}u^{\gamma} \nonumber \\
        & + 3g_{\alpha\beta}h^{\R1}_{\mu\nu}u^{\mu}u^{\nu} + 6h^{\R1}_{\mu\nu}u_{\alpha}u_{\beta}u^{\mu}u^{\nu})\phi^{\alpha\beta}\odif{\tau}, \label{eq:KEhdm} \\
    \MoveEqLeft[4] \lim_{R\to 0^+}\int_{\rho=R}K^{Q_{\Ric}}_\alpha[h^{\S1}]\odif{\Sigma}^\alpha \nonumber \\
        ={}& -\frac{2\pi m}{3}\int(2h^{\R1}_{\alpha\beta} - (2g_{\alpha\beta} + 3u_{\alpha}u_{\beta})h^{\R1}_{\gamma}{}^{\gamma} \nonumber \\
        & + 6 h^{\R1}_{\alpha\gamma}u_{\beta}u^{\gamma} - 3g_{\alpha\beta}h^{\R1}_{\mu\nu}u^{\mu}u^{\nu})\phi^{\alpha\beta}\odif{\tau} \label{eq:KQhS1}
\end{align}
at second order. Notice that the integral over $\tau$ is evaluated along the worldline at $\rho=0$, allowing us to replace $\phi^{\mu\nu}_s$ with $\phi^{\mu\nu}$ by virtue of $\phi_s^{\mu\nu}(\rho=0) = \phi^{\mu\nu}(\rho=0)$. Hence, our expressions are now independent of $s$, and we can take the limit $s\to0$ without consequence.

Returning to Eq.~\eqref{eq:int phi Tbar}, as in Ref.~\cite{Upton:2021oxf}, we can write the stress-energy tensor at each order as a sum of the boundary terms:
\begin{align}
    \int \TeffTR^1_{\mu\nu}\phi^{\mu\nu}\odif{V} ={}& \frac{1}{16\pi}\lim_{R\to 0^+}\int_{\rho=R}K^E_\alpha[h^{\S1}]\odif{\Sigma}^{\alpha}, \\
    \int \TeffTR^2_{\mu\nu}\phi^{\mu\nu}\odif{V} ={}& \frac{1}{16\pi}\lim_{R\to 0^+}\int_{\rho=R}(K^E_\alpha[h^{\S\R}] + K^E_{\alpha}[h^{\delta m}] \nonumber \\
        & - 4K^{Q_{\Ric}}_\alpha[h^{\S1}])\odif{\Sigma}^\alpha.
\end{align}
Appealing to Eqs.~\eqref{eq:KEhS1}--\eqref{eq:KQhS1} and noting that this holds true for arbitrary test fields $\phi^{\mu\nu}$, we find the final expressions for the first- and second-order pieces of the trace-reversed Detweiler stress-energy tensor are given by
\begin{align}
    \TeffTR^1_{\mu\nu} ={}& \frac{m}{2}\int(g_{\mu\nu} + 2u_{\mu}u_{\nu})\delta^{4}(x,z)\odif{\tau}, \label{eq:TeffTR_1} \\
    \TeffTR^{2}_{\mu\nu} ={}& \frac{m}{2}\int\bigl\{2\barh^{\R1}_{\mu\nu} - \bigl[g_{\mu\nu}u^{\alpha}u^{\beta} + 2(g^{\alpha\beta}-u^{\alpha}u^{\beta})u_{\mu}u_{\nu}\bigr] \nonumber \\
        & \times h^{\R1}_{\alpha\beta} + 8 h^{\R1}_{\alpha(\mu}u_{\nu)}u^{\alpha}\bigr\}\delta^{4}(x,z)\odif{\tau}. \label{eq:TeffTR_2}
\end{align}
We note, as expected, that Eq.~\eqref{eq:TeffTR_1} is just the trace reversal of the stress-energy tensor for a point mass.

When validating the punctures, it is useful to define terms corresponding to each puncture field's contribution to the final stress-energy tensor.
These are
\begin{align}
    \int \TeffTR^{\S\R}_{\mu\nu}\phi^{\mu\nu} \odif{V} ={}& \frac{1}{16\pi}\lim_{R\to 0^+}\int_{\rho=R}K^E_\alpha[h^{\S\R}]\odif{\Sigma}^\alpha, \\
    \int \TeffTR^{\delta m}_{\mu\nu}\phi^{\mu\nu} \odif{V} ={}& \frac{1}{16\pi}\lim_{R\to 0^+}\int_{\rho=R}K^E_\alpha[h^{\delta m}]\odif{\Sigma}^\alpha, \\
    \int \TeffTR^{Q_{\Ric}}_{\mu\nu}\phi^{\mu\nu} \odif{V} ={}& -\frac{1}{4\pi}\lim_{R\to 0^+}\int_{\rho=R}K^{Q_{\Ric}}_\alpha[h^{\S1}]\odif{\Sigma}^\alpha,
\end{align}
so that
\begin{equation}
    \TeffTR^2_{\mu\nu} = \TeffTR^{\S\R}_{\mu\nu} + \TeffTR^{\delta m}_{\mu\nu} + \TeffTR^{\text{Ric}}_{\mu\nu},
\end{equation}
where
\begin{align}
    \TeffTR^{\S\R}_{\mu\nu} ={}& 0, \label{eq:TeffTR_SR} \\
    \TeffTR^{\delta m}_{\mu\nu} ={}& \frac{1}{4}\int_{\gamma}\delta m_{\mu\nu}\delta^{4}(x,z)\odif{\tau}, \label{eq:TeffTR_dm} \\
    \TeffTR^{\text{Ric}}_{\mu\nu} ={}& \frac{m}{6}\int_\gamma(2h^{\R1}_{\mu\nu} - (2g_{\mu\nu} + 3u_{\mu}u_{\nu})h^{\R1}_{\gamma}{}^{\gamma} \nonumber \\
        & + 6 h^{\R1}_{\mu\gamma}u_{\nu}u^{\gamma} - 3g_{\mu\nu}h^{\R1}_{\alpha\beta}u^{\alpha}u^{\beta})\delta^{4}(x,z)\odif{\tau}.\label{eq:TeffTR_Q}
\end{align}
Here $\delta m_{\mu\nu}$ is given by Eq.~\eqref{eq:dm_sc}.
Equation~\eqref{eq:TeffTR_dm} does not appear here for the first time though.
It was previously defined by one of us in Ref.~\cite{Pound:2014xva} when discussing the split of the second-order singular field.
There, $h^{\delta m}_{\mu\nu} = \delta m_{\mu\nu}/\rho + \order(\rho^0)$ is defined as being a particular piece of the second-order singular field that satisfies the point-particle-like wave equation
\begin{equation}
    E_{\mu\nu}[h^{\delta m}] = -16\pi\bar{T}^{\delta m}_{\mu\nu},
\end{equation}
where $\bar{T}^{\delta m}_{\mu\nu}$ is given by Eq.~\eqref{eq:TeffTR_dm}.
Recovering this expression is an important check on our analysis.

\subsection{Effective metric trace-reversal}

The trace reversal of the Detweiler stress-energy with respect to the effective metric is given by
\begin{align}
    \TeffTR_{\mu\nu} ={}& \Teff_{\mu\nu} - \frac{1}{2}\geff_{\mu\nu}\geff^{\alpha\beta}\Teff_{\alpha\beta} + \order(\e^3) \nonumber \\
        ={}& \e T^1_{\mu\nu} + \e^2 T^2_{\mu\nu} - \frac{1}{2}(g_{\mu\nu} + \e h^{\R1}_{\mu\nu})(g^{\alpha\beta} - \e h_{\R1}^{\alpha\beta}) \nonumber \\
        & \times (\e T^1_{\alpha\beta} + \e^2 T^{2}_{\alpha\beta}) + \order(\e^3) \nonumber \\
        ={}& \e\barT^1_{\mu\nu} + \e^{2}\Bigl[\Bigl(g_\mu^{\ \alpha}g_\nu^{\ \beta} - \frac{1}{2}g_{\mu\nu}g^{\alpha\beta}\Bigr)T^2_{\alpha\beta} \nonumber\\
        & + \frac{1}{2}(g_{\mu\nu}h^{\alpha\beta}_{\R1}-g^{\alpha\beta}h^{\R1}_{\mu\nu})T^1_{\alpha\beta}\Bigr] + \order(\e^3). \label{eq:TeffTR}
\end{align}
Through some simple algebra, we can immediately show that this is equivalent to the expressions found via distributional analysis in Eqs.~\eqref{eq:TeffTR_1}--\eqref{eq:TeffTR_2}.
As expected, at first order we recover the trace reversal of the point-mass stress-energy tensor as in Eq.~\eqref{eq:TeffTR_1}.
We then find two terms at second order: the trace reversal of the Detweiler stress-energy tensor from Eq.~\eqref{eq:T_def} with respect to the \emph{background metric} and an additional term consisting of products of the first-order regular field and the first-order stress-energy tensor.

\subsection{Multiscale stress-energy tensor}\label{sec:multiscale T}

The derivation above has been based on the self-consistent scheme.
To obtain the stress-energy tensor in the multiscale scheme, we must account for the expansions of the worldline from Eq.~\eqref{z expansion} and the four-velocity from Eq.~\eqref{eq:u0def}.

After substituting in the expansion of the worldine and the four-velocity into Eq.~\eqref{eq:TeffTR}, we find at order $\e$,
\begin{align}
    \mathring{\barT}^1_{\mu\nu} = \frac{m}{2}\int_{\gamma_0}(g_{\mu\nu} + 2\mathring{u}_\mu^0\mathring{u}_\nu^0)\delta^{4}(x,z_0)\odif{\tau_0}.
\end{align}

Moving to terms at order $\e^2$, we start by looking at the term that appears as a result of expanding the four-velocity.
After substituting Eq.~\eqref{eq:u0def} into Eq.~\eqref{eq:TeffTR_1}, setting $z^\mu\to z^\mu_0$ and looking at order-$\e^2$ terms, we find
\begin{equation}
    \mathring{\barT}^{\ms}_{\mu\nu} = 2m\int_{\gamma_0} \mathring{u}^0_{(\mu}v_{\nu)}\delta^{4}(x,z_0)\odif{\tau_0}, \label{eq:Tms}
\end{equation}
which we have so-named as it acts as one of the sources for the $h^{\ms}_{\mu\nu}$ metric perturbation from Eq.~\eqref{eq:hms}.

The expansion of the worldline from Eq.~\eqref{z expansion} in the stress-energy tensor is slightly more subtle but has previously been calculated by one of us in Ref.~\cite{Pound:2015fma} and is given by
\begin{align}
    \barT^{z_1}_{\mu\nu} ={}& m\Bigl(g_{\mu\alpha}g_{\nu\beta}-\frac{1}{2}g_{\mu\nu}g_{\alpha\beta}\Bigr) \nonumber \\
        & \times \int_{\gamma_0}g^{\alpha}_{\alpha'}g^{\beta}_{\beta'}\bigl[2\mathring{u}_0^{(\alpha'}\dot{\mathring{z}}_{1\perp}^{\beta')}\delta^{4}(x,z_0) \nonumber \\
        & - \mathring{u}_0^{\alpha'}\mathring{u}_0^{\beta'}z^{\rho'}_{1\perp}g^\rho_{\rho'}\nabla_{\rho}\delta^{4}(x,z_0)\bigr]\odif{\tau_0},
\end{align}
where $g^\alpha_{\alpha'}$ is the parallel propagator.
This features two terms: the first accounts for the mass monopole correction absorbed into $\mathring{\delta m}_{\mu\nu}$ discussed in Sec.~\ref{sec:field eqs}, and the second sources the $\hring^{\delta z}_{\mu\nu}$ puncture.
As we did with the redefinition of $\hring^{\delta m}_{\mu\nu}$, we alter the definition of $\barT^{\delta m}_{\mu\nu}$ to account for this new term.
Explicitly, this is given by
\begin{equation}
    \mathring{\barT}^{\delta m}_{\mu\nu} = \frac{1}{4}\int_{\gamma_0}\mathring{\delta m}_{\mu\nu}\delta^{4}(x,z_0)\odif{\tau_0}, \label{eq:Tdm_ms}
\end{equation}
where $\mathring{\delta m}_{\mu\nu}$ is given by Eq.~\eqref{eq:dm_ms}.
The second term we then use to define
\begin{align}
    \mathring{\barT}^{\delta z}_{\mu\nu} ={}& -m\Bigl(g_{\mu\alpha}g_{\nu\beta}-\frac{1}{2}g_{\mu\nu}g_{\alpha\beta}\Bigr) \nonumber \\
        & \times \int_{\gamma_0}g^{\alpha}_{\alpha'}g^{\beta}_{\beta'}\mathring{u}_0^{\alpha'}\mathring{u}_0^{\beta'}z^{\rho'}_{1\perp}g^\rho_{\rho'}\nabla_{\rho}\delta^{4}(x,z_0)\odif{\tau_0}. \label{eq:Tdz}
\end{align}

Finally, we define
\begin{equation}
    \mathring{\barT}^{\Ric}_{\mu\nu} = \barT^{\Ric}_{\mu\nu}|_{\gamma\to\gamma_0}, \label{eq:TRic_ms}
\end{equation}
as we did in the definitions of $\hring^{\S\S}_{\mu\nu}$ and $\hring^{\S\R}_{\mu\nu}$ in Eq.~\eqref{eq:hS2_split}.
The order $\e^2$ piece of the multiscale Detweiler stress-energy tensor is then given by
\begin{equation}
    \mathring{\barT}^2_{\mu\nu} = \mathring{\barT}^{\delta m}_{\mu\nu} + \mathring{\barT}^{\Ric}_{\mu\nu} + \mathring{\barT}^{\delta z}_{\mu\nu} + \mathring{\barT}^{\ms}_{\mu\nu}. \label{eq:TDet_ms}
\end{equation}

\section{BLS operators without gauge damping, \texorpdfstring{\(E^{0}_{ij\ell m}\)}{E⁰{\textunderscore}ijℓm} and \texorpdfstring{\(E^{1}_{ij\ell m}\)}{E¹{\textunderscore}ijℓm}}\label{sec:E1Form}

When testing that the punctures satisfy the correct field equations in Sec.~\ref{sec:validation}, we make use of the Lorenz-gauge wave operator without gauge damping, $E_{ij\ell m}$.
This operator is expanded in multiscale form by applying the chain rule from Eq.~\eqref{eq:chain rule},  so that
\begin{equation}
     E_{ij\ell m} = E^0_{ij\ell m} + \e E^1_{ij\ell m} + \order(\e^2).
 \end{equation}

The leading-order term, $E^0_{ij\ell m}$, is the standard  Lorenz-gauge wave operator in mode-decomposed form. It can be extracted from, e.g.,\ the \textsc{PerturbationEquations} package~\cite{PerturbationEquations} (applying the appropriate rescaling discussed in footnote~\ref{fn:PertEqScal}).
Because it has never been presented with our specific conventions, for completeness we include it here. It is given by Eq.~\eqref{eq:E0 def}, with the coupling operators given by
\begingroup%
\allowdisplaybreaks%
\begin{subequations}
\begin{align}
    {\cal M}^0_{1j}h_j ={}& \frac{1}{2r^2}\bigl[2M\partial_{r^*}h_{1} + f^2(h_{1}-h_{5}) \nonumber \\*
        & - f^3(h_{3}+h_6) + 2iM \omega_m(h_2 + H h_1)\bigr], \\
    {\cal M}^0_{2j}h_j ={}& \frac{1}{2r^2}\bigl[2M\partial_{r^*}h_2 + f^2(h_2-h_4) \nonumber \\*
        & + 2iM \omega_m(h_1 + H h_2)\bigr], \\
    {\cal M}^0_{(3/6)j}h_j ={}& \frac{f}{2r^3}\bigl[r(h_5-h_1) + (r-4M)(h_3+h_6)\bigr], \\
    {\cal M}^0_{4j}h_j ={}& \frac{1}{2r^3}\bigl[M r\partial_{r^*}h_4 - f\bigl(3Mh_4+\ell(\ell+1)rh_2\bigr) \nonumber \\*
        & + iM \omega_m r(h_5+Hh_4)\bigr],  \\
    {\cal M}^0_{5j}h_j ={}& \frac{1}{2r^3}\bigl[Mr\partial_{r^*}h_5+ iM \omega_m r(h_4+Hh_5) \nonumber \\*
        & +f\bigl((2r-7M)h_5-\ell(\ell+1)rh_1\bigr) \nonumber \\*
        & + f^2r\bigl(\ell(\ell+1)(h_3+h_6)-h_7\bigr) \bigr],  \\
    {\cal M}^0_{7j}h_j ={}& -\frac{f}{2r^2}\bigl[(\ell-1)(\ell+2)h_5 + h_7\bigr], \label{eq:M07} \\
    {\cal M}^0_{8j}h_j ={}& \frac{M}{2r^3}\bigl[(r\partial_{r^*} - 3f)h_8 + i \omega_m r(h_9+Hh_8)\bigr], \\
    {\cal M}^0_{9j}h_j ={}& \frac{1}{2r^3}\bigl[(Mr\partial_{r^*} + f(2r-7M))h_9 - f^2rh_{10} \nonumber \\*
        & +iM \omega_m r(h_8+Hh_9)\bigr], \\
    {\cal M}^0_{10j}h_j ={}& \text{Eq.~\eqref{eq:M07} with } 5 \leftrightarrow 9 \text{ and } 7 \to 10
\end{align}
\end{subequations}%
\endgroup%
where $\omega_m=m\Omega$.

The first subleading term is 
\beq\label{E1ijlm}
E^1_{ij\ell m} h_{j\ell m} := \Box^1_{\ell m} h_{i\ell m} + {\cal M}_{ij}^1 h_{j\ell m},
\eeq
with $\Box^1_{\ell m}$ given by Eq.~\eqref{tildeBox1}. The coupling operators are given by
\begin{subequations}
\begin{align}
    {\cal M}^1_{1j}h_j ={}& -\frac{M}{r^2}(\pdvel h_2 + H\pdvel h_1), \label{eq:M11} \\
    {\cal M}^1_{2j}h_j ={}& \text{Eq.~\eqref{eq:M11} with } 1 \leftrightarrow 2, \\
    {\cal M}^1_{(3/6/7/10)j}h_j ={}& 0, \\
    {\cal M}^1_{4j}h_j ={}& -\frac{M}{2r^2}(\pdvel h_5 + H\pdvel h_4), \label{eq:M14} \\
    {\cal M}^1_{5j}h_j ={}& \text{Eq.~\eqref{eq:M14} with } 4 \leftrightarrow 5, \\
    {\cal M}^1_{8j}h_j ={}& \text{Eq.~\eqref{eq:M14} with } 4 \to 8 \text{ and } 5 \to 9, \\
    {\cal M}^1_{9j}h_j ={}& \text{Eq.~\eqref{eq:M14} with } 4 \to 9 \text{ and } 5 \to 8.
\end{align}
\end{subequations}

\bibliography{bibfile}

@article{Miller:2020bft,
    author = "Miller, Jeremy and Pound, Adam",
    title = "{Two-timescale evolution of extreme-mass-ratio inspirals: waveform generation scheme for quasicircular orbits in Schwarzschild spacetime}",
    eprint = "2006.11263",
    archivePrefix = "arXiv",
    primaryClass = "gr-qc",
    doi = "10.1103/PhysRevD.103.064048",
    journal = "Phys. Rev. D",
    volume = "103",
    number = "6",
    pages = "064048",
    year = "2021"
}

@article{Pound:2014xva,
    author = "Pound, Adam and Miller, Jeremy",
    title = "{Practical, covariant puncture for second-order self-force calculations}",
    eprint = "1403.1843",
    archivePrefix = "arXiv",
    primaryClass = "gr-qc",
    doi = "10.1103/PhysRevD.89.104020",
    journal = "Phys. Rev. D",
    volume = "89",
    number = "10",
    pages = "104020",
    year = "2014"
}

@misc{PerturbationEquations,
  author       = "Spiers, Andrew and Wardell, Barry and Pound, Adam and Upton, Samuel D. and Warburton, Niels",
  title        = "{PerturbationEquations}",
  month        = "may",
  year         = "2024",
  publisher    = "Zenodo",
  version      = "0.4.0",
  doi          = "10.5281/zenodo.11199718",
  url          = "https://doi.org/10.5281/zenodo.11199718"
}

@article{Spiers:2023mor,
    author = "Spiers, Andrew and Pound, Adam and Wardell, Barry",
    title = "{Second-order perturbations of the Schwarzschild spacetime: Practical, covariant, and gauge-invariant formalisms}",
    eprint = "2306.17847",
    archivePrefix = "arXiv",
    primaryClass = "gr-qc",
    doi = "10.1103/PhysRevD.110.064030",
    journal = "Phys. Rev. D",
    volume = "110",
    number = "6",
    pages = "064030",
    year = "2024"
}

@article{Upton:2021oxf,
    author = "Upton, Samuel D. and Pound, Adam",
    title = "{Second-order gravitational self-force in a highly regular gauge}",
    eprint = "2101.11409",
    archivePrefix = "arXiv",
    primaryClass = "gr-qc",
    doi = "10.1103/PhysRevD.103.124016",
    journal = "Phys. Rev. D",
    volume = "103",
    number = "12",
    pages = "124016",
    year = "2021"
}

@article{Pound:2015fma,
    author = "Pound, Adam",
    title = "{Gauge and motion in perturbation theory}",
    eprint = "1506.02894",
    archivePrefix = "arXiv",
    primaryClass = "gr-qc",
    doi = "10.1103/PhysRevD.92.044021",
    journal = "Phys. Rev. D",
    volume = "92",
    number = "4",
    pages = "044021",
    year = "2015"
}

@article{Barack:2002mha,
    author = "Barack, Leor and Ori, Amos",
    title = "{Regularization parameters for the selfforce in Schwarzschild space-time. 1. Scalar case}",
    eprint = "gr-qc/0204093",
    archivePrefix = "arXiv",
    doi = "10.1103/PhysRevD.66.084022",
    journal = "Phys. Rev. D",
    volume = "66",
    pages = "084022",
    year = "2002"
}

@article{Barack:2001gx,
    author = "Barack, Leor and Mino, Yasushi and Nakano, Hiroyuki and Ori, Amos and Sasaki, Misao",
    title = "{Calculating the gravitational selfforce in Schwarzschild space-time}",
    eprint = "gr-qc/0111001",
    archivePrefix = "arXiv",
    doi = "10.1103/PhysRevLett.88.091101",
    journal = "Phys. Rev. Lett.",
    volume = "88",
    pages = "091101",
    year = "2002"
}

@article{Detweiler:2002gi,
    author = "Detweiler, Steven L. and Messaritaki, Eirini and Whiting, Bernard F.",
    title = "{Selfforce of a scalar field for circular orbits about a Schwarzschild black hole}",
    eprint = "gr-qc/0205079",
    archivePrefix = "arXiv",
    doi = "10.1103/PhysRevD.67.104016",
    journal = "Phys. Rev. D",
    volume = "67",
    pages = "104016",
    year = "2003"
}

@article{Barack:2002bt,
    author = "Barack, Leor and Ori, Amos",
    title = "{Regularization parameters for the selfforce in Schwarzschild space-time. 2. Gravitational and electromagnetic cases}",
    eprint = "gr-qc/0209072",
    archivePrefix = "arXiv",
    doi = "10.1103/PhysRevD.67.024029",
    journal = "Phys. Rev. D",
    volume = "67",
    pages = "024029",
    year = "2003"
}

@article{Haas:2006ne,
    author = "Haas, Roland and Poisson, Eric",
    title = "{Mode-sum regularization of the scalar self-force: Formulation in terms of a tetrad decomposition of the singular field}",
    eprint = "gr-qc/0605077",
    archivePrefix = "arXiv",
    doi = "10.1103/PhysRevD.74.044009",
    journal = "Phys. Rev. D",
    volume = "74",
    pages = "044009",
    year = "2006"
}

@article{Heffernan:2012vj,
    author = "Heffernan, Anna and Ottewill, Adrian and Wardell, Barry",
    title = "{High-order expansions of the Detweiler-Whiting singular field in Kerr spacetime}",
    eprint = "1211.6446",
    archivePrefix = "arXiv",
    primaryClass = "gr-qc",
    doi = "10.1103/PhysRevD.89.024030",
    journal = "Phys. Rev. D",
    volume = "89",
    number = "2",
    pages = "024030",
    year = "2014"
}

@article{Heffernan:2012su,
    author = "Heffernan, Anna and Ottewill, Adrian and Wardell, Barry",
    title = "{High-order expansions of the Detweiler-Whiting singular field in Schwarzschild spacetime}",
    eprint = "1204.0794",
    archivePrefix = "arXiv",
    primaryClass = "gr-qc",
    doi = "10.1103/PhysRevD.86.104023",
    journal = "Phys. Rev. D",
    volume = "86",
    pages = "104023",
    year = "2012"
}

@Book{Wigner1931,
  author    = {Wigner, Eugen},
  publisher = {Vieweg+Teubner Verlag},
  title     = {Gruppentheorie und ihre Anwendung auf die Quantenmechanik der Atomspektren},
  year      = {1931},
  isbn      = {9783663025559},
  doi       = {10.1007/978-3-663-02555-9},
  address   = {Braunschweig},
}

@article{Ottewill:2008uu,
    author = "Ottewill, Adrian C. and Wardell, Barry",
    title = "{Quasi-local contribution to the scalar self-force: Non-geodesic Motion}",
    eprint = "0810.1961",
    archivePrefix = "arXiv",
    primaryClass = "gr-qc",
    doi = "10.1103/PhysRevD.79.024031",
    journal = "Phys. Rev. D",
    volume = "79",
    pages = "024031",
    year = "2009"
}

@article{Upton:2023tcv,
    author = "Upton, Samuel D.",
    title = "{Second-order gravitational self-force in a highly regular gauge: Covariant and coordinate punctures}",
    eprint = "2309.03778",
    archivePrefix = "arXiv",
    primaryClass = "gr-qc",
    doi = "10.1103/PhysRevD.109.044021",
    journal = "Phys. Rev. D",
    volume = "109",
    number = "4",
    pages = "044021",
    year = "2024"
}

@article{Wardell:2015ada,
    author = "Wardell, Barry and Warburton, Niels",
    title = "{Applying the effective-source approach to frequency-domain self-force calculations: Lorenz-gauge gravitational perturbations}",
    eprint = "1505.07841",
    archivePrefix = "arXiv",
    primaryClass = "gr-qc",
    doi = "10.1103/PhysRevD.92.084019",
    journal = "Phys. Rev. D",
    volume = "92",
    number = "8",
    pages = "084019",
    year = "2015"
}

@article{Durkan:2022fvm,
    author = "Durkan, Leanne and Warburton, Niels",
    title = "{Slow evolution of the metric perturbation due to a quasicircular inspiral into a Schwarzschild black hole}",
    eprint = "2206.08179",
    archivePrefix = "arXiv",
    primaryClass = "gr-qc",
    doi = "10.1103/PhysRevD.106.084023",
    journal = "Phys. Rev. D",
    volume = "106",
    number = "8",
    pages = "084023",
    year = "2022"
}

@article{Akcay:2013wfa,
    author = "Akcay, Sarp and Warburton, Niels and Barack, Leor",
    title = "{Frequency-domain algorithm for the Lorenz-gauge gravitational self-force}",
    eprint = "1308.5223",
    archivePrefix = "arXiv",
    primaryClass = "gr-qc",
    doi = "10.1103/PhysRevD.88.104009",
    journal = "Phys. Rev. D",
    volume = "88",
    number = "10",
    pages = "104009",
    year = "2013"
}

@misc{h1Lorenz,
    author = "Warburton, Niels and Akcay, Sarp",
    title = "{h1Lorenz}",
    url = "https://github.com/BlackHolePerturbationToolkit/h1Lorenz",
    year = "2023"
}

@misc{SecondOrderRicci,
    author = "Barry Wardell",
    title = "{SecondOrderRicci}",
    url = "https://github.com/BlackHolePerturbationToolkit/SecondOrderRicci",
    year = "2023"
}

@misc{SecondOrderRicciSS,
    author = "Barry Wardell",
    title = "{SecondOrderRicciSS}",
    url = "https://github.com/BlackHolePerturbationToolkit/SecondOrderRicciSS",
    year = "2025"
}

@misc{PuncturesRepository,
    author = {Miller, Jeremy and Pound, Adam and Upton, Samuel D. and Wardell, Barry},
    license = {MIT},
    title = {{Punctures}},
    url = {https://github.com/BlackHolePerturbationToolkit/Punctures},
    year = "2025"
}

@article{Barack:2005nr,
    author = "Barack, Leor and Lousto, Carlos O.",
    title = "{Perturbations of Schwarzschild black holes in the Lorenz gauge: Formulation and numerical implementation}",
    eprint = "gr-qc/0510019",
    archivePrefix = "arXiv",
    doi = "10.1103/PhysRevD.72.104026",
    journal = "Phys. Rev. D",
    volume = "72",
    pages = "104026",
    year = "2005"
}

@article{Barack:2007tm,
    author = "Barack, Leor and Sago, Norichika",
    title = "{Gravitational self force on a particle in circular orbit around a Schwarzschild black hole}",
    eprint = "gr-qc/0701069",
    archivePrefix = "arXiv",
    doi = "10.1103/PhysRevD.75.064021",
    journal = "Phys. Rev. D",
    volume = "75",
    pages = "064021",
    year = "2007"
}

@article{Warburton:2013lea,
    author = "Warburton, Niels and Wardell, Barry",
    title = "{Applying the effective-source approach to frequency-domain self-force calculations}",
    eprint = "1311.3104",
    archivePrefix = "arXiv",
    primaryClass = "gr-qc",
    doi = "10.1103/PhysRevD.89.044046",
    journal = "Phys. Rev. D",
    volume = "89",
    number = "4",
    pages = "044046",
    year = "2014"
}

@article{Gralla:2012db,
    author = "Gralla, Samuel E.",
    title = "{Second Order Gravitational Self Force}",
    eprint = "1203.3189",
    archivePrefix = "arXiv",
    primaryClass = "gr-qc",
    reportNumber = "UMD-94-11",
    doi = "10.1103/PhysRevD.85.124011",
    journal = "Phys. Rev. D",
    volume = "85",
    pages = "124011",
    year = "2012"
}

@article{Gralla:2008fg,
    author = "Gralla, Samuel E. and Wald, Robert M.",
    title = "{A Rigorous Derivation of Gravitational Self-force}",
    eprint = "0806.3293",
    archivePrefix = "arXiv",
    primaryClass = "gr-qc",
    doi = "10.1088/0264-9381/25/20/205009",
    journal = "Class. Quant. Grav.",
    volume = "25",
    pages = "205009",
    year = "2008",
}

@article{Miller:2023ers,
    author = "Miller, Jeremy and Leather, Benjamin and Pound, Adam and Warburton, Niels",
    title = "{Worldtube puncture scheme for first- and second-order self-force calculations in the Fourier domain}",
    eprint = "2401.00455",
    archivePrefix = "arXiv",
    primaryClass = "gr-qc",
    doi = "10.1103/PhysRevD.109.104010",
    journal = "Phys. Rev. D",
    volume = "109",
    number = "10",
    pages = "104010",
    year = "2024"
}

@article{Detweiler:2011tt,
    author = "Detweiler, Steven",
    title = "{Gravitational radiation reaction and second order perturbation theory}",
    eprint = "1107.2098",
    archivePrefix = "arXiv",
    primaryClass = "gr-qc",
    doi = "10.1103/PhysRevD.85.044048",
    journal = "Phys. Rev. D",
    volume = "85",
    pages = "044048",
    year = "2012"
}

@article{Pound:2012nt,
    author = "Pound, Adam",
    title = "{Second-order gravitational self-force}",
    eprint = "1201.5089",
    archivePrefix = "arXiv",
    primaryClass = "gr-qc",
    doi = "10.1103/PhysRevLett.109.051101",
    journal = "Phys. Rev. Lett.",
    volume = "109",
    pages = "051101",
    year = "2012"
}

@article{Pound:2009sm,
    author = "Pound, Adam",
    title = "{Self-consistent gravitational self-force}",
    eprint = "0907.5197",
    archivePrefix = "arXiv",
    primaryClass = "gr-qc",
    doi = "10.1103/PhysRevD.81.024023",
    journal = "Phys. Rev. D",
    volume = "81",
    pages = "024023",
    year = "2010"
}

@article{Pound:2012dk,
    author = "Pound, Adam",
    title = "{Nonlinear gravitational self-force. I. Field outside a small body}",
    eprint = "1206.6538",
    archivePrefix = "arXiv",
    primaryClass = "gr-qc",
    doi = "10.1103/PhysRevD.86.084019",
    journal = "Phys. Rev. D",
    volume = "86",
    pages = "084019",
    year = "2012"
}

@incollection{Pound:2015tma,
    author = "Pound, Adam",
    editor = {Puetzfeld, Dirk and L\"ammerzahl, Claus and Schutz, Bernard},
    booktitle = "{Equations of Motion in Relativistic Gravity}",
    title = "{Motion of small objects in curved spacetimes: An introduction to gravitational self-force}",
    eprint = "1506.06245",
    archivePrefix = "arXiv",
    primaryClass = "gr-qc",
    doi = "10.1007/978-3-319-18335-0_13",
    series = "Fundamental Theories of Physics",
    volume = "179",
    pages = "399--486",
    year = "2015",
    publisher = "Springer"
}

@article{Miller:2016hjv,
    author = "Miller, Jeremy and Wardell, Barry and Pound, Adam",
    title = "{Second-order perturbation theory: the problem of infinite mode coupling}",
    eprint = "1608.06783",
    archivePrefix = "arXiv",
    primaryClass = "gr-qc",
    doi = "10.1103/PhysRevD.94.104018",
    journal = "Phys. Rev. D",
    volume = "94",
    number = "10",
    pages = "104018",
    year = "2016"
}

@article{Barack:2018yvs,
    author = "Barack, Leor and Pound, Adam",
    title = "{Self-force and radiation reaction in general relativity}",
    eprint = "1805.10385",
    archivePrefix = "arXiv",
    primaryClass = "gr-qc",
    doi = "10.1088/1361-6633/aae552",
    journal = "Rept. Prog. Phys.",
    volume = "82",
    number = "1",
    pages = "016904",
    year = "2019"
}

@article{Abac:2025saz,
    author = "Abac, Adrian and others",
    title = "{The Science of the Einstein Telescope}",
    eprint = "2503.12263",
    archivePrefix = "arXiv",
    primaryClass = "gr-qc",
    reportNumber = "ET-0036C-25",
    month = "3",
    year = "2025",
    journal = ""
}

@article{LISAConsortiumWaveformWorkingGroup:2023arg,
    author = "Afshordi, Niayesh and others",
    collaboration = "LISA Consortium Waveform Working Group",
    title = "{Waveform modelling for the Laser Interferometer Space Antenna}",
    eprint = "2311.01300",
    archivePrefix = "arXiv",
    primaryClass = "gr-qc",
    doi = "10.1007/s41114-025-00056-1",
    journal = "Living Rev. Rel.",
    volume = "28",
    number = "1",
    pages = "9",
    year = "2025"
}

@article{Albertini:2022rfe,
    author = "Albertini, Angelica and Nagar, Alessandro and Pound, Adam and Warburton, Niels and Wardell, Barry and Durkan, Leanne and Miller, Jeremy",
    title = "{Comparing second-order gravitational self-force, numerical relativity, and effective one body waveforms from inspiralling, quasicircular, and nonspinning black hole binaries}",
    eprint = "2208.01049",
    archivePrefix = "arXiv",
    primaryClass = "gr-qc",
    doi = "10.1103/PhysRevD.106.084061",
    journal = "Phys. Rev. D",
    volume = "106",
    number = "8",
    pages = "084061",
    year = "2022"
}

@article{Nasipak:2023kuf,
    author = "Nasipak, Zachary",
    title = "{Adiabatic gravitational waveform model for compact objects undergoing quasicircular inspirals into rotating massive black holes}",
    eprint = "2310.19706",
    archivePrefix = "arXiv",
    primaryClass = "gr-qc",
    doi = "10.1103/PhysRevD.109.044020",
    journal = "Phys. Rev. D",
    volume = "109",
    number = "4",
    pages = "044020",
    year = "2024"
}

@article{Islam:2022laz,
    author = "Islam, Tousif and Field, Scott E. and Hughes, Scott A. and Khanna, Gaurav and Varma, Vijay and Giesler, Matthew and Scheel, Mark A. and Kidder, Lawrence E. and Pfeiffer, Harald P.",
    title = "{Surrogate model for gravitational wave signals from nonspinning, comparable-to large-mass-ratio black hole binaries built on black hole perturbation theory waveforms calibrated to numerical relativity}",
    eprint = "2204.01972",
    archivePrefix = "arXiv",
    primaryClass = "gr-qc",
    doi = "10.1103/PhysRevD.106.104025",
    journal = "Phys. Rev. D",
    volume = "106",
    number = "10",
    pages = "104025",
    year = "2022"
}

@article{Katz:2021yft,
    author = "Katz, Michael L. and Chua, Alvin J. K. and Speri, Lorenzo and Warburton, Niels and Hughes, Scott A.",
    title = "{Fast extreme-mass-ratio-inspiral waveforms: New tools for millihertz gravitational-wave data analysis}",
    eprint = "2104.04582",
    archivePrefix = "arXiv",
    primaryClass = "gr-qc",
    doi = "10.1103/PhysRevD.104.064047",
    journal = "Phys. Rev. D",
    volume = "104",
    number = "6",
    pages = "064047",
    year = "2021"
}

@article{Burke:2023lno,
    author = "Burke, Ollie and Piovano, Gabriel Andres and Warburton, Niels and Lynch, Philip and Speri, Lorenzo and Kavanagh, Chris and Wardell, Barry and Pound, Adam and Durkan, Leanne and Miller, Jeremy",
    title = "{Assessing the importance of first postadiabatic terms for small-mass-ratio binaries}",
    eprint = "2310.08927",
    archivePrefix = "arXiv",
    primaryClass = "gr-qc",
    doi = "10.1103/PhysRevD.109.124048",
    journal = "Phys. Rev. D",
    volume = "109",
    number = "12",
    pages = "124048",
    year = "2024"
}

@article{Pound:2017psq,
    author = "Pound, Adam",
    title = "{Nonlinear gravitational self-force: second-order equation of motion}",
    eprint = "1703.02836",
    archivePrefix = "arXiv",
    primaryClass = "gr-qc",
    doi = "10.1103/PhysRevD.95.104056",
    journal = "Phys. Rev. D",
    volume = "95",
    number = "10",
    pages = "104056",
    year = "2017"
}

@article{Pound:2015wva,
    author = "Pound, Adam",
    title = "{Second-order perturbation theory: problems on large scales}",
    eprint = "1510.05172",
    archivePrefix = "arXiv",
    primaryClass = "gr-qc",
    doi = "10.1103/PhysRevD.92.104047",
    journal = "Phys. Rev. D",
    volume = "92",
    number = "10",
    pages = "104047",
    year = "2015"
}

@article{Rosenthal:2006iy,
    author = "Rosenthal, Eran",
    title = "{Second-order gravitational self-force}",
    eprint = "gr-qc/0609069",
    archivePrefix = "arXiv",
    doi = "10.1103/PhysRevD.74.084018",
    journal = "Phys. Rev. D",
    volume = "74",
    pages = "084018",
    year = "2006"
}

@article{Hinderer:2008dm,
    author = "Hinderer, Tanja and Flanagan, Eanna E.",
    title = "{Two timescale analysis of extreme mass ratio inspirals in Kerr. I. Orbital Motion}",
    eprint = "0805.3337",
    archivePrefix = "arXiv",
    primaryClass = "gr-qc",
    doi = "10.1103/PhysRevD.78.064028",
    journal = "Phys. Rev. D",
    volume = "78",
    pages = "064028",
    year = "2008"
}

@article{Harte:2025tmd,
    author = "Harte, Abraham I. and Blanco, Francisco M. and Flanagan, Eanna E.",
    title = "{Nonlinearly Self-Interacting Extended Bodies Move as Test Bodies in Effective External Fields}",
    eprint = "2504.11912",
    archivePrefix = "arXiv",
    primaryClass = "gr-qc",
    doi = "10.1103/k74d-dj7y",
    journal = "Phys. Rev. Lett.",
    volume = "135",
    number = "15",
    pages = "151401",
    year = "2025"
}

@article{Harte:2011ku,
    author = "Harte, Abraham I.",
    title = "{Mechanics of extended masses in general relativity}",
    eprint = "1103.0543",
    archivePrefix = "arXiv",
    primaryClass = "gr-qc",
    doi = "10.1088/0264-9381/29/5/055012",
    journal = "Class. Quant. Grav.",
    volume = "29",
    pages = "055012",
    year = "2012"
}

@article{Cunningham:2024dog,
    author = "Cunningham, Kevin and Kavanagh, Chris and Pound, Adam and Trestini, David and Warburton, Niels and Neef, Jakob",
    title = "{Gravitational memory: new results from post-Newtonian and self-force theory}",
    eprint = "2410.23950",
    archivePrefix = "arXiv",
    primaryClass = "gr-qc",
    doi = "10.1088/1361-6382/adbc3d",
    journal = "Class. Quant. Grav.",
    volume = "42",
    number = "13",
    pages = "135009",
    year = "2025"
}

@article{Ramos-Buades:2022lgf,
    author = {Ramos-Buades, Antoni and van de Meent, Maarten and Pfeiffer, Harald P. and R\"uter, Hannes R. and Scheel, Mark A. and Boyle, Michael and Kidder, Lawrence E.},
    title = "{Eccentric binary black holes: Comparing numerical relativity and small mass-ratio perturbation theory}",
    eprint = "2209.03390",
    archivePrefix = "arXiv",
    primaryClass = "gr-qc",
    doi = "10.1103/PhysRevD.106.124040",
    journal = "Phys. Rev. D",
    volume = "106",
    number = "12",
    pages = "124040",
    year = "2022"
}

@article{vandeMeent:2020xgc,
    author = "van de Meent, Maarten and Pfeiffer, Harald P.",
    title = "{Intermediate mass-ratio black hole binaries: Applicability of small mass-ratio perturbation theory}",
    eprint = "2006.12036",
    archivePrefix = "arXiv",
    primaryClass = "gr-qc",
    doi = "10.1103/PhysRevLett.125.181101",
    journal = "Phys. Rev. Lett.",
    volume = "125",
    number = "18",
    pages = "181101",
    year = "2020"
}

@article{LeTiec:2014oez,
    author = "Le Tiec, Alexandre",
    title = "{The Overlap of Numerical Relativity, Perturbation Theory and Post-Newtonian Theory in the Binary Black Hole Problem}",
    eprint = "1408.5505",
    archivePrefix = "arXiv",
    primaryClass = "gr-qc",
    doi = "10.1142/S0218271814300225",
    journal = "Int. J. Mod. Phys. D",
    volume = "23",
    number = "10",
    pages = "1430022",
    year = "2014"
}

@article{Wardell:2021fyy,
    author = "Wardell, Barry and Pound, Adam and Warburton, Niels and Miller, Jeremy and Durkan, Leanne and Le Tiec, Alexandre",
    title = "{Gravitational Waveforms for Compact Binaries from Second-Order Self-Force Theory}",
    eprint = "2112.12265",
    archivePrefix = "arXiv",
    primaryClass = "gr-qc",
    doi = "10.1103/PhysRevLett.130.241402",
    journal = "Phys. Rev. Lett.",
    volume = "130",
    number = "24",
    pages = "241402",
    year = "2023"
}

@article{Pound:2019lzj,
    author = "Pound, Adam and Wardell, Barry and Warburton, Niels and Miller, Jeremy",
    title = "{Second-Order Self-Force Calculation of Gravitational Binding Energy in Compact Binaries}",
    eprint = "1908.07419",
    archivePrefix = "arXiv",
    primaryClass = "gr-qc",
    doi = "10.1103/PhysRevLett.124.021101",
    journal = "Phys. Rev. Lett.",
    volume = "124",
    number = "2",
    pages = "021101",
    year = "2020"
}

@article{Warburton:2021kwk,
    author = "Warburton, Niels and Pound, Adam and Wardell, Barry and Miller, Jeremy and Durkan, Leanne",
    title = "{Gravitational-Wave Energy Flux for Compact Binaries through Second Order in the Mass Ratio}",
    eprint = "2107.01298",
    archivePrefix = "arXiv",
    primaryClass = "gr-qc",
    doi = "10.1103/PhysRevLett.127.151102",
    journal = "Phys. Rev. Lett.",
    volume = "127",
    number = "15",
    pages = "151102",
    year = "2021"
}

@article{Mathews:2025nyb,
    author = "Mathews, Josh and Pound, Adam",
    title = "{Post-adiabatic waveform-generation framework for asymmetric precessing binaries}",
    eprint = "2501.01413",
    archivePrefix = "arXiv",
    primaryClass = "gr-qc",
    month = "1",
    year = "2025",
    journal = ""
}

@article{Wei:2025lva,
    author = "Wei, Yi-Xiang and Zhu, Xian-Long and Zhang, Jian-dong and Mei, Jianwei",
    title = "{Toward second-order self-force for eccentric extreme-mass ratio inspirals in Schwarzschild spacetimes}",
    eprint = "2504.09640",
    archivePrefix = "arXiv",
    primaryClass = "gr-qc",
    doi = "10.1103/42fh-sw5h",
    journal = "Phys. Rev. D",
    volume = "112",
    number = "6",
    pages = "064048",
    year = "2025"
}

@article{Hughes:2021exa,
    author = "Hughes, Scott A. and Warburton, Niels and Khanna, Gaurav and Chua, Alvin J. K. and Katz, Michael L.",
    title = "{Adiabatic waveforms for extreme mass-ratio inspirals via multivoice decomposition in time and frequency}",
    eprint = "2102.02713",
    archivePrefix = "arXiv",
    primaryClass = "gr-qc",
    doi = "10.1103/PhysRevD.103.104014",
    journal = "Phys. Rev. D",
    volume = "103",
    number = "10",
    pages = "104014",
    year = "2021",
}

@InCollection{Pound:2021qin,
  author        = {Adam Pound and Barry Wardell},
  booktitle     = {Handbook of Gravitational Wave Astronomy},
  publisher     = {Springer},
  title         = {Black hole perturbation theory and gravitational self-force},
  year          = {2022},
  address       = {Singapore},
  editor        = {Cosimo Bambi and Stavros Katsanevas and Konstantinos D. Kokkotas},
  month         = mar,
  archiveprefix = {arXiv},
  doi           = {10.1007/978-981-15-4702-7_38-1},
  eprint        = {2101.04592},
  primaryclass  = {gr-qc},
}

@article{Poisson:2011nh,
    author = "Poisson, Eric and Pound, Adam and Vega, Ian",
    title = "{The Motion of point particles in curved spacetime}",
    eprint = "1102.0529",
    archivePrefix = "arXiv",
    primaryClass = "gr-qc",
    doi = "10.12942/lrr-2011-7",
    journal = "Living Rev. Rel.",
    volume = "14",
    pages = "7",
    year = "2011"
}

@article{Bonetto:2021exn,
    author = "Bonetto, Riccardo and Pound, Adam and Sam, Zeyd",
    title = "{Deformed Schwarzschild horizons in second-order perturbation theory: Mass, geometry, and teleology}",
    eprint = "2109.09514",
    archivePrefix = "arXiv",
    primaryClass = "gr-qc",
    doi = "10.1103/PhysRevD.105.024048",
    journal = "Phys. Rev. D",
    volume = "105",
    number = "2",
    pages = "024048",
    year = "2022"
}

@article{Lewis:2025ydo,
    author = "Lewis, Jack and Kakehi, Takafumi and Pound, Adam and Tanaka, Takahiro",
    title = "{Post-adiabatic dynamics and waveform generation in self-force theory: an invariant pseudo-Hamiltonian framework}",
    eprint = "2507.08081",
    archivePrefix = "arXiv",
    primaryClass = "gr-qc",
    month = "7",
    year = "2025",
    journal = ""
}

@article{Vega:2009qb,
    author = "Vega, Ian and Diener, Peter and Tichy, Wolfgang and Detweiler, Steven L.",
    title = "{Self-force with (3+1) codes: A Primer for numerical relativists}",
    eprint = "0908.2138",
    archivePrefix = "arXiv",
    primaryClass = "gr-qc",
    doi = "10.1103/PhysRevD.80.084021",
    journal = "Phys. Rev. D",
    volume = "80",
    pages = "084021",
    year = "2009"
}

@misc{NIST:DLMF,
    key = "{\relax DLMF}",
    title = "{\it NIST Digital Library of Mathematical Functions}",
    howpublished = "\url{https://dlmf.nist.gov/}, Release 1.2.4 of 2025-03-15",
    url = "https://dlmf.nist.gov/",
    note = "{F}.~W.~J. Olver, A.~B. {Olde Daalhuis}, D.~W. Lozier, B.~I. Schneider, R.~F. Boisvert, C.~W. Clark, B.~R. Miller, B.~V. Saunders, H.~S. Cohl, and M.~A. McClain, eds.",
    year = "2025"
}

@unpublished{Leather:2025InPrep,
    title = {Second-order {T}eukolsky calculations for nonspinning, quasicircular binaries},
    author = {Benjamin Leather and Andrew Spiers and Samuel D. Upton and Adam Pound and Niels Warburton and Barry Wardell},
    note = {in preparation}
}

@article{Trestini:2025nzr,
    author = "Trestini, David",
    title = "{Schott term in the binding energy for compact binaries on circular orbits at fourth post-Newtonian order}",
    eprint = "2504.13245",
    archivePrefix = "arXiv",
    primaryClass = "gr-qc",
    doi = "10.1103/lsbb-sv45",
    journal = "Phys. Rev. D",
    volume = "112",
    number = "2",
    pages = "024076",
    year = "2025"
}

@unpublished{Grant:2025InPrep,
    author = {Alexander M. Grant and Alexandre Le Tiec and Adam Pound},
    title = {Energy Balance and the First Law of Binary Black Hole Mechanics},
    note = {in preparation}
}

@article{PanossoMacedo:2024pox,
    author = "Panosso Macedo, Rodrigo and Bourg, Patrick and Pound, Adam and Upton, Samuel D.",
    title = "{Multidomain spectral method for self-force calculations}",
    eprint = "2404.10083",
    archivePrefix = "arXiv",
    primaryClass = "gr-qc",
    doi = "10.1103/PhysRevD.110.084008",
    journal = "Phys. Rev. D",
    volume = "110",
    number = "8",
    pages = "084008",
    year = "2024"
}

@article{Bourg:2024cgh,
    author = "Bourg, Patrick and Pound, Adam and Upton, Samuel D. and Panosso Macedo, Rodrigo",
    title = "{Simple, efficient method of calculating the Detweiler-Whiting singular field to very high order}",
    eprint = "2404.10082",
    archivePrefix = "arXiv",
    primaryClass = "gr-qc",
    doi = "10.1103/PhysRevD.110.084007",
    journal = "Phys. Rev. D",
    volume = "110",
    number = "8",
    pages = "084007",
    year = "2024"
}

@article{Osburn:2022bby,
    author = "Osburn, Thomas and Nishimura, Nami",
    title = "{New self-force method via elliptic partial differential equations for Kerr inspiral models}",
    eprint = "2206.07031",
    archivePrefix = "arXiv",
    primaryClass = "gr-qc",
    doi = "10.1103/PhysRevD.106.044056",
    journal = "Phys. Rev. D",
    volume = "106",
    number = "4",
    pages = "044056",
    year = "2022"
}

@article{Dolan:2023enf,
    author = "Dolan, Sam R. and Durkan, Leanne and Kavanagh, Chris and Wardell, Barry",
    title = "{Metric perturbations of Kerr spacetime in Lorenz gauge: circular equatorial orbits}",
    eprint = "2306.16459",
    archivePrefix = "arXiv",
    primaryClass = "gr-qc",
    doi = "10.1088/1361-6382/ad52e3",
    journal = "Class. Quant. Grav.",
    volume = "41",
    number = "15",
    pages = "155011",
    year = "2024"
}

@article{Spiers:2023cip,
    author = "Spiers, Andrew and Pound, Adam and Moxon, Jordan",
    title = "{Second-order Teukolsky formalism in Kerr spacetime: Formulation and nonlinear source}",
    eprint = "2305.19332",
    archivePrefix = "arXiv",
    primaryClass = "gr-qc",
    doi = "10.1103/PhysRevD.108.064002",
    journal = "Phys. Rev. D",
    volume = "108",
    number = "6",
    pages = "064002",
    year = "2023"
}

@article{Leather:2024mls,
    author = "Leather, Benjamin",
    title = "{Gravitational self-force with hyperboloidal slicing and spectral methods}",
    eprint = "2411.14976",
    archivePrefix = "arXiv",
    primaryClass = "gr-qc",
    doi = "10.1007/s10714-025-03443-9",
    journal = "Gen. Rel. Grav.",
    volume = "57",
    number = "7",
    pages = "112",
    year = "2025"
}

@article{PanossoMacedo:2022fdi,
    author = "Panosso Macedo, Rodrigo and Leather, Benjamin and Warburton, Niels and Wardell, Barry and Zengino\u{g}lu, An\i{}l",
    title = "{Hyperboloidal method for frequency-domain self-force calculations}",
    eprint = "2202.01794",
    archivePrefix = "arXiv",
    primaryClass = "gr-qc",
    doi = "10.1103/PhysRevD.105.104033",
    journal = "Phys. Rev. D",
    volume = "105",
    number = "10",
    pages = "104033",
    year = "2022"
}

@article{Barack:2007jh,
    author = "Barack, Leor and Golbourn, Darren A.",
    title = "{Scalar-field perturbations from a particle orbiting a black hole using numerical evolution in 2+1 dimensions}",
    eprint = "0705.3620",
    archivePrefix = "arXiv",
    primaryClass = "gr-qc",
    doi = "10.1103/PhysRevD.76.044020",
    journal = "Phys. Rev. D",
    volume = "76",
    pages = "044020",
    year = "2007"
}

@article{Vega:2007mc,
    author = "Vega, Ian and Detweiler, Steven L.",
    title = "{Regularization of fields for self-force problems in curved spacetime: Foundations and a time-domain application}",
    eprint = "0712.4405",
    archivePrefix = "arXiv",
    primaryClass = "gr-qc",
    doi = "10.1103/PhysRevD.77.084008",
    journal = "Phys. Rev. D",
    volume = "77",
    pages = "084008",
    year = "2008"
}

@phdthesis{Berndtson:2007gsc,
    author = "Berndtson, Mark V.",
    title = "{Harmonic gauge perturbations of the Schwarzschild metric}",
    eprint = "0904.0033",
    archivePrefix = "arXiv",
    primaryClass = "gr-qc",
    year = "2007",
    school = "University of Colorado, Boulder"
}

@article{Porto:2024cwd,
    author = "Porto, Rafael A. and Riva, Massimiliano M. and Yang, Zixin",
    title = "{Nonlinear gravitational radiation reaction: failed tail, memories {\&} squares}",
    eprint = "2409.05860",
    archivePrefix = "arXiv",
    primaryClass = "gr-qc",
    reportNumber = "DESY 24-133",
    doi = "10.1007/JHEP04(2025)050",
    journal = "JHEP",
    number = "4",
    pages = "50",
    year = "2025"
}

@article{Barack:1998bw,
    author = "Barack, Leor",
    title = "{Late time dynamics of scalar perturbations outside black holes. 2. Schwarzschild geometry}",
    eprint = "gr-qc/9811028",
    archivePrefix = "arXiv",
    doi = "10.1103/PhysRevD.59.044017",
    journal = "Phys. Rev. D",
    volume = "59",
    pages = "044017",
    year = "1999"
}

\end{document}